\def\PsfigVersion{1.10}
\def\setDriver{\DvipsDriver} 
\ifx\undefined\psfig\else\endinput\fi
%

\let\LaTeXAtSign=\@
\let\@=\relax
\edef\psfigRestoreAt{\catcode`\@=\number\catcode`@\relax}
\catcode`\@=11\relax
\newwrite\@unused
\def\ps@typeout#1{{\let\protect\string\immediate\write\@unused{#1}}}

\def\DvipsDriver{
	\ps@typeout{psfig/tex \PsfigVersion -dvips}
\def\PsfigSpecials{\DvipsSpecials} 	\def\ps@dir{/}
\def\ps@predir{} }
\def\OzTeXDriver{
	\ps@typeout{psfig/tex \PsfigVersion -oztex}
	\def\PsfigSpecials{\OzTeXSpecials}
	\def\ps@dir{:}
	\def\ps@predir{:}
	\catcode`\^^J=5
}


\def\figurepath{./:}

\def\DoPaths#1{\expandafter\EachPath#1\stoplist}
\def\leer{}
\def\EachPath#1:#2\stoplist{
  \ExistsFile{#1}{\SearchedFile}
  \ifx#2\leer
  \else
    \expandafter\EachPath#2\stoplist
  \fi}
%
%
\def\ps@dir{/}
\def\ExistsFile#1#2{%
   \openin1=\ps@predir#1\ps@dir#2
   \ifeof1
       \closein1
   \else
       \closein1
        \ifx\ps@founddir\leer
           \edef\ps@founddir{#1}
        \fi
   \fi}
%
%
\def\get@dir#1{%
  \def\ps@founddir{}
  \def\SearchedFile{#1}
  \DoPaths\figurepath
}

%
%
\def\@nnil{\@nil}
\def\@empty{}
\def\@psdonoop#1\@@#2#3{}
\def\@psdo#1:=#2\do#3{\edef\@psdotmp{#2}\ifx\@psdotmp\@empty \else
    \expandafter\@psdoloop#2,\@nil,\@nil\@@#1{#3}\fi}
\def\@psdoloop#1,#2,#3\@@#4#5{\def#4{#1}\ifx #4\@nnil \else
       #5\def#4{#2}\ifx #4\@nnil \else#5\@ipsdoloop #3\@@#4{#5}\fi\fi}
\def\@ipsdoloop#1,#2\@@#3#4{\def#3{#1}\ifx #3\@nnil 
       \let\@nextwhile=\@psdonoop \else
      #4\relax\let\@nextwhile=\@ipsdoloop\fi\@nextwhile#2\@@#3{#4}}
\def\@tpsdo#1:=#2\do#3{\xdef\@psdotmp{#2}\ifx\@psdotmp\@empty \else
    \@tpsdoloop#2\@nil\@nil\@@#1{#3}\fi}
\def\@tpsdoloop#1#2\@@#3#4{\def#3{#1}\ifx #3\@nnil 
       \let\@nextwhile=\@psdonoop \else
      #4\relax\let\@nextwhile=\@tpsdoloop\fi\@nextwhile#2\@@#3{#4}}
%
\ifx\undefined\fbox
\newdimen\fboxrule
\newdimen\fboxsep
\newdimen\ps@tempdima
\newbox\ps@tempboxa
\fboxsep = 3pt
\fboxrule = .4pt
\long\def\fbox#1{\leavevmode\setbox\ps@tempboxa\hbox{#1}\ps@tempdima\fboxrule
    \advance\ps@tempdima \fboxsep \advance\ps@tempdima \dp\ps@tempboxa
   \hbox{\lower \ps@tempdima\hbox
  {\vbox{\hrule height \fboxrule
          \hbox{\vrule width \fboxrule \hskip\fboxsep
          \vbox{\vskip\fboxsep \box\ps@tempboxa\vskip\fboxsep}\hskip 
                 \fboxsep\vrule width \fboxrule}
                 \hrule height \fboxrule}}}}
\fi
%
%
\newread\ps@stream
\newif\ifnot@eof       
\newif\if@noisy        
\newif\if@atend        
\newif\if@psfile       
%
%
{\catcode`\%=12\global\gdef\epsf@start{
\def\epsf@PS{PS}
\def\epsf@getbb#1{%
%
%
\openin\ps@stream=\ps@predir#1
\ifeof\ps@stream\ps@typeout{Error, File #1 not found}\else
%
%
   {\not@eoftrue \chardef\other=12
    \def\do##1{\catcode`##1=\other}\dospecials \catcode`\ =10
    \loop
       \if@psfile
	  \read\ps@stream to \epsf@fileline
       \else{
	  \obeyspaces
          \read\ps@stream to \epsf@tmp\global\let\epsf@fileline\epsf@tmp}
       \fi
       \ifeof\ps@stream\not@eoffalse\else
%
%
       \if@psfile\else
       \expandafter\epsf@test\epsf@fileline:. \\%
       \fi
%
%
          \expandafter\epsf@aux\epsf@fileline:. \\%
       \fi
   \ifnot@eof\repeat
   }\closein\ps@stream\fi}%
%
%
\long\def\epsf@test#1#2#3:#4\\{\def\epsf@testit{#1#2}
			\ifx\epsf@testit\epsf@start\else
\ps@typeout{Warning! File does not start with `\epsf@start'.  It may not be a PostScript file.}
			\fi
			\@psfiletrue} 
%
%
{\catcode`\%=12\global\let\epsf@percent=
%
%
%
\long\def\epsf@aux#1#2:#3\\{\ifx#1\epsf@percent
   \def\epsf@testit{#2}\ifx\epsf@testit\epsf@bblit
	\@atendfalse
        \epsf@atend #3 . \\%
	\if@atend	
	   \if@verbose{
		\ps@typeout{psfig: found `(atend)'; continuing search}
	   }\fi
        \else
        \epsf@grab #3 . . . \\%
        \not@eoffalse
        \global\no@bbfalse
        \fi
   \fi\fi}%
%
%
\def\epsf@grab #1 #2 #3 #4 #5\\{%
   \global\def\epsf@llx{#1}\ifx\epsf@llx\empty
      \epsf@grab #2 #3 #4 #5 .\\\else
   \global\def\epsf@lly{#2}%
   \global\def\epsf@urx{#3}\global\def\epsf@ury{#4}\fi}%
%
%
\def\epsf@atendlit{(atend)} 
\def\epsf@atend #1 #2 #3\\{%
   \def\epsf@tmp{#1}\ifx\epsf@tmp\empty
      \epsf@atend #2 #3 .\\\else
   \ifx\epsf@tmp\epsf@atendlit\@atendtrue\fi\fi}


\chardef\psletter = 11 
\chardef\other = 12

\newif \ifdebug 
\newif\ifc@mpute 
\c@mputetrue 

\let\then = \relax
\def\r@dian{pt }
\let\r@dians = \r@dian
\let\dimensionless@nit = \r@dian
\let\dimensionless@nits = \dimensionless@nit
\def\internal@nit{sp }
\let\internal@nits = \internal@nit
\newif\ifstillc@nverging
\def \Mess@ge #1{\ifdebug \then \message {#1} \fi}

{ 
	\catcode `\@ = \psletter
	\gdef \nodimen {\expandafter \n@dimen \the \dimen}
	\gdef \term #1 #2 #3%
	       {\edef \t@ {\the #1}
		\edef \t@@ {\expandafter \n@dimen \the #2\r@dian}%
		\t@rm {\t@} {\t@@} {#3}%
	       }
	\gdef \t@rm #1 #2 #3%
	       {{%
		\count 0 = 0
		\dimen 0 = 1 \dimensionless@nit
		\dimen 2 = #2\relax
		\Mess@ge {Calculating term #1 of \nodimen 2}%
		\loop
		\ifnum	\count 0 < #1
		\then	\advance \count 0 by 1
			\Mess@ge {Iteration \the \count 0 \space}%
			\Multiply \dimen 0 by {\dimen 2}%
			\Mess@ge {After multiplication, term = \nodimen 0}%
			\Divide \dimen 0 by {\count 0}%
			\Mess@ge {After division, term = \nodimen 0}%
		\repeat
		\Mess@ge {Final value for term #1 of 
				\nodimen 2 \space is \nodimen 0}%
		\xdef \Term {#3 = \nodimen 0 \r@dians}%
		\aftergroup \Term
	       }}
	\catcode `\p = \other
	\catcode `\t = \other
	\gdef \n@dimen #1pt{#1} 
}

\def \Divide #1by #2{\divide #1 by #2} 

\def \Multiply #1by #2
       {{
	\count 0 = #1\relax
	\count 2 = #2\relax
	\count 4 = 65536
	\Mess@ge {Before scaling, count 0 = \the \count 0 \space and
			count 2 = \the \count 2}%
	\ifnum	\count 0 > 32767 
	\then	\divide \count 0 by 4
		\divide \count 4 by 4
	\else	\ifnum	\count 0 < -32767
		\then	\divide \count 0 by 4
			\divide \count 4 by 4
		\else
		\fi
	\fi
	\ifnum	\count 2 > 32767 
	\then	\divide \count 2 by 4
		\divide \count 4 by 4
	\else	\ifnum	\count 2 < -32767
		\then	\divide \count 2 by 4
			\divide \count 4 by 4
		\else
		\fi
	\fi
	\multiply \count 0 by \count 2
	\divide \count 0 by \count 4
	\xdef \product {#1 = \the \count 0 \internal@nits}%
	\aftergroup \product
       }}

\def\r@duce{\ifdim\dimen0 > 90\r@dian \then   
		\multiply\dimen0 by -1
		\advance\dimen0 by 180\r@dian
		\r@duce
	    \else \ifdim\dimen0 < -90\r@dian \then  
		\advance\dimen0 by 360\r@dian
		\r@duce
		\fi
	    \fi}

\def\Sine#1%
       {{%
	\dimen 0 = #1 \r@dian
	\r@duce
	\ifdim\dimen0 = -90\r@dian \then
	   \dimen4 = -1\r@dian
	   \c@mputefalse
	\fi
	\ifdim\dimen0 = 90\r@dian \then
	   \dimen4 = 1\r@dian
	   \c@mputefalse
	\fi
	\ifdim\dimen0 = 0\r@dian \then
	   \dimen4 = 0\r@dian
	   \c@mputefalse
	\fi
	\ifc@mpute \then
		\divide\dimen0 by 180
		\dimen0=3.141592654\dimen0
		\dimen 2 = 3.1415926535897963\r@dian 
		\divide\dimen 2 by 2 
		\Mess@ge {Sin: calculating Sin of \nodimen 0}%
		\count 0 = 1 
		\dimen 2 = 1 \r@dian 
		\dimen 4 = 0 \r@dian 
		\loop
			\ifnum	\dimen 2 = 0 
			\then	\stillc@nvergingfalse 
			\else	\stillc@nvergingtrue
			\fi
			\ifstillc@nverging 
			\then	\term {\count 0} {\dimen 0} {\dimen 2}%
				\advance \count 0 by 2
				\count 2 = \count 0
				\divide \count 2 by 2
				\ifodd	\count 2 
				\then	\advance \dimen 4 by \dimen 2
				\else	\advance \dimen 4 by -\dimen 2
				\fi
		\repeat
	\fi		
			\xdef \sine {\nodimen 4}%
       }}

\def\Cosine#1{\ifx\sine\UnDefined\edef\Savesine{\relax}\else
		             \edef\Savesine{\sine}\fi
	{\dimen0=#1\r@dian\advance\dimen0 by 90\r@dian
	 \Sine{\nodimen 0}
	 \xdef\cosine{\sine}
	 \xdef\sine{\Savesine}}}	      

\def\psdraft{
	\def\@psdraft{0}
}
\def\psfull{
	\def\@psdraft{100}
}

\psfull

\newif\if@scalefirst
\def\psscalefirst{\@scalefirsttrue}
\def\psrotatefirst{\@scalefirstfalse}
\psrotatefirst

\newif\if@draftbox
\def\psnodraftbox{
	\@draftboxfalse
}
\def\psdraftbox{
	\@draftboxtrue
}
\@draftboxtrue

\newif\if@prologfile
\newif\if@postlogfile
\def\pssilent{
	\@noisyfalse
}
\def\psnoisy{
	\@noisytrue
}
\psnoisy
\newif\if@bbllx
\newif\if@bblly
\newif\if@bburx
\newif\if@bbury
\newif\if@height
\newif\if@width
\newif\if@rheight
\newif\if@rwidth
\newif\if@angle
\newif\if@clip
\newif\if@verbose
\def\@p@@sclip#1{\@cliptrue}
\newif\if@decmpr
\def\@p@@sfigure#1{\def\@p@sfile{null}\def\@p@sbbfile{null}\@decmprfalse
   \openin1=\ps@predir#1
   \ifeof1
	\closein1
	\get@dir{#1}
	\ifx\ps@founddir\leer
		\openin1=\ps@predir#1.bb
		\ifeof1
			\closein1
			\get@dir{#1.bb}
			\ifx\ps@founddir\leer
				\ps@typeout{Can't find #1 in \figurepath}
			\else
				\@decmprtrue
				\def\@p@sfile{\ps@founddir\ps@dir#1}
				\def\@p@sbbfile{\ps@founddir\ps@dir#1.bb}
			\fi
		\else
			\closein1
			\@decmprtrue
			\def\@p@sfile{#1}
			\def\@p@sbbfile{#1.bb}
		\fi
	\else
		\def\@p@sfile{\ps@founddir\ps@dir#1}
		\def\@p@sbbfile{\ps@founddir\ps@dir#1}
	\fi
   \else
	\closein1
	\def\@p@sfile{#1}
	\def\@p@sbbfile{#1}
   \fi
}
\def\@p@@sfile#1{\@p@@sfigure{#1}}
\def\@p@@sbbllx#1{
		\@bbllxtrue
		\dimen100=#1
		\edef\@p@sbbllx{\number\dimen100}
}
\def\@p@@sbblly#1{
		\@bbllytrue
		\dimen100=#1
		\edef\@p@sbblly{\number\dimen100}
}
\def\@p@@sbburx#1{
		\@bburxtrue
		\dimen100=#1
		\edef\@p@sbburx{\number\dimen100}
}
\def\@p@@sbbury#1{
		\@bburytrue
		\dimen100=#1
		\edef\@p@sbbury{\number\dimen100}
}
\def\@p@@sheight#1{
		\@heighttrue
		\dimen100=#1
   		\edef\@p@sheight{\number\dimen100}
}
\def\@p@@swidth#1{
		\@widthtrue
		\dimen100=#1
		\edef\@p@swidth{\number\dimen100}
}
\def\@p@@srheight#1{
		\@rheighttrue
		\dimen100=#1
		\edef\@p@srheight{\number\dimen100}
}
\def\@p@@srwidth#1{
		\@rwidthtrue
		\dimen100=#1
		\edef\@p@srwidth{\number\dimen100}
}
\def\@p@@sangle#1{
		\@angletrue
		\edef\@p@sangle{#1} 
}
\def\@p@@ssilent#1{ 
		\@verbosefalse
}
\def\@p@@sprolog#1{\@prologfiletrue\def\@prologfileval{#1}}
\def\@p@@spostlog#1{\@postlogfiletrue\def\@postlogfileval{#1}}
\def\@cs@name#1{\csname #1\endcsname}
\def\@setparms#1=#2,{\@cs@name{@p@@s#1}{#2}}
%
%
\def\ps@init@parms{
		\@bbllxfalse \@bbllyfalse
		\@bburxfalse \@bburyfalse
		\@heightfalse \@widthfalse
		\@rheightfalse \@rwidthfalse
		\def\@p@sbbllx{}\def\@p@sbblly{}
		\def\@p@sbburx{}\def\@p@sbbury{}
		\def\@p@sheight{}\def\@p@swidth{}
		\def\@p@srheight{}\def\@p@srwidth{}
		\def\@p@sangle{0}
		\def\@p@sfile{} \def\@p@sbbfile{}
		\def\@p@scost{10}
		\def\@sc{}
		\@prologfilefalse
		\@postlogfilefalse
		\@clipfalse
		\if@noisy
			\@verbosetrue
		\else
			\@verbosefalse
		\fi
}
%
%
\def\parse@ps@parms#1{
	 	\@psdo\@psfiga:=#1\do
		   {\expandafter\@setparms\@psfiga,}}
%
%
\newif\ifno@bb
\def\bb@missing{
	\if@verbose{
		\ps@typeout{psfig: searching \@p@sbbfile \space  for bounding box}
	}\fi
	\no@bbtrue
	\epsf@getbb{\@p@sbbfile}
        \ifno@bb \else \bb@cull\epsf@llx\epsf@lly\epsf@urx\epsf@ury\fi
}	
\def\bb@cull#1#2#3#4{
	\dimen100=#1 bp\edef\@p@sbbllx{\number\dimen100}
	\dimen100=#2 bp\edef\@p@sbblly{\number\dimen100}
	\dimen100=#3 bp\edef\@p@sbburx{\number\dimen100}
	\dimen100=#4 bp\edef\@p@sbbury{\number\dimen100}
	\no@bbfalse
}
\newdimen\p@intvaluex
\newdimen\p@intvaluey
\def\rotate@#1#2{{\dimen0=#1 sp\dimen1=#2 sp
		  \global\p@intvaluex=\cosine\dimen0
		  \dimen3=\sine\dimen1
		  \global\advance\p@intvaluex by -\dimen3
		  \global\p@intvaluey=\sine\dimen0
		  \dimen3=\cosine\dimen1
		  \global\advance\p@intvaluey by \dimen3
		  }}
\def\compute@bb{
		\no@bbfalse
		\if@bbllx \else \no@bbtrue \fi
		\if@bblly \else \no@bbtrue \fi
		\if@bburx \else \no@bbtrue \fi
		\if@bbury \else \no@bbtrue \fi
		\ifno@bb \bb@missing \fi
		\ifno@bb \ps@typeout{FATAL ERROR: no bb supplied or found}
			\no-bb-error
		\fi
		%
%
		\count203=\@p@sbburx
		\count204=\@p@sbbury
		\advance\count203 by -\@p@sbbllx
		\advance\count204 by -\@p@sbblly
		\edef\ps@bbw{\number\count203}
		\edef\ps@bbh{\number\count204}
		\if@angle 
			\Sine{\@p@sangle}\Cosine{\@p@sangle}
	        	{\dimen100=\maxdimen\xdef\r@p@sbbllx{\number\dimen100}
					    \xdef\r@p@sbblly{\number\dimen100}
			                    \xdef\r@p@sbburx{-\number\dimen100}
					    \xdef\r@p@sbbury{-\number\dimen100}}
%
                        \def\minmaxtest{
			   \ifnum\number\p@intvaluex<\r@p@sbbllx
			      \xdef\r@p@sbbllx{\number\p@intvaluex}\fi
			   \ifnum\number\p@intvaluex>\r@p@sbburx
			      \xdef\r@p@sbburx{\number\p@intvaluex}\fi
			   \ifnum\number\p@intvaluey<\r@p@sbblly
			      \xdef\r@p@sbblly{\number\p@intvaluey}\fi
			   \ifnum\number\p@intvaluey>\r@p@sbbury
			      \xdef\r@p@sbbury{\number\p@intvaluey}\fi
			   }
			\rotate@{\@p@sbbllx}{\@p@sbblly}
			\minmaxtest
			\rotate@{\@p@sbbllx}{\@p@sbbury}
			\minmaxtest
			\rotate@{\@p@sbburx}{\@p@sbblly}
			\minmaxtest
			\rotate@{\@p@sbburx}{\@p@sbbury}
			\minmaxtest
			\edef\@p@sbbllx{\r@p@sbbllx}\edef\@p@sbblly{\r@p@sbblly}
			\edef\@p@sbburx{\r@p@sbburx}\edef\@p@sbbury{\r@p@sbbury}
		\fi
		\count203=\@p@sbburx
		\count204=\@p@sbbury
		\advance\count203 by -\@p@sbbllx
		\advance\count204 by -\@p@sbblly
		\edef\@bbw{\number\count203}
		\edef\@bbh{\number\count204}
}
%
%
\def\in@hundreds#1#2#3{\count240=#2 \count241=#3
		     \count100=\count240	
		     \divide\count100 by \count241
		     \count101=\count100
		     \multiply\count101 by \count241
		     \advance\count240 by -\count101
		     \multiply\count240 by 10
		     \count101=\count240	
		     \divide\count101 by \count241
		     \count102=\count101
		     \multiply\count102 by \count241
		     \advance\count240 by -\count102
		     \multiply\count240 by 10
		     \count102=\count240	
		     \divide\count102 by \count241
		     \count200=#1\count205=0
		     \count201=\count200
			\multiply\count201 by \count100
		 	\advance\count205 by \count201
		     \count201=\count200
			\divide\count201 by 10
			\multiply\count201 by \count101
			\advance\count205 by \count201
		     \count201=\count200
			\divide\count201 by 100
			\multiply\count201 by \count102
			\advance\count205 by \count201
		     \edef\@result{\number\count205}
}
\def\compute@wfromh{
		\in@hundreds{\@p@sheight}{\@bbw}{\@bbh}
		\edef\@p@swidth{\@result}
}
\def\compute@hfromw{
	        \in@hundreds{\@p@swidth}{\@bbh}{\@bbw}
		\edef\@p@sheight{\@result}
}
\def\compute@handw{
		\if@height 
			\if@width
			\else
				\compute@wfromh
			\fi
		\else 
			\if@width
				\compute@hfromw
			\else
				\edef\@p@sheight{\@bbh}
				\edef\@p@swidth{\@bbw}
			\fi
		\fi
}
\def\compute@resv{
		\if@rheight \else \edef\@p@srheight{\@p@sheight} \fi
		\if@rwidth \else \edef\@p@srwidth{\@p@swidth} \fi
}
%
\def\compute@sizes{
	\compute@bb
	\if@scalefirst\if@angle
	\if@width
	   \in@hundreds{\@p@swidth}{\@bbw}{\ps@bbw}
	   \edef\@p@swidth{\@result}
	\fi
	\if@height
	   \in@hundreds{\@p@sheight}{\@bbh}{\ps@bbh}
	   \edef\@p@sheight{\@result}
	\fi
	\fi\fi
	\compute@handw
	\compute@resv}
\def\OzTeXSpecials{
	\special{empty.ps /@isp {true} def}
	\special{empty.ps \@p@swidth \space \@p@sheight \space
			\@p@sbbllx \space \@p@sbblly \space
			\@p@sbburx \space \@p@sbbury \space
			startTexFig \space }
	\if@clip{
		\if@verbose{
			\ps@typeout{(clip)}
		}\fi
		\special{empty.ps doclip \space }
	}\fi
	\if@angle{
		\if@verbose{
			\ps@typeout{(rotate)}
		}\fi
		\special {empty.ps \@p@sangle \space rotate \space} 
	}\fi
	\if@prologfile
	    \special{\@prologfileval \space } \fi
	\if@decmpr{
		\if@verbose{
			\ps@typeout{psfig: Compression not available
			in OzTeX version \space }
		}\fi
	}\else{
		\if@verbose{
			\ps@typeout{psfig: including \@p@sfile \space }
		}\fi
		\special{epsf=\ps@predir\@p@sfile \space }
	}\fi
	\if@postlogfile
	    \special{\@postlogfileval \space } \fi
	\special{empty.ps /@isp {false} def}
}
\def\DvipsSpecials{
	\special{ps::[begin] 	\@p@swidth \space \@p@sheight \space
			\@p@sbbllx \space \@p@sbblly \space
			\@p@sbburx \space \@p@sbbury \space
			startTexFig \space }
	\if@clip{
		\if@verbose{
			\ps@typeout{(clip)}
		}\fi
		\special{ps:: doclip \space }
	}\fi
	\if@angle
		\if@verbose{
			\ps@typeout{(clip)}
		}\fi
		\special {ps:: \@p@sangle \space rotate \space} 
	\fi
	\if@prologfile
	    \special{ps: plotfile \@prologfileval \space } \fi
	\if@decmpr{
		\if@verbose{
			\ps@typeout{psfig: including \@p@sfile.Z \space }
		}\fi
		\special{ps: plotfile "`zcat \@p@sfile.Z" \space }
	}\else{
		\if@verbose{
			\ps@typeout{psfig: including \@p@sfile \space }
		}\fi
		\special{ps: plotfile \@p@sfile \space }
	}\fi
	\if@postlogfile
	    \special{ps: plotfile \@postlogfileval \space } \fi
	\special{ps::[end] endTexFig \space }
}
%
%
\def\psfig#1{\vbox {
	%
	\ps@init@parms
	\parse@ps@parms{#1}
	\compute@sizes
	\ifnum\@p@scost<\@psdraft{
		\PsfigSpecials 
		\vbox to \@p@srheight sp{
			\hbox to \@p@srwidth sp{
				\hss
			}
		\vss
		}
	}\else{
		\if@draftbox{		
			\hbox{\fbox{\vbox to \@p@srheight sp{
			\vss
			\hbox to \@p@srwidth sp{ \hss 
			 \hss }
			\vss
			}}}
		}\else{
			\vbox to \@p@srheight sp{
			\vss
			\hbox to \@p@srwidth sp{\hss}
			\vss
			}
		}\fi

	}\fi
}}
\psfigRestoreAt
\setDriver
\let\@=\LaTeXAtSign

%
%
%
%
\font\twelverm = cmr10 scaled\magstep1 \font\tenrm = cmr10
       \font\sevenrm = cmr7
\font\twelvei = cmmi10 scaled\magstep1 
       \font\teni = cmmi10 \font\seveni = cmmi7
\font\twelveit = cmti10 scaled\magstep1 \font\tenit = cmti10
       \font\sevenit = cmti7
\font\twelvesy = cmsy10 scaled\magstep1 
       \font\tensy = cmsy10 \font\sevensy = cmsy7
\font\twelvebf = cmbx10 scaled\magstep1 \font\tenbf = cmbx10
       \font\sevenbf = cmbx7
\font\twelvesl = cmsl10 scaled\magstep1
\font\twelveit = cmti10 scaled\magstep1
\font\twelvett = cmtt10 scaled\magstep1
\font\klein=cmr10                                                                
\font\mittel=cmr10                                                               
\font\gross=cmbx12 at 1132462sp 
\font\normal=cmr10                                                               
\font\mathnormal=cmmi10                                                          
\font\mathgross=cmmi10 at 1132462sp
\font\mathklein=cmmi9                                                            
\font\mathmit=cmmi10                                                             
\font\kleinind=cmr7                                                              
\font\grossind=cmbx10 at 12truept                                                  
\font\midind=cmmi10                                                              
%
\textfont0 = \twelverm \scriptfont0 = \twelverm 
       \scriptscriptfont0 = \tenrm
       \def\rm{\fam0 \twelverm}
\textfont1 = \twelvei \scriptfont1 = \twelvei 
       \scriptscriptfont1 = \teni
       \def\mit{\fam1 }
\textfont2 = \twelvesy \scriptfont2 = \twelvesy 
       \scriptscriptfont2 = \tensy
       \def\cal{\fam2 }
\newfam\itfam \def\it{\fam\itfam \twelveit} \textfont\itfam=\twelveit
\newfam\slfam \def\sl{\fam\slfam \twelvesl} \textfont\slfam=\twelvesl
\newfam\bffam \def\bf{\fam\bffam \twelvebf} \textfont\bffam=\twelvebf
       \scriptfont\bffam=\twelvebf \scriptscriptfont\bffam=\tenbf
\newfam\ttfam \def\tt{\fam\ttfam \twelvett} \textfont\ttfam=\twelvett
\rm
\hsize=6in
\hoffset=.05in
\vsize=9in
\baselineskip=24pt
%
%
\dimen1=\baselineskip \multiply\dimen1 by 3 \divide\dimen1 by 4
\dimen2=\dimen1 \divide\dimen2 by 2
%
\nopagenumbers
\headline={\ifnum\pageno=1 \hss\thinspace\hss 
     \else\hss\folio\hss \fi}
%
\def\title#1 {\centerline{\gross \textfont1=\gross \textfont0=\gross
\scriptfont1=\gross \scriptfont0=\grossind #1}  }
%
\def\begfig#1 {\midinsert \vskip #1}
\def\figure#1#2{ \baselineskip=16pt {
 \noindent Figure #1: #2}}
\def\endfig {\endinsert}
\def\begtab{\midinsert}
\def\table#1#2{ \bigskip \baselineskip=11pt {\klein \textfont1=\mathklein \textfont0=\mathklein
\scriptfont1=\mathklein \scriptfont0=\kleinind \noindent Table #1: #2}  }
\def\endtab{\endinsert \bigskip}
\def\begtabsmall{\midinsert 
\klein \textfont1=\mathklein \textfont0=\klein \scriptfont1=\mathklein \scriptfont0=\kleinind }
%
\def\heading#1{\vfill\eject \vbox to \dimen1 {\vfill}
     \centerline{\bf #1} 
     \vskip \dimen1} 
%
\newcount\sectcount
\newcount\subcount
\newcount\subsubcount
\global\sectcount=0
\global\subcount=0
\global\subsubcount=0
\def\section#1{\vfill\eject \vbox to \dimen1 {\vfill}
    \global\advance\sectcount by 1
    \centerline{\bf \the\sectcount.\ \ {#1}}
    \global\subcount=0  
    \global\subsubcount=0  
    \vskip \dimen1}
%
\def\subsection#1{\global\advance\subcount by 1 
    \vskip \parskip  \vskip \dimen2
    \centerline{{\it \the\sectcount.\the\subcount.\ \ #1}} 
    \global\subsubcount=0  
    \vskip \dimen2}
%
\def\subsubsection#1{\global\advance\subsubcount by 1 
    \vskip \parskip  \vskip \dimen2
    \centerline{{\it \the\sectcount.\the\subcount.\the\subsubcount.\ \ #1}} 
    \vskip \dimen2}
%
%
\def\refindent{\advance\leftskip by 24pt \parindent=-24pt}
%
\def\journal#1#2#3#4#5{{\refindent
                      {#1}        
                      {#2},       
                      {#3},       
                      {#4},       
                      {#5}        
                      \par }}
%
\def\infuture#1#2#3#4{{\refindent
                  {#1}         
                  {#2},        
                  {#3},        
                  {#4}         
                  \par }}
%
\def\inbook#1#2#3#4#5#6#7{{\refindent
                         {#1}         
                         {#2},        
                      in {\it #3\/},  
                     ed. {#4}         
                        ({#5}:        
                         {#6}),       
                       p.{#7}         
                         \par }}
%
\def\infutbook#1#2#3#4#5#6#7{{\refindent
                            {#1}         
                            {#2},        
                         in {\it #3\/},  
                        ed. {#4}         
                           ({#5}:        
                            {#6}),       
                            {#7}         
                            \par }}
%
\def\book#1#2#3#4#5{{\refindent
                   {#1}         
                   {#2},        
                   {\it #3\/}   
                  ({#4}:        
                   {#5})        
                   \par }}
%
\def\privcom#1#2#3{{\refindent
                  {#1}        
                  {#2},       
                  {#3}        
                  \par }}
%
\def\phdthesis#1#2#3{{\refindent
                    {#1}                 
                    {#2}, Ph.D. thesis,  
                    {#3}                 
                    \par}}
%
\def\circular#1#2#3#4{{\refindent
                     {#1}          
                     {#2},         
                     {#3},         
                     {#4}          
                     \par}}
%
\def\figcap#1#2{{\refindent
                   Fig. {#1}.---   
                        {#2}       
                        \par}}
%
\def\etal{{\it et al.\/\ }}
\def\eg{{\it e.g.\/}}
\def\sun{_\odot}
\def\samename{\vrule height0.4pt depth0.0pt width1.0in \thinspace.}
%
%
%
\def\footstrut{\vbox to \baselineskip{}}
\newcount\notenumber
\notenumber=0
\def\hline{\vskip 3pt \hrule\vskip 5pt}
\def\note{\global\advance\notenumber by 1 
	  \footnote{$^{\the\notenumber}$}}
\def\apjnote#1{\global\advance \notenumber by 1 
	       $^{\the\notenumber}$ \kern -.6em
               \vadjust{\midinsert
                        \hbox to \hsize{\hrulefill}
                        $^{\the\notenumber}${#1} \hfill \break
                        \hbox to \hsize{\hrulefill}
                        \endinsert
                       }
              }

\overfullrule=0pt
\baselineskip=15pt
%
%
 
\def\MPA#1#2{Max-Planck-Institut f\"ur Astrophysik 19{#1},
             Preprint {$\underline{#2}$} }
 \def\z{\phantom 1}
 \def\Hbar{$\overline H\ $}
 \def\Htil{$\widetilde H\ $}
 \def\Etilbv{$\widetilde E_{B-V}\ $}
 \def\etal{{et al.} \thinspace}
 \def\eg{{e.g.,} \thinspace}
 \def\ie{{i.e.,} \thinspace}
 \def\eck#1{\left\lbrack #1 \right\rbrack}
 \def\eqck#1{$\bigl\lbrack$ #1 $\bigr\rbrack$}
 \def\rund#1{\left( #1 \right)}
 \def\ave#1{\langle #1 \rangle}
 \def\:{\mskip\medmuskip}                         
 \def\lb{\lbrack} \def\rb{\rbrack}                
 \def\unit#1{\nobreak{\:{\rm#1}}}                 
 \def\inunits#1{\nobreak{\:\lb{\rm#1}\rb}}        
 \def\gcc{gcm$^{-3}$}
 \def\mstar{ M_{\ast} }
 \def\msol{ M_\odot }            
 \def\ms{ M_\odot }              
 \def\lsol{ L_\odot }            
 \def\ni{$^{56}{\rm Ni}\ $}      
 \def\Ni{$^{56}{\rm Ni}$}      
 \def\co{$^{56}{\rm Co}\ $}      
 \def\Co{$^{56}{\rm Co}$}      
 \def\fe{$^{56}{\rm Fe}\ $}      
 \def\kelvin{\thinspace\rm{\sp{o}{\kern-.08333em }K}\ }
%
%
%
\title{Explosion Models for Type Ia Supernovae:}
\title{A comparison with observed light curves,}
\title{distances, $H_o$ and $q_o$}
\bigskip
\bigskip
\centerline{P.~H\"oflich$^1$ and A. Khokhlov $^2$}
\bigskip
\leftline{1. Center for Astrophysics, Harvard University, 60 Garden Street,
Cambridge, MA 02138}
\leftline{2. Department of Astronomy, University of Texas, Austin, TX 07871}
\bigskip
\bigskip

\heading{Abstract}
 
\noindent
Theoretical monochromatic light curves and photospheric expansion 
velocities   are compared with observations             
 of 27 Type~Ia supernovae (SNe~Ia). 
 A set of 37 models 
has been considered which encompasses all currently discussed explosion 
scenarios for Type~Ia supernovae  including deflagrations, detonations, delayed
detonations, pulsating delayed detonations and tamped detonations of Chandrasekhar
mass, and Helium detonations of low mass white dwarfs.
The explosions are calculated using one-dimensional Lagrangian hydro and
radiation-hydro codes with incorporated nuclear networks. Subsequently, light 
curves are constructed using our LC scheme which includes an implicit radiation 
transport, expansion opacities,
 a Monte-Carlo $\gamma $-ray transport, and
molecular and dust formation. For some supernovae, results of detailed non-LTE
calculations have been considered.
 
 Observational properties of our series of models are  discussed, in particular,
the relation between the absolute brightness, post-maximum decline rates, 
 the colors at several moments of time, etc. All models  with a \ni production
 larger than $\approx 0.4 M_\odot $ produce light curves of similar brightness.           
 The influence of the cosmological red shift
on the light curves and on the correction for interstellar reddening is discussed. 
 
 Based on data rectification 
of the standard deviation, a quantitative  procedure to fit the observations  has been used 
to the determine the free parameters, i.e. the distance, the reddening, and the
time of the explosion.
 Fast rising light curves (\eg
 SN~1981B and SN~1994D) can be reproduced  by delayed
detonation models or deflagration models similar to W7. 
Slowly rising ($t_{max} \geq 16$~days) light
curves (\eg SN~1984A and SN~1990N) cannot be reproduced by standard
detonation, deflagration, or delayed detonation models.  To
obtain an acceptable agreement with observations,
 models are required where the C/O white dwarf is
surrounded by an unburnt extended envelope of typically 0.2 to
0.4~$M_{\odot}$ which may either be pre-existing or produced during the explosion.
  Our interpretation of the light curves is also supported by
     the photospheric expansion velocities. Mainly due to the fast increase 
of the $\gamma $ radiation produced by the outer \ni,
 the post maximum decline of Helium detonations tends to be  
faster compared to  observations of normal bright SNe~Ia.
 
 Strongly subluminous SNe~Ia can be understood in the framework of
pulsating delayed detonations, both from the absolute brightness and
 the colors. Alternatively,  subluminosity can be produced 
 within the scenario  of helium detonations in low mass white dwarfs 
of about 0.6 to 0.8 $M_\odot $ if the explosion occurs when rather little
Helium has been accreted.  However, even subluminous Helium detonation 
models are very blue at maximum light owing to
heating in the outer layers and  brighter models show a fast 
postmaximum decline, in contradiction to the observations.

 We find evidence for a correlation between the type of 
host galaxy and the explosion mechanism. In spiral galaxies, about
the same amount of prompt explosions (delayed detonations and W7) 
and pulsating delayed detonations seems to occur.
 In contrast, in ellipticals, the latter type is strongly favored.
 This difference may provide a hint about the stellar evolution of the progenitors.
 
 Based on a comparison of theoretical light curves and observational data,
 the distances of the parent galaxies
 are       determined independent from secondary distance indicators.
A comparison with theoretical models allows for a consistent 
determination of the interstellar reddening and the cosmological red shift.
For the example of SN~1988U, we
show the need for a simultaneous  use  of both spectral and light curve data if the data
set is incomplete. Based on the models,
 SNe~Ia allow for a measurement of the value
 of   the Hubble constant $H_o$.
 $H_o$ is found to be $ 67 \pm 9~km/(sec Mpc)$ 
within a 95~\% probability for  distances up to 1.3 Gpc. 
 SN1988U at 1.3 Gpc is consistent with a  deceleration parameter 
$q_o$ of $0.7 \pm 0.5 $.

\bigskip \noindent
{\it Subject headings:} Supernovae and supernovae remnants: general -- hydrodynamics -- radiation transfer
 -- light curves -- Hubble constant -- deceleration parameter $q_o$
%
%
%
%
%
\section{Introduction}
Type Ia Supernovae (SNe~Ia)
may reach the same brightness as the entire parent galaxy. In  
 principle, this allows for the measurements of extragalactic distances and
cosmological parameters such as the Hubble constant $H_o$ and the deceleration 
parameter $q_o$. However, the absolute brightness must be known either
by using distance calibrators or theoretical models for the light curves
and spectra. Type Ia events are  major contributors to the production
of heavy elements. An understanding of the underlying physics is   important
for our picture of the chemical evolution of galaxies. 
Improvements in the observations and theoretical calculations have resulted in 
the rapid growth of our knowledge of these objects, but have also raised several
new questions on the explosion mechanism and the homogeneity of
SNe~Ia.
 
During the last two decades, the observational database for Type~Ia
supernovae  has significantly increased.  Both the similarity of
their light curves and spectra favored the idea that SNe~Ia form a
homogeneous class and, thus, can be used as standard candles (for a recent
review see Branch \& Tammann  1992). However,
several authors have raised doubts in the past about the
homogeneity of SNe~Ia and, nowadays, a spread in the properties of 
SNe~Ia is widely accepted.
  Based on a large number of photometric light
curves, Barbon, Ciatti \& Rosino (1978) divided SNe~Ia into ``fast" and ``slow"
events depending on the rate of decline from maximum, the contrast from
peak to tail, and the rate of decline of the tail.  Pskovskii (1970,
1977) and Branch (1981) argued that SNe~Ia form a continuous sequence,
the expansion velocity and the peak absolute magnitude being correlated
with the rate of decline from maximum. The $B$ and $V$ light curves of SN~1986G
declined faster after maximum than those of SN~1981B (Phillips \etal
1987; Cristiani \etal 1992), the infrared light curve of SN~1986G was
unusual (Frogel \etal 1987), the optical spectra of SN~1986G (Phillips
\etal 1987) and of SN~1990N (Leibundgut \etal 1991) showed small but
significant anomalies, and pre-maximum observations of SN~1991T
exhibited large deviations from the ``standard" behaviour of SNe~Ia
(Filippenko \etal 1992a; Phillips \etal 1992; Hamuy \etal  1995).
 Finally,  the ``unusual" Type~Ia supernova SN~1991bg clearly proved the
existence of a wide diversity among SNe~Ia.
Besides showing other peculiarities, 
SN1991bg was dimmer than SN~1957B, another SN~Ia in
the same galaxy, by $\sim 2.5$~mag in $B$ and by $\sim 1.5$~mag in $V$
(Filippenko \etal 1992ab; Leibundgut \etal 1993). 
 
 Despite all uncertainties, it is widely accepted that SNe~Ia are thermonuclear
explosions of carbon-oxygen white dwarfs (Hoyle \& Fowler 1960; for discussions
of various theoretical aspects see Woosley and Weaver 1986, 1995, 
Wheeler \& Harkness 1990, Canal 1995, Nomoto et al. 1995, Nomoto 1995,
and Wheeler et al. 1995). Three main scenarios can be distinguished.
 
 A   first group consists of massive carbon-oxygen white
dwarfs (WDs) with a mass close to the Chandrasekhar mass which accrete mass
through Roche-lobe overflow from an evolved companion star (Nomoto \&
Sugimoto 1977; Nomoto 1982).  In these accretion models, the explosion is
triggered by compressional heating. 
From the theoretical standpoint, the key question is 
how the flame propagates through the white
dwarf. Several models of Type~Ia supernovae have been proposed in the
past, including detonation (Arnett 1969; Hansen \& Wheeler 1969), 
deflagration (Ivanova, Imshennik \& Chechetkin 1974; Nomoto, Sugimoto \&
Neo 1976) and the delayed detonation model,                     
which assumes that the flame starts as a deflagration and turns into a 
detonation later on (Khokhlov 1991ab, Woosley \& Weaver 1995, Yamaoka \etal 
1992).

 The second group of progenitor
models consists of two low-mass white dwarfs in a close orbit which
decays due to the emission of gravitational radiation and this, eventually,
leads to the merging of the two WDs. In an intermediate step,
these models form a low density WD surrounded by a CO envelope
 (Webbink 1984; Iben \& Tutukov 1984; Paczy\'nski 1985).

      Another class of models -- double detonation of a C/O-WD
triggered by detonation of helium layer in low-mass white dwarfs
 -- was explored by Nomoto (1980),
Woosley, Weaver \& Taam (1980), and most recently by Woosley and Weaver (1994,
hereafter WW94).
This scenario was also suggested for the explanation of 
subluminous Type~Ia (WW94). Note that the explosion of a low mass WD 
was also suggested by Ruiz-Lapuente et al. (1993) to explain subluminous SNe~Ia
but the mechanism for triggering of the central carbon detonation was not considered.
 Because low-mass WDs are much more common
than massive WDs,  the possible impact on our understanding of supernovae
statistics and, consequently, the chemical evolution of galaxies 
must be noted. Moreover,
 based on light curve calculations with constant line
opacities, Woosley and Weaver (1994) suspected that Helium detonations are commonly dimmer
by about $0.4^m$ with the obvious consequences for $H_o$.
 
 To clarify the possible scenarios, the theoretical models must be
tested by observations.
Light curves and spectra of different scenarios for normal bright SNe~Ia 
including detonations, delayed detonations,
pulsating delayed detonations, and envelope models
have been investigated in previous papers
and compared with the observations 
(HKM91ab, 
KMH92; HMK93; KMH93;
M\"uller \& H\"oflich  1994; H\"oflich, Khokhlov \& Wheeler  1995, H\"oflich 1995;
 hereafter HKM91, KMH92, HMK93, KMH93, MH94, HKW95 and H95 where H, K, M and W stand
for H\"oflich, Khokhlov, M\"uller and Wheeler, respectively).
 We have investigated the validity and influence of the physical assumptions 
made in light curve and spectral calculations. The importance 
of a consistent treatment of the explosion mechanism, light curves and
spectra became evident.   We found that these normal bright supernovae
show a   small spread in $M_V$.
  These papers argued that different models are needed
to explain the observations and $H_o$ was derived to be
$66 \pm  10 km/sec Mpc$.
 However, none
 of the previous models produced subluminous
SNe~Ia as required to account for SN1991bg (see above).
 In a recent paper 
(HKW95),
we have shown that the existence of a range of luminosities of Type Ia
 can be understood in 
the framework of pulsating delayed detonation scenario. On the other hand,
Woosley and Weaver suggested the class Helium detonations to be responsible
for Type Ia Supernovae (see above).
  Moreover, detailed NLTE analysis of  
SN1994D showed that classical delayed detonation models may provide 
better fits to both the light curves and spectra, if the C/O white dwarf had a somewhat
 smaller central density than anticipated in our previous analyses (H95).
So, in conclusion, the picture became somewhat confusing.
 
 New, well-defined data even of very distant SN~Ia have been published during the 
last two years. This allows for a much more comprehensive study than previously 
undertaken (MH94). New questions can also be investigated such as the suggestion by 
Bartlett et al. (1995) that $H_o$ may show variations on large scales with 
a global value  as low as 30 km/sec Mpc.
 
 To address the problems raised above, one-dimensional
models have been  considered which  encompass 
all currently discussed explosion scenarios.
 In section 2, our  set of models is discussed, in particular,
the Helium detonations. In section 3, general properties of the light curves
are studied. In the following section, the influence of the red shift on 
the observations is discussed. In section 5, the method of comparison of models
and observations and the individual comparisons are presented. Sections 6 and 7
 focus on the  relation between the type of the explosion and of the host galaxy and
on the cosmological constants
$H_o$ and $q_o$. Finally, the results are summarized,
the relation between type of explosion and the host galaxy is discussed 
 and conclusions  are drawn.

\section{Models}

 A set of 37 SNe~Ia parameterized explosion models has been considered which encompasses all
currently discussed explosion scenarios (Table 1). The explosions are calculated using  one-dimensional
Lagrangian hydro with artificial viscosity (Khokhlov, 1991ab) 
and  radiation-hydro codes including nuclear networks  (H\"oflich et al. 1995b).
 The latter code is based on our light curve code that solves the hydrodynamical equations explicitly
by the piecewise parabolic method (Collela and Woodward 1984) and includes the solution of the radiation transport
implicitly via the moment equations, expansion opacities similar to Karp, Lasher \& Salpeter (1977),
 and a  detailed equation of state. Radiation transport has been included   
to provide a smoother transition from the hydrodynamical explosion to the phase
of free expansion.  We omit $\gamma$-ray transport during the hydrodynamical phase  because the 
high optical depth of the inner layers and the negligible energy input due to radioactive decay of $^{56} Ni$
 (M\"uller \& H\"oflich 1991, HMK92, HMK93, HKW95).
 Nuclear burning is taken into account using Thielemann's  network
 (Thielemann,
Arnould \& Truran 1987, Cowan, Thielemann \& Truran  1991 Thielemann, Nomoto \&   Hashimoto 1994
 and references therein). During the hydro, 
 a network of 20 isotopes is considered to properly describe the energy release. Based on a network of 216
isotopes, the final chemical
structure is calculated  by postprocessing the hydrodynamical model.
 The accuracy of the energy release has been found
to be about 1 to 3 \% in the reduced network compared to about 10 \% for
the approximations used previously (Khokhlov, 1991ab).
  For SNe~Ia, the new code has
been compared to previous results for several models, such as deflagration
(DF1), classical delayed 
detonations (M35/37), and pulsating delayed detonations (PDD5/6).  
 We note that radiation transport effects remain unimportant and
 only affect small scales even in models (PDD5/6) which enter 
the phase of homologous expansion on time scales of 10 to 60 minutes.
 
Based on the explosion models,  further
hydrodynamical evolution, and  bolometric and monochromatic light
curves are calculated using a scheme recently developed and widely
applied to  SN Ia (e.g. HKM91,
 KMH93, MH94).  
 In principle, it is the same code as that described above, but 
 nuclear burning is neglected and
 $\gamma $ ray transport is included via a Monte Carlo scheme. The
monochromatic colors $B$, $V$, $R$, and $I$ 
are calculated using 100 discrete wavelength bands.  
 For several models,
detailed NLTE-spectra have been constructed (H95).
The colors based on the NLTE atmospheres and LC calculations have been compared.
Typically, the absolute errors in V, B, R, and I are below $0.05 $ to $0.1^m$ 
near maximum light and remain between 0.2 to 0.4$^m$ at late time. 
Another uncertainty which affects the comparison with observations is due to the filter functions.
For observations, any  filter
can be recalibrated to a set of standard stars, but the frequency response must be 
known   to translate theoretical monochromatic fluxes into colors.
 In the literature (Lamla 1982, Bessell 1990, Challis 1994), the transmission functions
show  wide variations, especially, for the infrared colors. For example, the { I} color
may or may not cut off the region of the strong Ca II-IR emission. This has little influence on measurements of
 standard stars, but it will vastly change the  color in  SNe~Ia, in particular, the strength of the 
secondary IR maximum.
For more details see MH94 and HKW94.
Previously,  we used  transmission functions as reconstructed at Vilnia  (Lamla 1982). However,
most of the modern measurements are based on  Cousin's functions as defined by
 Bessell (1990) which are used in this study.
 Small differences in $M_V$ and related quantities with respect to previous studies
can be understood by the change
of the filter calibrations.

\subsection{Explosions of massive white dwarfs}
  
The first group of models consists of thermonuclear explosions of carbon oxygen white
dwarfs close to the Chandrasekhar limit. We cover   all scenarios being suggested
including detonations (DET1/2, hereafter D-series), deflagrations (W7, DF1, DF1mix),
 delayed detonations (N21/32, M35-39;  N- and M-series) and 
 pulsating delayed detonations (PDD1-9;  P-series).
 The deflagration speed is parameterized as $D_{def} = \alpha a_s$, where $a_s$ is the local sound velocity
ahead of the flame and $\alpha $ is a free parameter. The speed of the detonation wave is given by the
sound-speed behind the front. For delayed detonation models, the transition to a detonation is given 
by another free parameter $\rho _{tr}$. When the density ahead of the deflagration front
reaches $\rho_{tr}$, the transition  to a detonation is forced by increasing $\alpha $ to 0.5 over 
5 time steps bringing
the speed well above the Chapman-Jouguet threshold for steady deflagration.
 For pulsating delayed detonation models,
 the initial phase of burning fails to release sufficient energy to disrupt the white dwarf.
During the subsequent contraction phase, compression of the mixed layer of 
products of burning and C/O formed 
at the dead deflagration front would give rise to a detonation via
compression and  spontaneous ignition (Khokhlov 1991b).
 In this
scenario, $\rho_{tr} $ represents  the density at which 
the detonation is initiated after the burning front dies out.
 Besides the description of the burning front, 
 the central density of the WD at the time of the explosion is another free parameter. For white
dwarfs close to the Chandrasekhar limit, it  depends sensitively on the   
chemistry and the accretion rate $\dot M $ at the time of the explosion.
 In a recent study on SN1994D (H95),           
evidence was found that the models with a somewhat smaller 
central density provide better agreement with 
both the observed spectra and light curves.
 Therefore, we have extended  our grid of PDD models starting with lower central
densities. 
 
\begtabsmall
\table{1}{
Overview of investigated theoretical SNe~Ia models.
The quantities given
in columns 3 to 11 are:  $M_\star$ white dwarf mass; $\rho_c$ central
density of the white dwarf; $\alpha$ ratio of deflagration velocity and
local sound speed; $\rho_{tr}$ transition density at which the
deflagration is assumed to turn into a detonation; $E_{kin}$ final
kinetic energy; $M_{Ni}$ mass of synthesized \ni.
 In Model DF1MIX the composition
was completely homogenized after burning had stopped. For the helium detonations,
the mass is of the C/O core, and He-layers of the hydrostatic WD are given separately.
}
 
\halign{#\hfil&&\quad#\hfil\cr}
\hline 
\+Model~~~~~~~~~ & Mode of~~~~~~~~ & $M_\star$ ~~~~~~~~~& $\rho_c$ ~~~~~~~~~ & $\alpha$ ~~~~~~~~~&
  $\rho_{tr}$~~~~~~~~~ &
 $E_{kin}$~~~~~~~~~  & $M_{Ni}$~~~~~~~~ & Symbol~~~~~~~~ \cr
\+& explosion & $[ M_\odot ]$ & $[10^9 g~cm^{-3} ]$ &  &   $[ 10^7 g~cm^{-3} ]$ & $[   10^{51} erg    ]$ & $[ M_\odot ]$ &     \cr
\hline                     
\+DET1   & detonation   &  1.4  &  3.5   &   ---  &  ---  &  1.75  &
                          0.92  &   $\ast     $ \cr
\+DF1    & deflagration &  1.4  &  3.5   &  0.30  &  ---  &  1.10  &
                          0.50  &   $\circ $ \cr
\+DF1MIX & deflagration &  1.4  &  3.5   &  0.30  &  ---  &  1.10  &
                          0.50  &   $\circ $ \cr
\+W7     & deflagration &  1.4  &  2.0   &  n.a.  &  ---  &  1.30  &
                          0.59  &   & $\circ $ \cr
\hline
\+N21     & delayed det. &  1.4  &  3.5   &  0.03  &  5.0  &  1.63  &
                           0.83  & $\bullet $  \cr
\+N32     & delayed det. &  1.4  &  3.5   &  0.03  &  2.6  &  1.52  &
                           0.56  & $\bullet $  \cr
\+M35     & delayed det. &  1.4  &  2.8   &  0.03  &  3.0  &  1.56  &
                           0.67  & $\bullet $  \cr
\+M36     & delayed det. &  1.4  &  2.8   &  0.03  &  2.4  &  1.52  &
                           0.60  & $\bullet $  \cr
\+M37     & delayed det. &  1.4  &  2.8   &  0.03  &  2.0  &  1.49  &
                           0.51  & $\bullet $  \cr
\+M38     & delayed det. &  1.4  &  2.8   &  0.03  &  1.7  &  1.44  &
                           0.43  & $\bullet $  \cr
\+M39     & delayed det. &  1.4  &  2.8   &  0.03  &  1.4  &  1.38  &
                           0.34  & $\bullet $  \cr
\+M312    & delayed det. &  1.4  &  2.8   &  0.03  &  1.0  &  1.35  &
                           0.20  & $\bullet $  \cr
%
\hline
\+PDD3    & pul.del.det. &  1.4  &  2.1   &  0.04  &  2.0  &  1.37  &
                           0.49  & $ bl.tr. $  \cr
\+PDD535  & pul.del.det. &  1.4  &  2.7   &  0.035 &  0.45 &  0.34  &
                           0.15  & $ bl.tr. $  \cr
\+PDD54   & pul.del.det. &  1.4  &  2.7   &  0.04  &  0.82 &  1.02  &
                           0.19  & $ bl.tr. $  \cr
\+PDD5    & pul.del.det. &  1.4  &  2.7   &  0.03  &  0.76 &  1.23  &
                           0.12  & $ bl.tr. $  \cr
\+PDD8    & pul.del.det. &  1.4  &  2.7   &  0.03  &  0.85 &  1.30  &
                           0.18  & $ bl.tr. $  \cr
\+PDD7    & pul.del.det. &  1.4  &  2.7   &  0.03  &  1.1  &  1.40  &
                           0.36  &   $ bl.tr. $  \cr
\+PDD9    & pul.del.det. &  1.4  &  2.7   &  0.03  &  1.7  &  1.49  &
                           0.66  &   $ bl.tr. $  \cr
\+PDD6    & pul.del.det. &  1.4  &  2.7   &  0.03  &  2.2  &  1.49  &
                           0.56  &   $ bl.tr. $  \cr
\+PDD1a   & pul.del.det. &  1.4  &  2.4   &  0.03  &  2.3  &  1.65  &
                           0.61  &   $ bl.tr. $  \cr
\+PDD1c   & pul.del.det. &  1.4  &  2.4   &  0.03  &  0.71 &  0.47  &
                           0.10  &   $ bl.tr. $  \cr
\hline
\+HeD2    & He-det.      &  0.6+0.22  &  .013   &  ---   &  ---  &  0.94  &
                           0.43  &   $\star     $  \cr
\+HeD4    & He-det.      &  1.0+0.18  &  .150   &  ---   &  ---  &  1.50  &
                           1.07  &   $\star     $  \cr
\+HeD6    & He-det.      &  0.6+0.172 &  .0091  &  ---   &  ---  &   0.72 &
                           0.252 &   $\star     $  \cr
\+HeD7    & He-det.      &  0.6+0.14  &  .0089  &  ---   &  ---  &   ---- &
                           ----  &   goes ''nova"     \cr
\+HeD8    & He-det.      &  0.8+0.16  &  .025  &  ---   &  ---  &   1.08 &
                           0.526 &   $\star $      \cr
\+HeD10   & He-det.      &  0.8+0.22  &  .036  &  ---   &  ---  &   1.24 &
                           0.75  &   $\star $      \cr
\+HeD11   & He-det.      &  0.9+0.16  &  .061  &  ---   &  ---  &   1.37 &
                           0.87  &   $\star $      \cr
\+HeD12   & He-det.      &  0.9+0.22  &  .083  &  ---   &  ---  &   1.45 &
                           0.92  &   $\star $      \cr
\hline
\+CO095   & detonation   &  0.95  &  .016   &  ---   &  ---  &  1.10  &
                           0.18  &   $\ast      $  \cr
\+CO10    & detonation   &  1.0  &  .020   &  ---   &  ---  &  1.22  &
                           0.32  &   $\ast      $  \cr
\+CO11    & detonation   &  1.1  &  .04   &  ---   &  ---  &  1.44  &
                           0.58  &   $\ast      $  \cr
\hline
\+DET2     & detonation       &  1.2        &  0.04  &  ---  &  ---  &
                             1.52  &  0.63  &   $\triangle $  \cr
\+DET2ENV2 & det.+envelope  &  1.2 + 0.2  &  0.04  &  ---  &  ---  &
                             1.52  &  0.63  &   $ \triangle $  \cr
\+DET2ENV4 & det.+envelope  &  1.2 + 0.4  &  0.04  &  ---  &  ---  &
                             1.52  &  0.63  &   $\triangle $  \cr
\+DET2ENV6 & det.+envelope  &  1.2 + 0.6  &  0.04  &  ---  &  ---  &
                             1.52  &  0.63  &   $ \triangle $ \cr
\hline         
\endtab          
 
\subsection{Merging white dwarfs}
 
 A second group of models is the outcome of the merger scenario (Webbink 1984; Iben \& Tutukov 1984;
Paczynski 1985). After the initial merging process,
one low density white dwarf is surrounded by an extended envelope
 (Hachisu, Eriguchi \& Nomoto 1986ab,  Benz,   Thielemann \& Hills     1989).
 This scenario is mimicked by our envelope models DET2ENV2...6 
 in which we consider the detonation in a
low mass white dwarf surrounded by a compact envelope between 0.2 and 0.6 $M_\odot $. 
 For more details on these models see HKM92 and  KMH93.

\vfill\eject
 
\subsection{Explosions of sub-Chandrasekhar mass white dwarfs}
 We have included explosions of sub-Chandrasekhar mass white dwarfs
(hereafter HeD-series),  previously considered  by several other authors 
  (e.g. Nomoto 1980, 1982,  Woosley, Weaver \& Taam 1980, WW94, Livne 1990, Livne \& Glasner 1990).
 Until now, no monochromatic LCs have been
 constructed for this class of models. Only WW94
calculated bolometric light curves, but under several restrictions, namely,
omission of a detailed $\gamma $-ray transport,
and, more importantly, the assumption that the line opacities are independent
of the structure and time evolution of the expanding envelope. Note, that the influence  of the 
opacities on the resulting LCs is comparable to the changes caused by different hydrodynamical models
(HMK93).
 \begfig 0.1cm
\psfig{figure=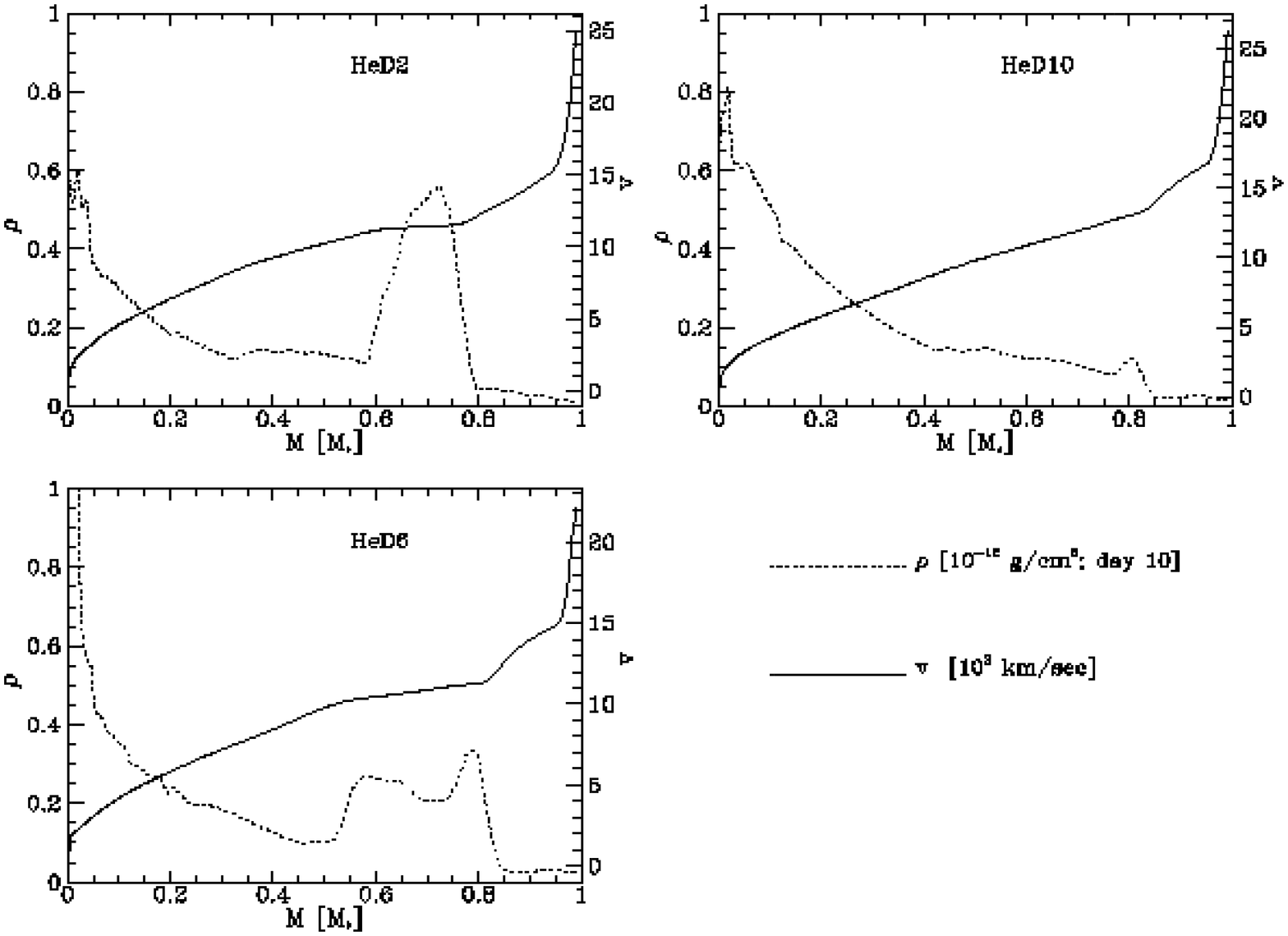,width=12.6cm,rwidth=9.5cm,angle=270}
\figure{1}{
Density and velocity as a function of  mass
for the models HeD2, HeD6, and HeD10.}
%
%
\vskip 0.1cm
\psfig{figure=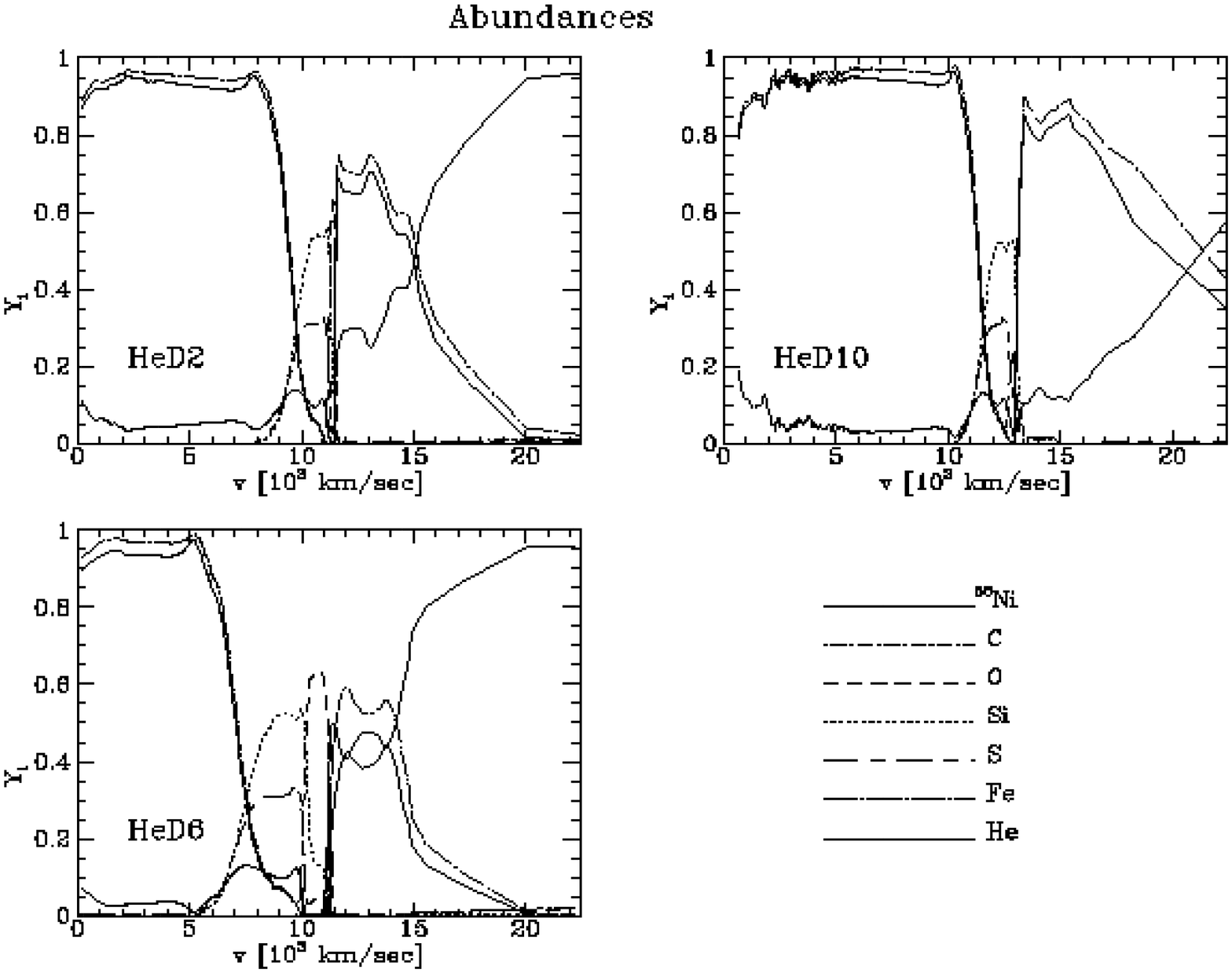,width=12.6cm,rwidth=9.5cm,clip=,angle=270}
\figure{2}{ Final composition as a function of velocity
for the models HeD2, HeD6, and HeD10.}
\endfig

 For ease of comparison,
 we have  used parameters  close  to those suggested by Woosley and Weaver (1994).
 To prevent repetition,
we refer to the latter work for a detailed discussion of this 
 class of models. Here, the basic assumptions and some technical details are outlined.
About 278 depth points are used for the CO core. For the ignition of the carbon detonation,
 a good mapping  is critical. To provide a resolution of
roughly $2.~10^{-5}~M_\odot $, the discretization in mass was increased logarithmically.
 This allowed for a higher resolution in the center
 by  a factor of 10  compared to the outer
 layers.
 The initial chemical composition was 70\% $^{16}O$, 29 \%
$^{12}C$, 1 \% $^{22}Ne$ by mass, and solar mass fractions of all other elements. 
We began with C/O white dwarfs of 0.6, 0.8, 0.9, and 1.0 $M_\odot $ with temperatures according to WW94        
              for the lower masses and Schaller et al.
 (1992) for the high mass. The accretion of
Helium is calculated by our radiation-hydro code, where we neglect partial time derivatives to evolve
a hydrostatic structure. The accretion rates are tuned such that about 0.16 and
0.22 $M_\odot$ of Helium are accreted when the thermonuclear runaway occurs at the bottom of the outer He layer.
In general, our models resembled those of WW94 but we want to note that 
the final outcome depends very sensitively on details such as the exact amount of Helium.
 For the lowest mass white dwarf,
 the thermonuclear runaway occurred already at a time when only 0.15 $M_\odot$ had 
been accreted; whereas about $0.17 M_\odot$ of He 
is required  to trigger the central carbon detonation (Table 1).  The discrepancy with WW94
may be explained by different discretizations, the numerical schemes, energy transport
during the accretion phase of the white dwarf,
or by nuclear rates, because the actual time of the explosion depends sensitively 
on the $^{14}N(e^-,\nu) ^{14}C$ rate. Within the uncertainties, we cannot rule out that, in reality, a larger 
amount of He can be accreted during the stellar evolution.
  Therefore, for this mass, we suppressed the  runaway accordingly. 
 The final velocity and density profiles and the chemical profile for the most abundant 
elements for several of our models are given in Figs. 1 and 2.
 Helium detonations show a qualitatively different structure in comparison to all models with a 
 Chandrasekhar mass white dwarf. The intermediate mass elements
are sandwiched by  Ni and He/Ni rich layers at the inner and outer regions, respectively. 
 Generally,  the density   smoothly decreases  with
mass because partial burning produces almost the same amount of kinetic energy as  the total
burning, but a moderate  shell-like structure is formed just below the former Helium layers.
 Observationally,  a distinguishing feature of this scenario  
 is the presence of Helium and Ni  with expansion velocities above 11,000 to
14,000 $km~ s^{-1}$.
 Typically, about 0.07 to 0.13 of Ni are produced in the outer layers, mainly depending on the mass of the
Helium shell. Note   the existence of 
a lower limit for the amount of Ni in the outer 
layers because of the requirements on the energy release
to trigger the central carbon detonation.
With increasing Helium mass, the subsequent
central compression and the size of the central region increase where,
 starting from an C/O rich mixture, the density becomes
sufficiently high to burn carbon and oxygen up to \ni .
For the same reason, i.e. the higher densities,   the Ni production rises with increasing mass. 
Already the model HeD10  (Table 1) 
 produces predominantly \ni and only little intermediate mass elements at velocities 
$\leq 10000 km/sec $ (Fig. 2).
Therefore, models with a CO-core of more than $0.9 M_\odot$    
 can be ruled out from the observations.                
 \begfig 0.1cm
\psfig{figure=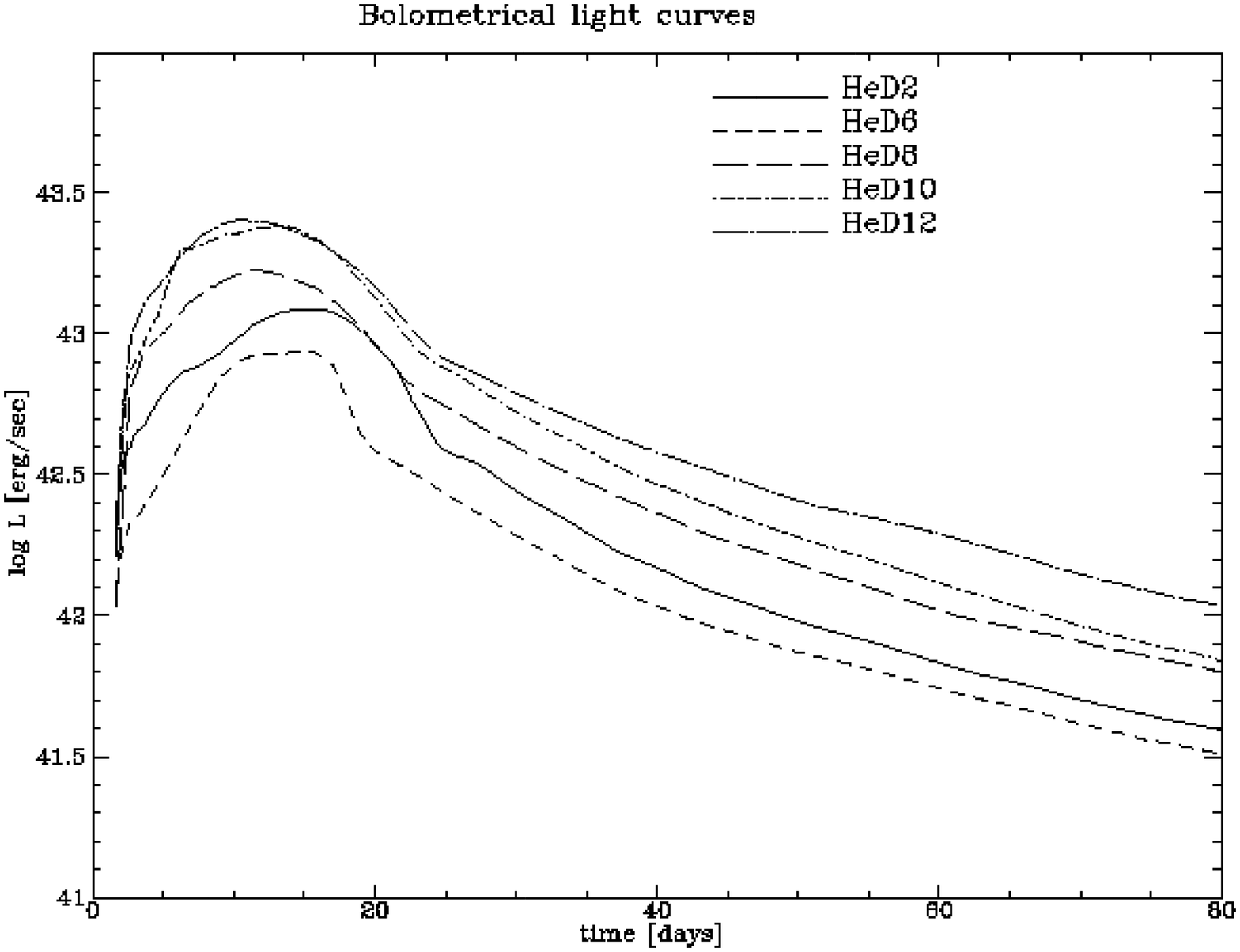,width=12.6cm,rwidth=9.5cm,angle=270}
\figure{3}{
Bolometric light curves of some of the Helium detonation models (see Table 1).}
\endfig
 \begfig 0.1cm
\psfig{figure=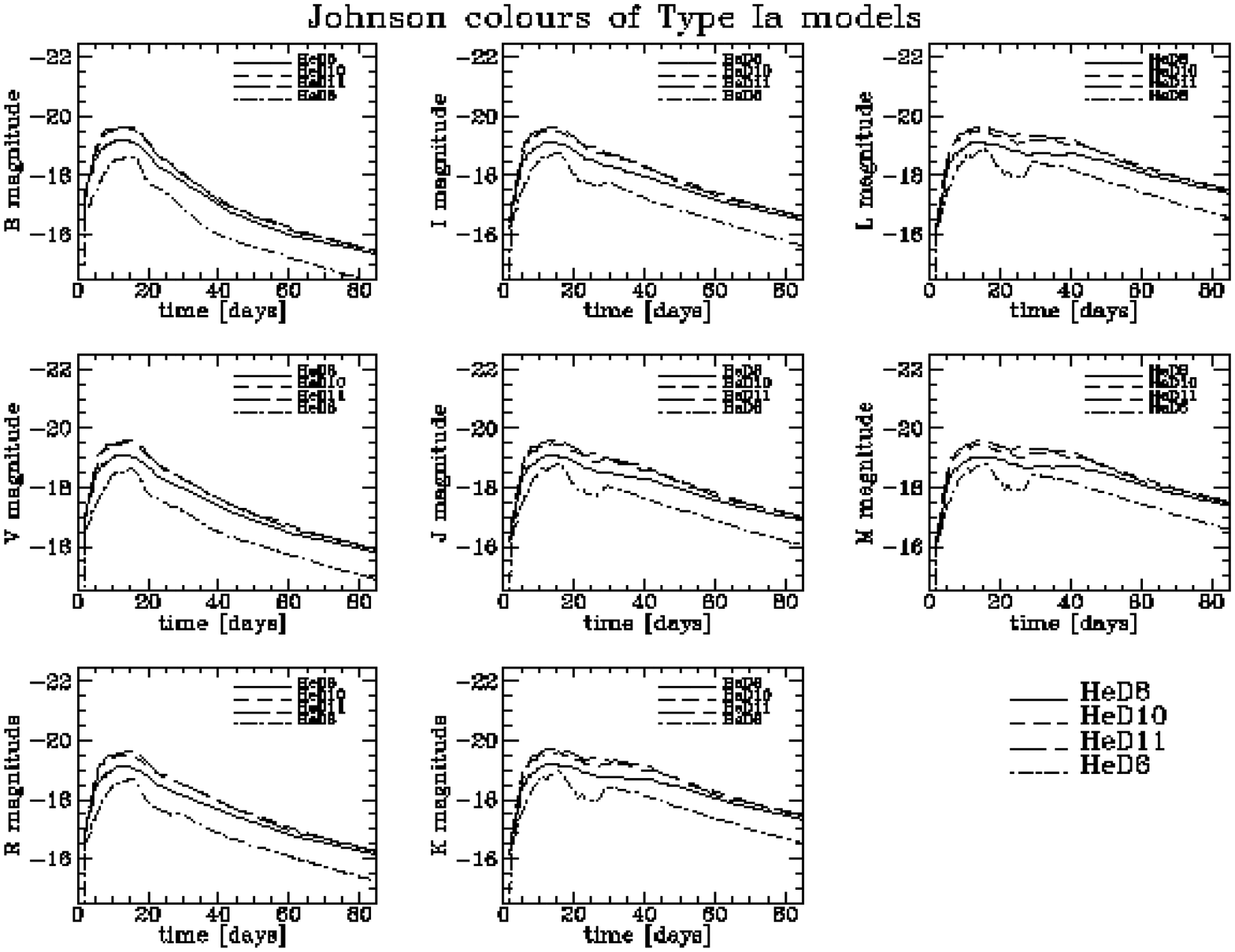,width=12.6cm,rwidth=9.5cm,angle=270}
\figure{4}{
Monochromatic light curves in Johnson's color system.}
\endfig
 
 Bolometric and monochromatic light curves are shown in Figs. 3 and 4. Helium detonations
are capable of producing both  bright and subluminous models. The characteristic
of the light curves can be mainly explained by the presence of \ni at expansion velocities
comparable to the photospheric velocities at about maximum light. 
 Initially, the brightness  increases rapidly.
 Often, a small plateau is  formed   because the contribution
of the radioactive heating from the deeper layers is delayed due to the diffusion time scales.
 The first plateau is most pronounced in models with little outer helium,
because those produce more intermediate mass elements.
 For the same reason, the final incline
to maximum light is rather slow.
       Both for bright and subluminous models, \ni  heating results in high color 
temperatures at maximum light (see table 2 and section 3). 
 In general,  the postmaximum decline is steep 
both in the bolometric and  optical LCs due to
the fast  increase of the escape probability for gamma-ray photons and due to the rapid 
decay rate of \ni compared to \co . 
 Note that   the radioactive decay of the outer \ni 
contributes to the  heating of the envelope
well after the corresponding layers have passed the photosphere
because  the $\gamma $ opacities are significantly lower than the optical opacity ($1/35 cm^{2}/g$ vs 
$0.1 cm^{2}/g$). Consequently, 
 layers well beyond the optical photosphere  receive energy.
 This additional energy source keeps the temperature and, consequently, the opacity at the
photosphere sufficiently high to allow $R_{ph}$ to increase well after maximum light.
 Consequently (H\"oflich et al. 1995), the monochromatic IR-light curves show secondary maxima even 
for subluminous models.
\begtabsmall
\table{2}{
Some observable characteristics of the models of Table 1.
    The quantities given in columns 2 to 9  are: The maximum
bolometric luminosity $L_{bol}$ in ergs/s,
 the rise time to bolometric
maximum $t_{bol}$ in days, the bolometric luminosity relatively to the 
energy generation by radioactive decay at $t_{bol}$,
 the
absolute visual magnitude  $M_V$,    the rise time to
visual maximum $t_V$ in days, the color index B-V at  maximum light, and
the post-maximum decline rates $dM_{15}$, $dM_{60}$, and $dM_{100}$ in V over
15, 60, and 100 days. $\Delta M_V$(15) provides an estimate of 
the possible error (see text).}
\bigskip
\hline 
\halign{#\hfil&&\quad#\hfil\cr}
\hline
\+Model~~~~~~~~~~~ & $ \lg L_{bol}$~~~ & $t_{bol}$~~~~  & ~~~~Q ~~~~&~~~~ $M_V$ ~~~~ & $t_V $~~~~
 & B-V~~~~ & $dM_{15} \pm \Delta dM_{15} $~~~~~~~~~~~&   $dM_{60}$ ~~~~~~&$ dM_{100}$~~~\cr
\hline 
\+DET1   & 43.51 &  7.5 & 0.76 & -19.99 &  8.5 &   -.06 & .071 -.008+.011 & .072 &  .045 \cr
\+DF1    & 43.18 & 13.5 & 1.03 & -19.33 & 15.5 &   0.14 & .080 -.003+.013 & .062 &  .040 \cr
\+DF1MIX & 43.17 &  7.3 & 0.62 & -19.10 & 13.2 &   0.11 & .059 -.033+.015 & .057 &  .039 \cr
\+W7     & 43.30 & 14.0 & 1.20 & -19.63 & 15.5 &   0.11 & .079 -.016+.011 & .069 &  .041 \cr
\hline
\+N21    & 43.41 & 10.7 & 0.87 & -19.81 & 12.7 &   0.07 & .052 -.008+.008 & .052 &  .042 \cr
\+N32    & 43.27 & 14.5 & 1.21 & -19.54 & 15.2 &   0.12 & .083 -.007+.012 & .068 &  .041  \cr
\+M35    & 43.30 & 13.5 & 1.03 & -19.51 & 14.8 &   0.03 & .053 -.015+.013 & .056 &  .041 \cr
\+M36    & 43.31 & 13.3 & 1.15 & -19.39 & 15.2 &   0.06 & .059 -.009+.004 & .060 &  .041 \cr
\+M37    & 43.24 & 13.5 & 1.17 & -19.26 & 15.1 &   0.10 & .064 -.007+.004 & .058 &  .040 \cr
\+M38    & 43.15 & 13.8 & 1.15 & -19.13 & 15.3 &   0.12 & .067 -.011+.004 & .059 &  .040 \cr
\+M39    & 43.06 & 14.3 & 1.22 & -19.01 & 15.4 &   0.15 & .073 -.013+.008 & .059 &  .039 \cr
\+M312   & 42.75 & 12.0 & 0.87 & -18.36 & 13.0 &   0.23 & .052 -.007+.012 & .054 &  .038 \cr
\hline 
\+PDD3   & 43.28 & 15.0 & 1.46 & -19.42 & 15.9 &   0.12 & .054 -.003+.005 & .051 &  .040 \cr
\+PDD535 & 42.51 & 17.8 & 0.97 & -17.77 & 21.3 &   0.60 & .046 -.010+.008 & .038 &  .025 \cr
\+PDD54  & 42.71 & 12.3 & 0.84 & -18.27 & 15.8 &   0.36 & .058 -.008+.008 & .057 &  .039 \cr
\+PDD5   & 42.54 & 11.2 & 0.84 & -17.99 & 13.0 &   0.44 & .087 -.010+.005 & .067 &  .043 \cr
\+PDD8   & 42.71 & 11.5 & 0.85 & -18.29 & 13.0 &   0.35 & .064 -.012+.017 & .066 &  .042 \cr
\+PDD7   & 43.01 & 11.3 & 0.83 & -18.93 & 13.5 &   0.22 & .056 -.013+.018 & .065 &  .042 \cr
\+PDD9   & 43.27 & 10.3 & 0.76 & -19.47 & 14.3 &   0.05 & .056 -.018+.008 & .058 &  .042 \cr 
\+PDD6   & 43.21 & 12.7 & 0.94 & -19.43 & 13.8 &   0.11 & .054 -.012+.012 & .059 &  .042 \cr
\+PDD1a  & 43.19 & 12.8 & 0.86 & -19.38 & 14.1 &   0.09 & .053 -.011+.005 & .057 &  .042 \cr
\+PDD1b  & 42.73 & 11.8 & 0.87 & -18.35 & 14.2 &   0.33 & .073 -.012+.005 & .067 &  .042 \cr
\+PDD1c  & 42.41 & 11.6 & 0.71 & -17.73 & 13.8 &   0.53 & .093 -.017+.005 & .068 &  .044 \cr
\hline 
\+HeD2     & 43.01 & 13.9 & 0.85 &  -18.99 & 15.9 &  +.06 & .108 -.008+.008 & .075 &  .047 \cr
\+HeD6     & 42.91 & 14.4 & 1.17 & -18.56 & 14.7 &  +.05 & .101 -.010+.004 & .063 &  .047 \cr
\+HeD8     & 43.22 & 13.4 & 1.08 & -19.21 & 13.8 &  -.02 & .071 -.009+.009 & .058 &  .042 \cr
\+HeD10    & 43.38 & 13.2 & 1.07 & -19.59 & 13.9 &  -.01 & .073 -.007+.007 & .066 &  .044 \cr
\+HeD11    & 43.38 & 13.7 & 0.97 & -19.62 & 14.1 &  -.01 & .075 -.008+.012 & .067 &  .045 \cr
\+HeD12    & 43.40 & 11.5 & 0.80 & -19.61 & 12.6 &  -.04 & .059 -.005+.005 & .052 &  .039 \cr
\hline 
\+  CO095  & 42.82 & 11.7 & 1.11 & -18.60 & 14.2 &   0.37 & .092 -.012+.014 & .074 &  .045 \cr
\+  CO10   & 43.12 & 11.8 & 1.26 & -19.00 & 14.0 &   0.25 & .071 -.011+.005 & .069 &  .044 \cr
\+  CO11   & 43.24 & 11.9 & 0.93 & -19.47 & 13.8 &   0.14 & .061 -.008+.007 & .065 &  .042 \cr
\hline
\+ DET2    & 43.32 & 13.7 & 1.16 & -19.67 & 14.2 &   0.14 & .088 -.014+.011 & .076 &  .045 \cr
\+ DET2ENV2& 43.27 & 16.6 & 1.24 & -19.41 & 19.5 &   0.16 & .049 -.011+.005 & .042 &  .041 \cr
\+ DET2ENV4& 43.25 & 19.0 & 1.34 & -19.31 & 21.1 &   0.20 & .044 -.007+.009 & .041 &  .040 \cr
\+ DET2ENV6& 43.21 & 19.3:& 1.26 & -19.21 & 21.8 &   0.23 & .042 -.006+.010 & .040 &  .039 \cr
\hline         
\endtab          
 
 Note that a significant  amount of the line emission at late
phases should be powered  by $^{56}Co$ at  high velocities.
Even very subluminous SNe~Ia must produce about  $0.1 M_\odot$
of \ni in the outer layers. Thus, in
 the subluminous models, about 1/2 to 1/4 of the radioactive material is ejected at high
velocities. This should  show up in late-time spectra. In the layers where He and \co coexist,
 we suspect that
 Helium is partially ionized by secondary, high energy electrons from the $\beta^+$-decay
of \co . Consequently, late
time spectra may have a superposed  Helium recombination
spectrum. Whether Helium can recombine by charge exchange or collisions 
needs further investigations.

\section{Comparison of theoretical light curves and observable relations}              
 Although our light curve calculations are most reliable for the bolometric fluxes, 
this investigation
will deal with  monochromatic LCs. $L_{bol}$ is uncertain 
from the observations for epochs earlier than $\approx $ 5 days after maximum light, because a
significant amount of the flux is emitted shortwards of the visual 
wavelength range (Branch \& Tammann 1992). Different  scenarios have been
studied in previous works, and the new ones, including  the Helium detonations,
can be understood in full analogy. Therefore, we  restrict our discussion to
some particular points and the display of some useful relations 
based on the entire sample.             
 
 Models can be  discriminated by  their
 monochromatic light curves (Fig. 5 and 6), in particular, by the shape before         
day 60.
 As a general trend, the maxima of subluminous supernovae are more pronounced.
The reason becomes obvious if we consider the following related point.
In the literature (Arnett, Branch and Wheeler 1985), the so called Arnett's law (1982) has been
 used to determine $H_o$ under the assumption that SN~Ia are standard 
candles. This law states that 
the bolometric luminosity $L_{bol}$ is equal to the instantaneous release of energy by radioactive
decay $L_{\gamma}$, i.e. $ Q= L_{bol} / \dot E_{\gamma} \approx 1$ independent of  the underlying physical 
model. Consistent with our earlier works (KMH93), we find 
  that  $ Q$ depends on the model because the opacities vary strongly with  
temperature  (HMK93). Typically, $Q$ varies from  0.7 to 1.4 (Table 2 and Fig. 7).
 Note that, in previous works, the related quantity
$ \tilde Q$  was defined as $L_{bol}$ compared to the instantaneous energy deposition 
by radioactive decay. $\tilde Q $ is larger by  about 5 to 25 \% compared to Q depending on the 
escape probability of $\gamma $ rays.
 Whereas $ \tilde Q $ is a measure of  the energy gain due to the opacity effect, Q contains
both the effects of the opacity and the increasing escape probability for $\gamma $ photons which
distinguish Arnett's analytical approach from our calculations.
 For models with fast rising LCs, the opacity  stays high and 
the photosphere receds mainly by the geometrical dilution of matter. For models with a slower
rise time or little \ni, the opacity drops strongly at about maximum light. The photosphere
receds quickly in mass, and thermal energy can be released from a region, being larger than in the 
first case. Consequently, the maximum is more pronounced, and the $\tilde Q$ exceeds unity.
 Because no additional energy is gained, the energy reservoir is exhausted faster, and the
post-maximum  decline becomes steeper.
 
 Independent of details of the explosion mechanism, the current models
have rise times between 11 to 14 and
13 to 16 days in $L_{bol}$ and V, respectively. Longer rise times of up to 22 days require 
 pulsating delayed detonations and envelope models which have a dense, shell-like structure. 
The low-mass WDs close to 0.6 $M_\odot$ can also produce some shell structure (Fig.1).
Models  with $   M_{Ni} \geq 0.4 M_\odot $ are similarly bright within
a  narrow strip of $\approx 0.5^m$ (Fig. 8).
 $M_V$ is rather insensitive to $M_{Ni}$ and the spread in 
$M_V$ at a given $M_{Ni}$ is comparable to the change of $M_V$ with $M_{Ni}$.
 If less \ni is produced, $M_V$  strongly declines   
with  $M_{Ni}$ with a similar rate both for massive and low mass WDs.
 Although as a general tendency, the brightness seems to 
increase within each series of models, the variations are  not monotonic     
even within a given type of explosion. E.g. the delayed detonation model N32 is brighter than
M36 despite its lesser \ni.  Below $0.4 M_\odot$, the overall spread in $M_V$
remains about the same.
 
            The subluminous models tend to be much redder at maximum light than normal bright SNe~Ia.
This is generally independent of the explosion mechanism (Fig. 8) because the radioactive heating by 
\ni is highly reduced.
 The exceptions are  Helium detonations of low-mass WDs (see last section).
Note that \ni  heating occurs independently from details of the explosion, because
a Helium detonation will produce predominantly \ni even if 
multidimensional effects become important (Livne  1990).  This
provides a clear distinction between explosions of sub- and Chandrasekhar mass white dwarfs. 
 The effect on the late time spectra was discussed in the previous section.
 For possible consequences for CO formation, see HKW 95.
 \begfig 0.5cm
\psfig{figure=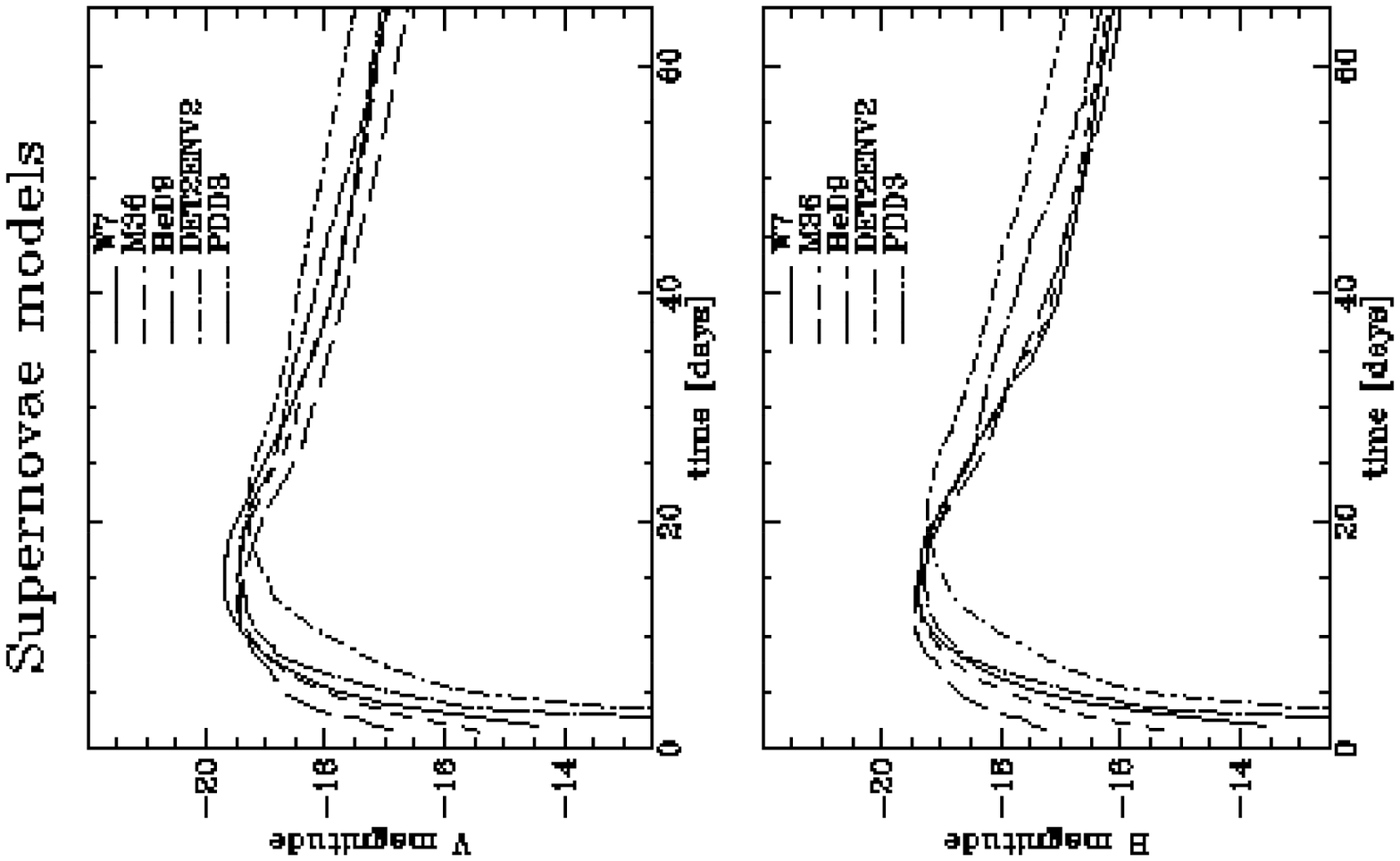,width=12.6cm,rwidth=9.5cm,clip=,angle=270}
\vskip -8.4cm
\psfig{figure=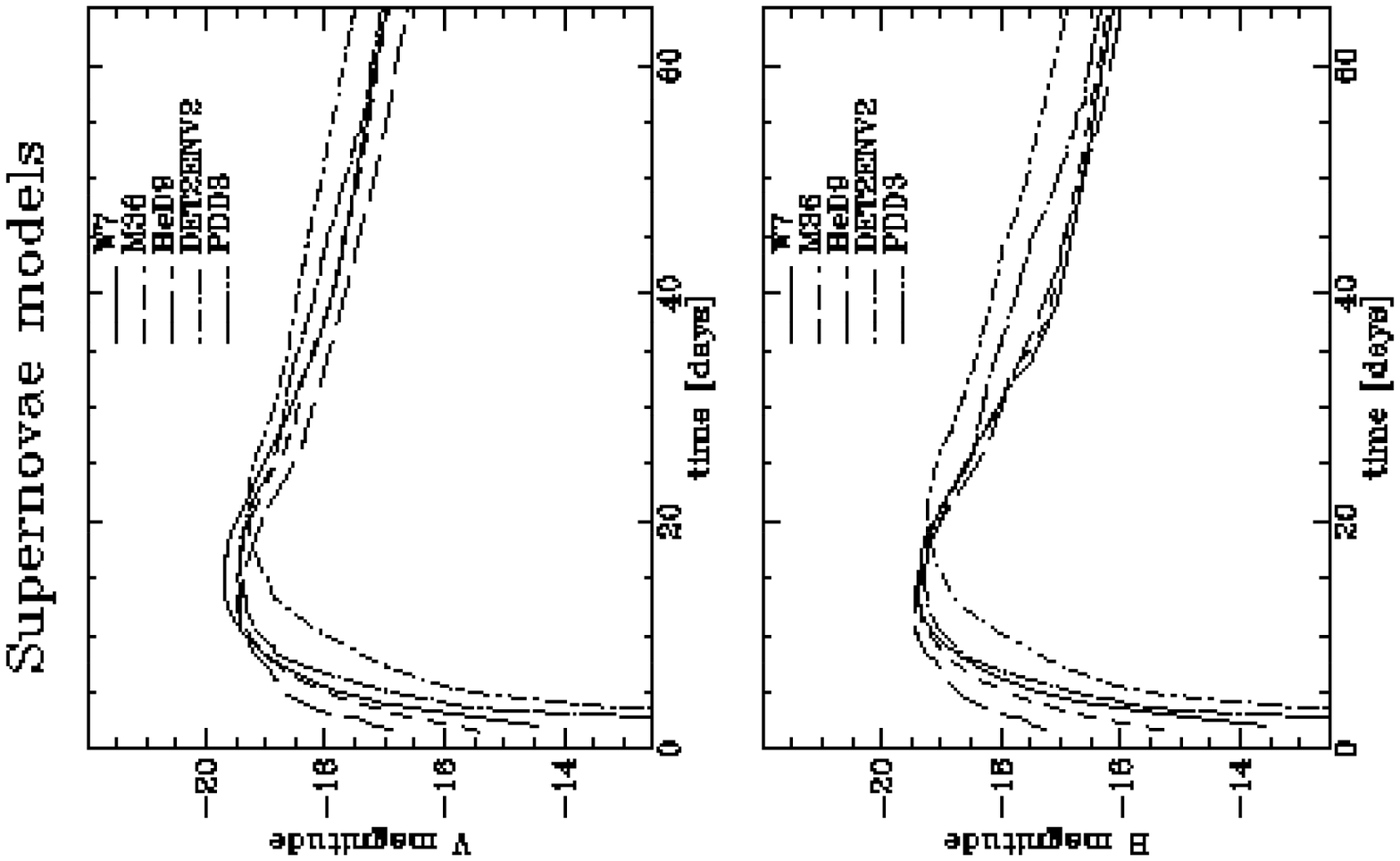,width=6.3cm,rwidth=9.5cm,clip=,angle=270}
\figure{5}{
Monochromatic B and V light curves of normal bright supernovae of deflagration
(W7), classical delayed detonation (M36), Helium detonation (HeD9), envelope
 (DET2ENV2), and pulsating delayed detonation (PDD3) models.}
\endfig
 \begfig 0.5cm
\psfig{figure=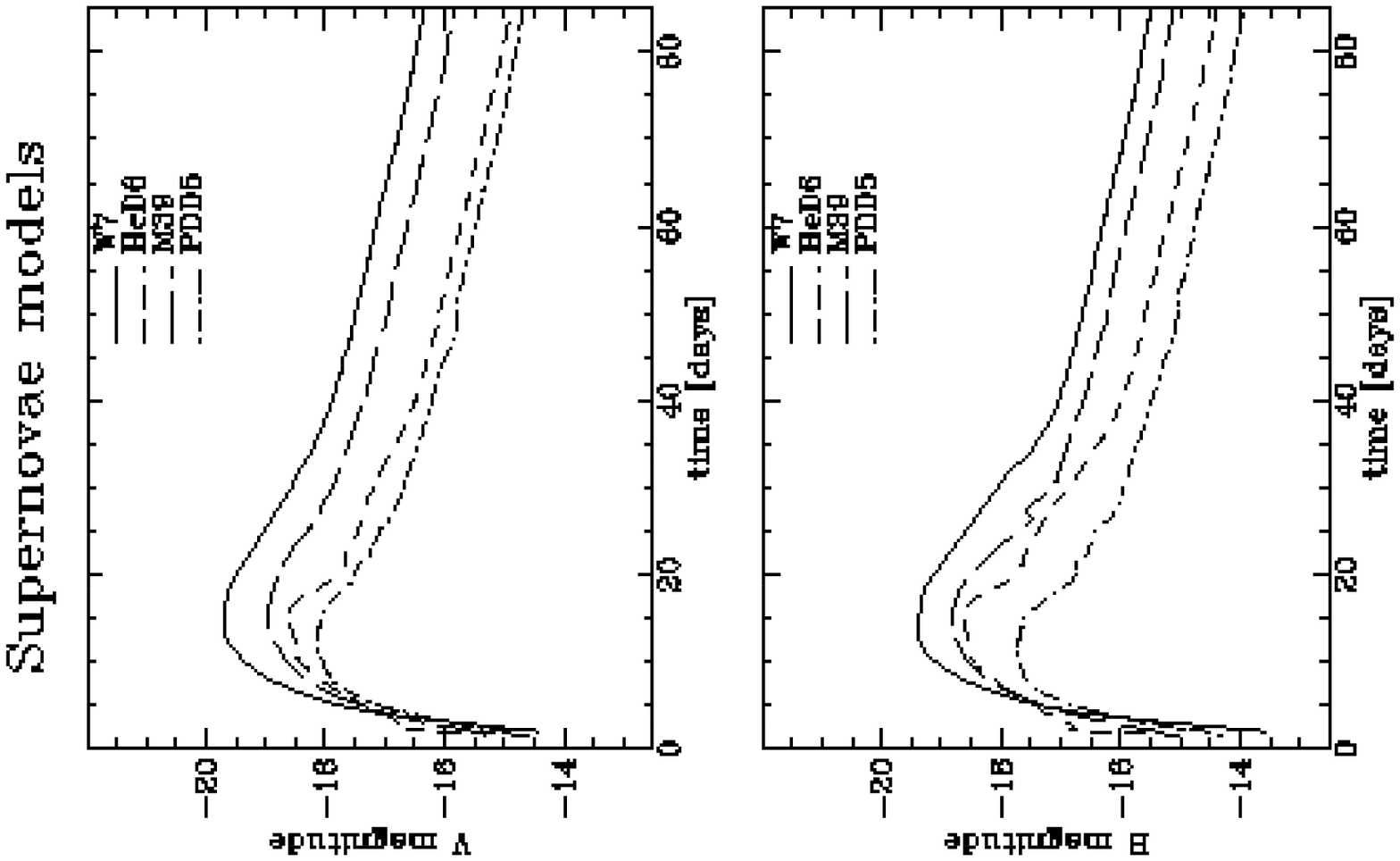,width=12.6cm,clip=,angle=270}
\vskip -4.05cm
\psfig{figure=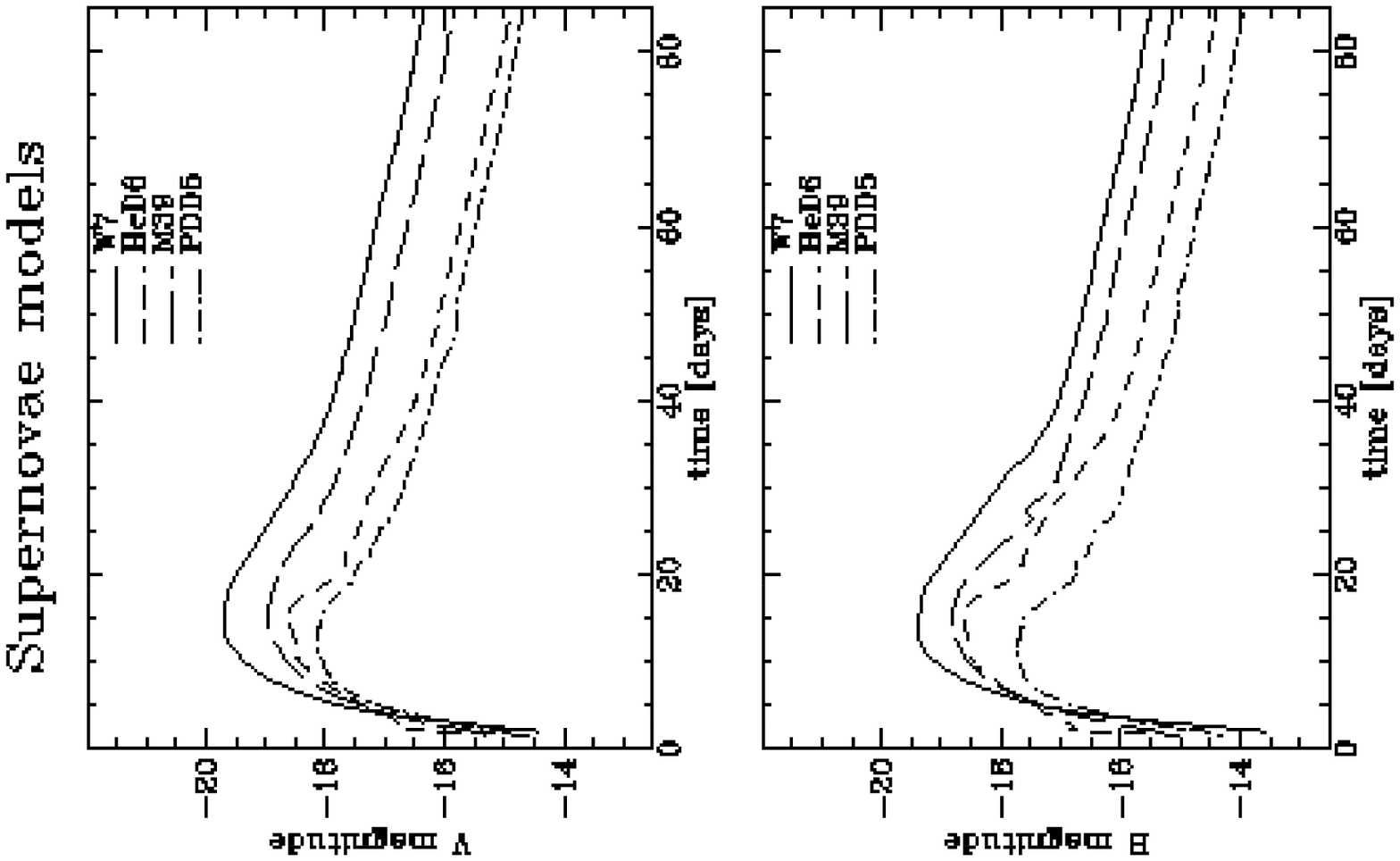,width=6.3cm,clip=,angle=270}
\figure{6}{
Monochromatic B and V light curves of subluminous  supernovae for         
pulsating delayed detonation (PDD5), classical delayed detonation (M39), and
Helium detonation (HeD6) models. For comparison, the deflagration model W7
is given.}
\endfig
 \begfig 0.1cm
\psfig{figure=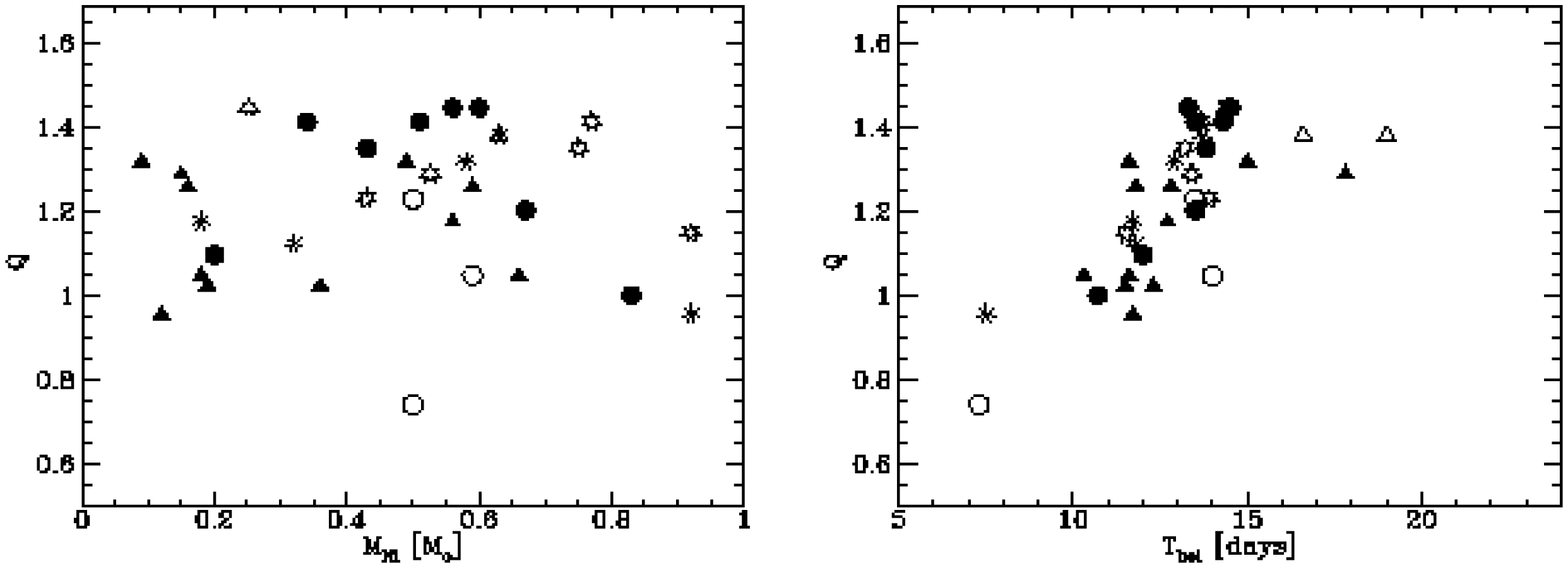,width=12.6cm,rwidth=9.5cm,angle=270}
\figure{7}{
Ratio Q between the bolometric luminosity and the energy release by 
 gamma decay at $t_{bol} $
 as a function of the $^{56}Ni $ mass (left) and 
rise time to bolometric maximum $t_{bol}$. The different
symbols correspond to different explosion scenarios (see Table 1).}
\endfig
 \begfig 0.1cm
\psfig{figure=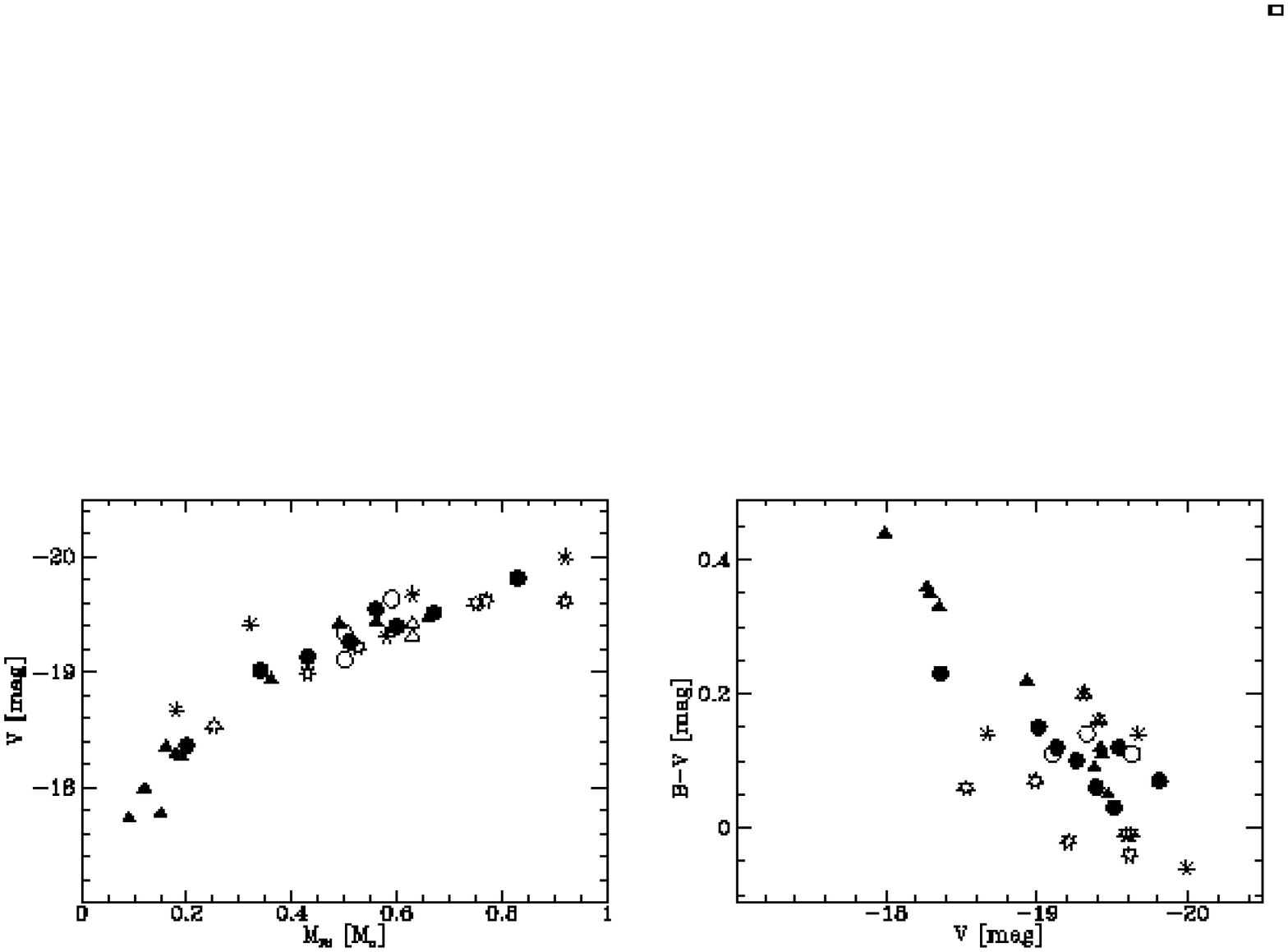,width=12.6cm,rwidth=9.5cm,angle=270}
\figure{8}{
Absolute visual brightness V  as a 
function of the $^{56}Ni$ mass (left) and intrinsic color B-V 
 as a function of V(right).}
\endfig
 \begfig 0.1cm
\psfig{figure=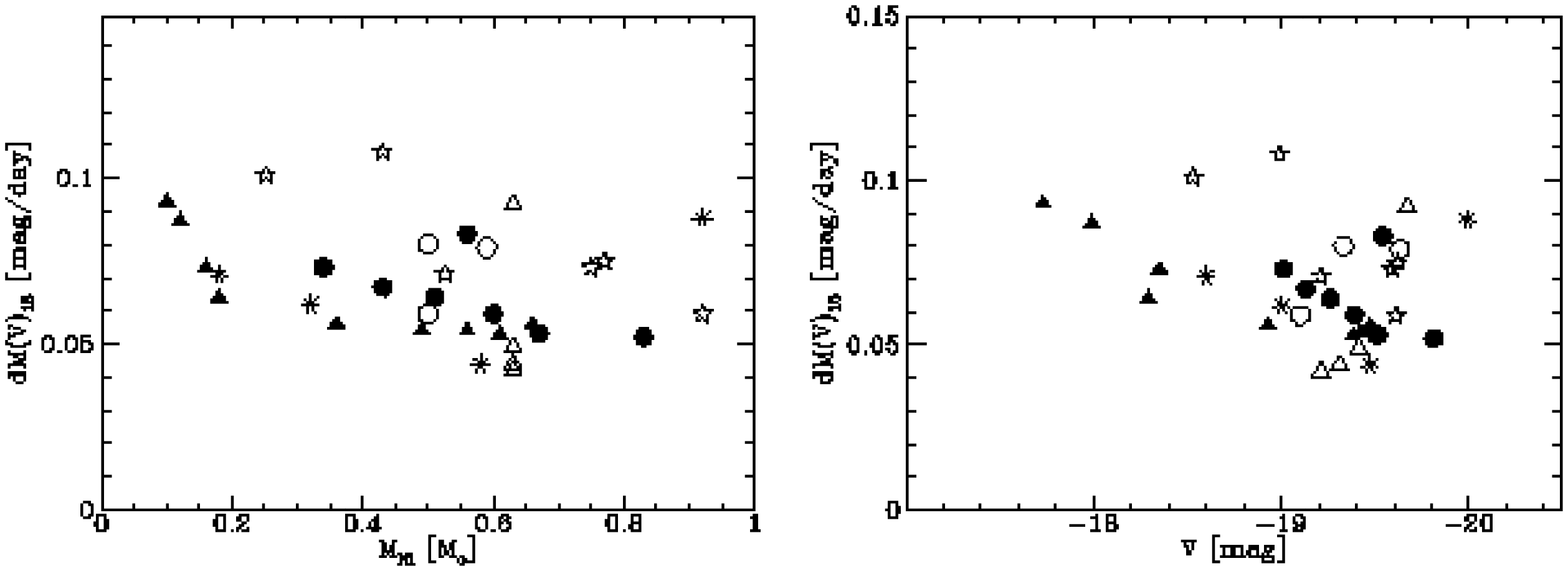,width=12.6cm,rwidth=9.5cm,angle=270}
\figure{9}{
Decline rate $dM_{15}(V)$ as a 
function of the $^{56}Ni$ mass (left) and as a function of V    
at maximum light.} 
\endfig
 \begfig 0.1cm
\psfig{figure=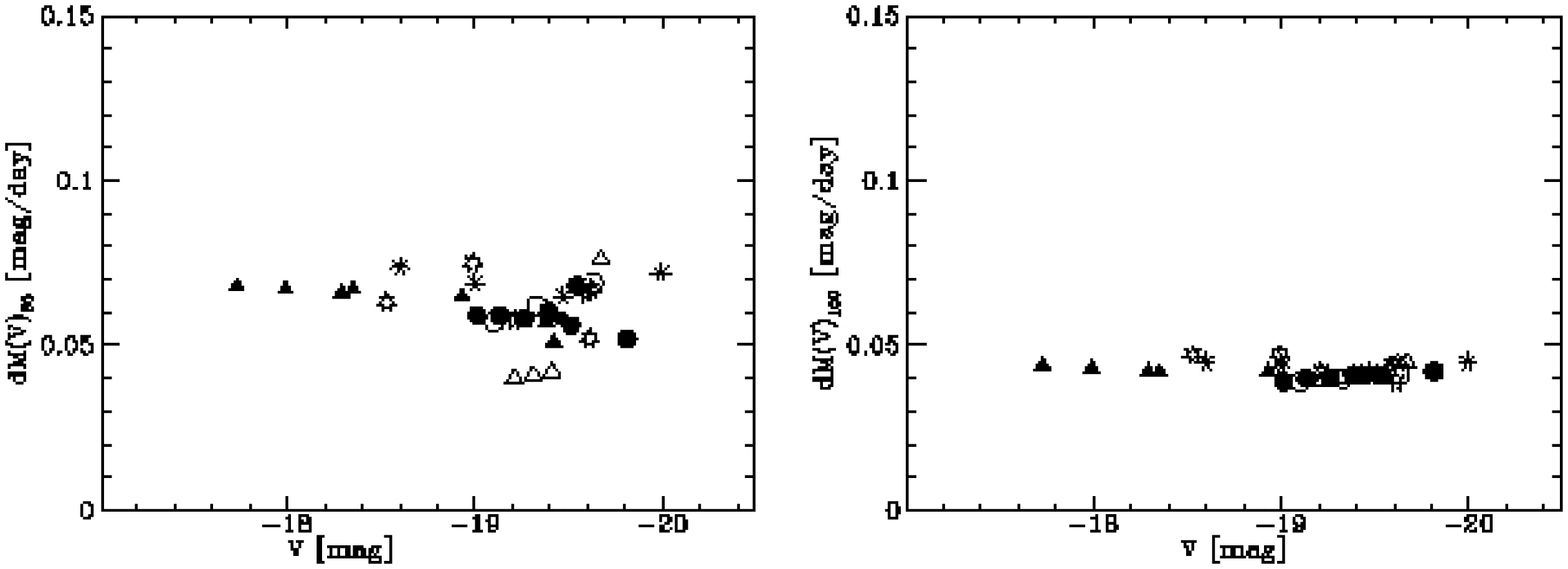,width=12.6cm,rwidth=9.5cm,angle=270}
\figure{10}{
Decline rate $dM_{60}(V)$ and  $dM_{100}(V)$ as   
functions of V at maximum light.}
\endfig

 The relations between the post-maximum declines $\Delta m_{15}(V)$,     
$\Delta M_{60}(V)$, and $\Delta M_{100}(V)$ and the Ni mass and absolute brightness $M_V$
are shown in Figs. 9 and 10. We use the decline in V instead of B as suggested by 
Phillips (1993) because B depends  sensitively on the filter functions and the metallicity of 
the progenitor. Because several of the light curves have flat maxima,  $\Delta M_V(15)$
is sensitive to errors in the time of maximum light. The latter may be affected by differences in   
 the filter functions, cosmological red shifts, and the uncertainties inherent to our light curve  
calculations. To estimate the size of  errors, we allowed for a shift in time corresponding to a 
change of $0.05^m$ in $M_V$.
 The early decline is mainly governed by the receding
photosphere. Within each series of models,  $dM_V(15)$  becomes larger with decreasing
\ni mass and, thus, with increasing $M_V$.  In total, a
spread of roughly 0.5$^m$ must be expected for the $M_V(dM_V(15))$ relation. 
 For $\Delta M_{60}$,                  
 the time base always reaches the linear tail produced by the \co decay,  but the
relative contribution of the linear tail varies. This breaks the overall monotonicity in the trends.
Finally,
 $\Delta m_{100}$ is  insensitive to both the underlying model and the absolute 
brightness because it is governed by the radioactive decay of \co.

\section {Cosmological  redshift}                          
 
 Before the comparison with observations, we want to address the
question of the influence of the cosmological red shift both on
the shape of  light curves and on the 
correction factors for the interstellar reddening.
\begfig 0.1cm
\psfig{figure=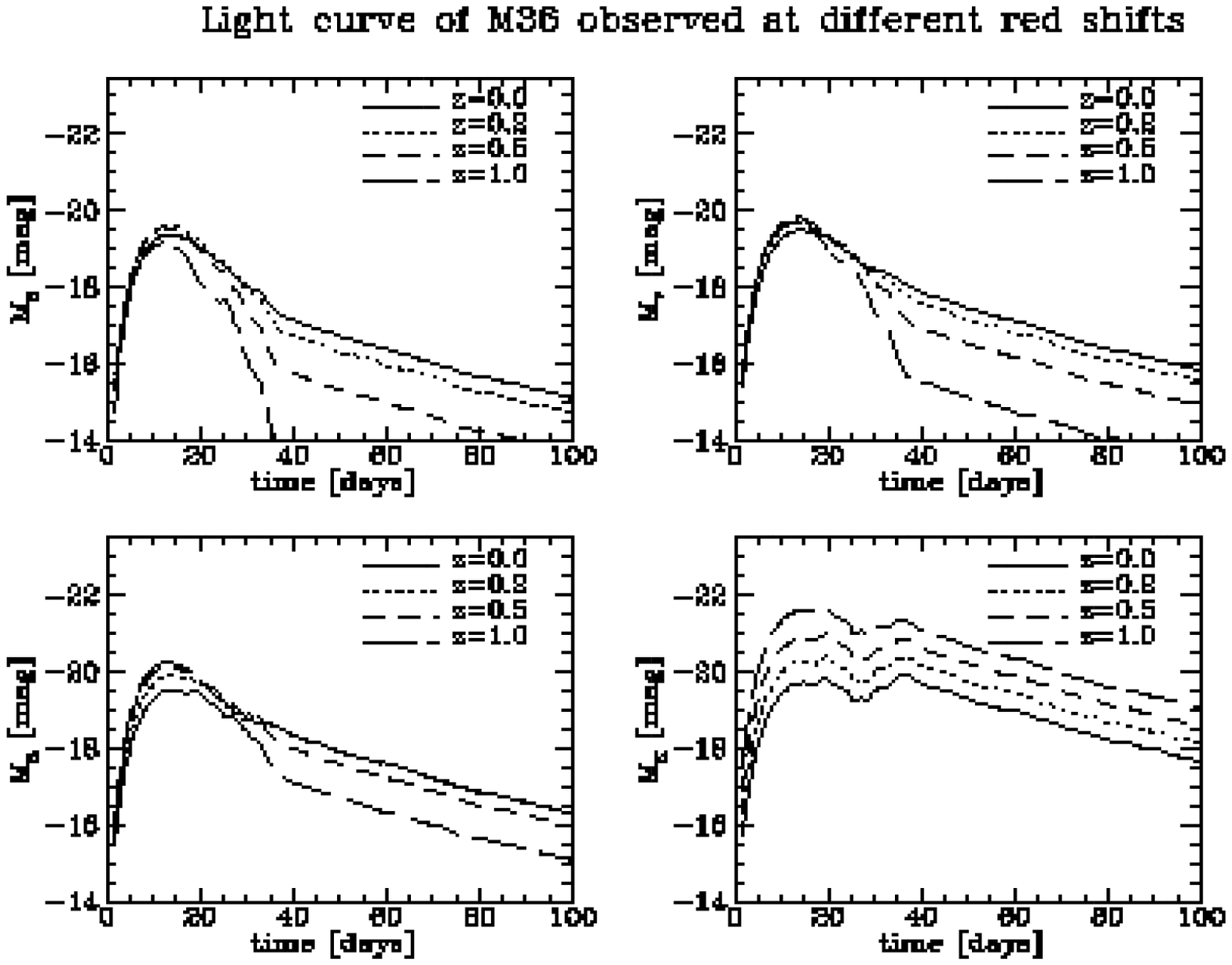,width=20.0cm,rwidth=9.5cm,clip=,angle=270}
\figure{11}{Monochromatic optical and infrared light curves 
of the delayed detonation model M36 at different cosmological red shifts.}
\endfig
 \begfig 0.1cm
\psfig{figure=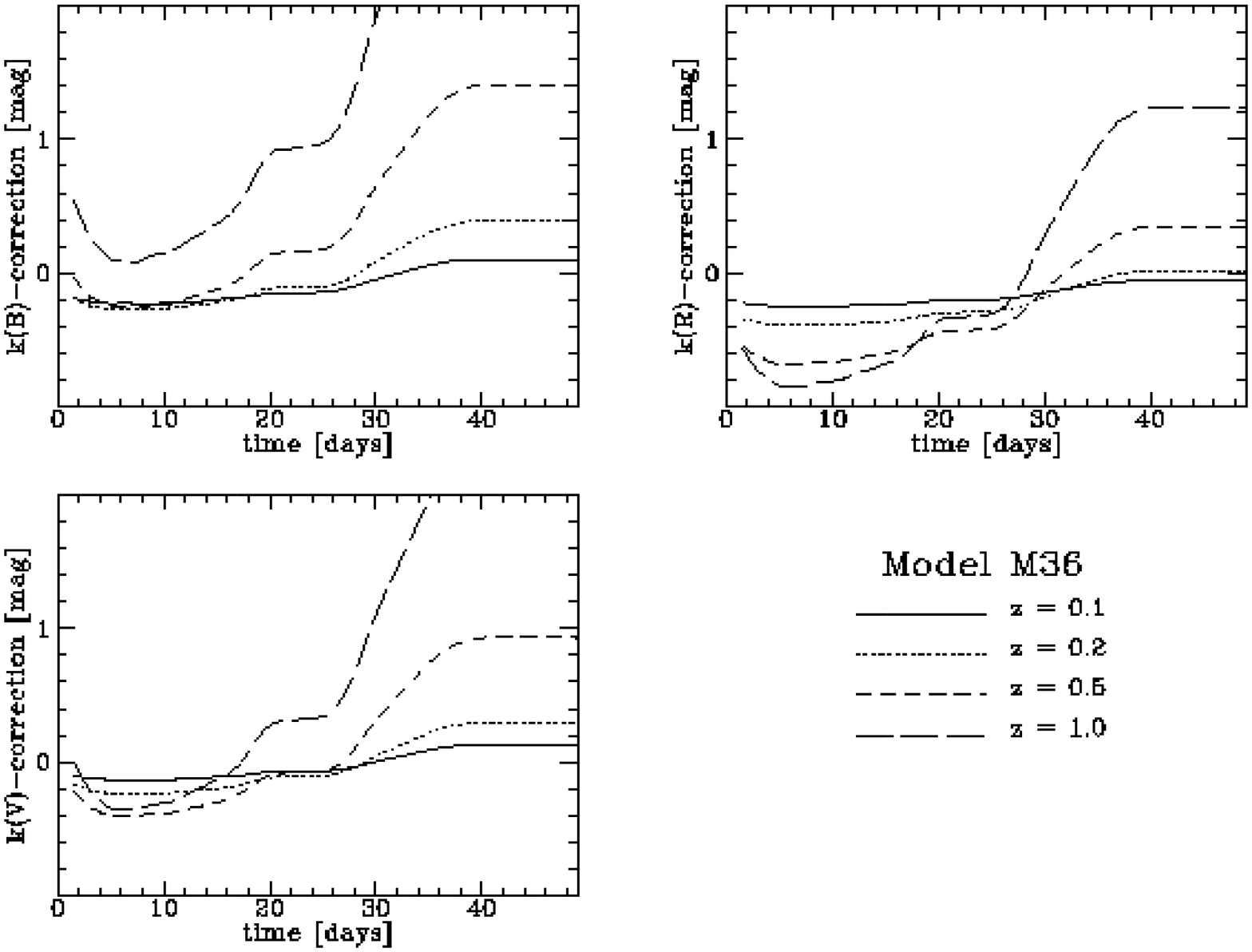,width=12.6cm,rwidth=9.5cm,angle=270}
\figure{12}{k-correction for different red shifts z as a function of time for M36.}
\endfig
\begfig 0.1cm
\psfig{figure=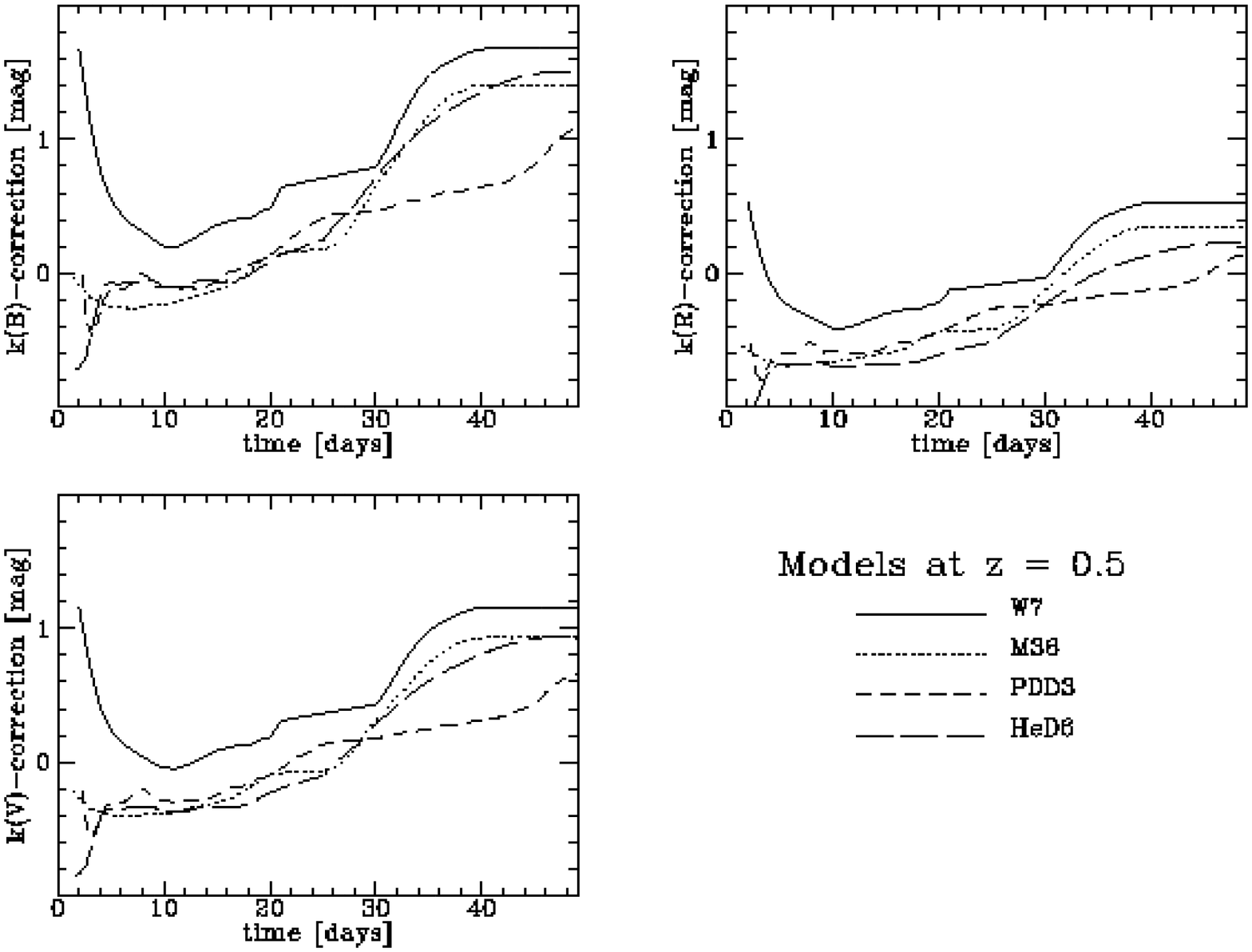,width=12.6cm,rwidth=9.5cm,angle=270}
\figure{13}{k-correction as a function of time for different models at z=0.5}.
\endfig
\begfig 0.1cm
\psfig{figure=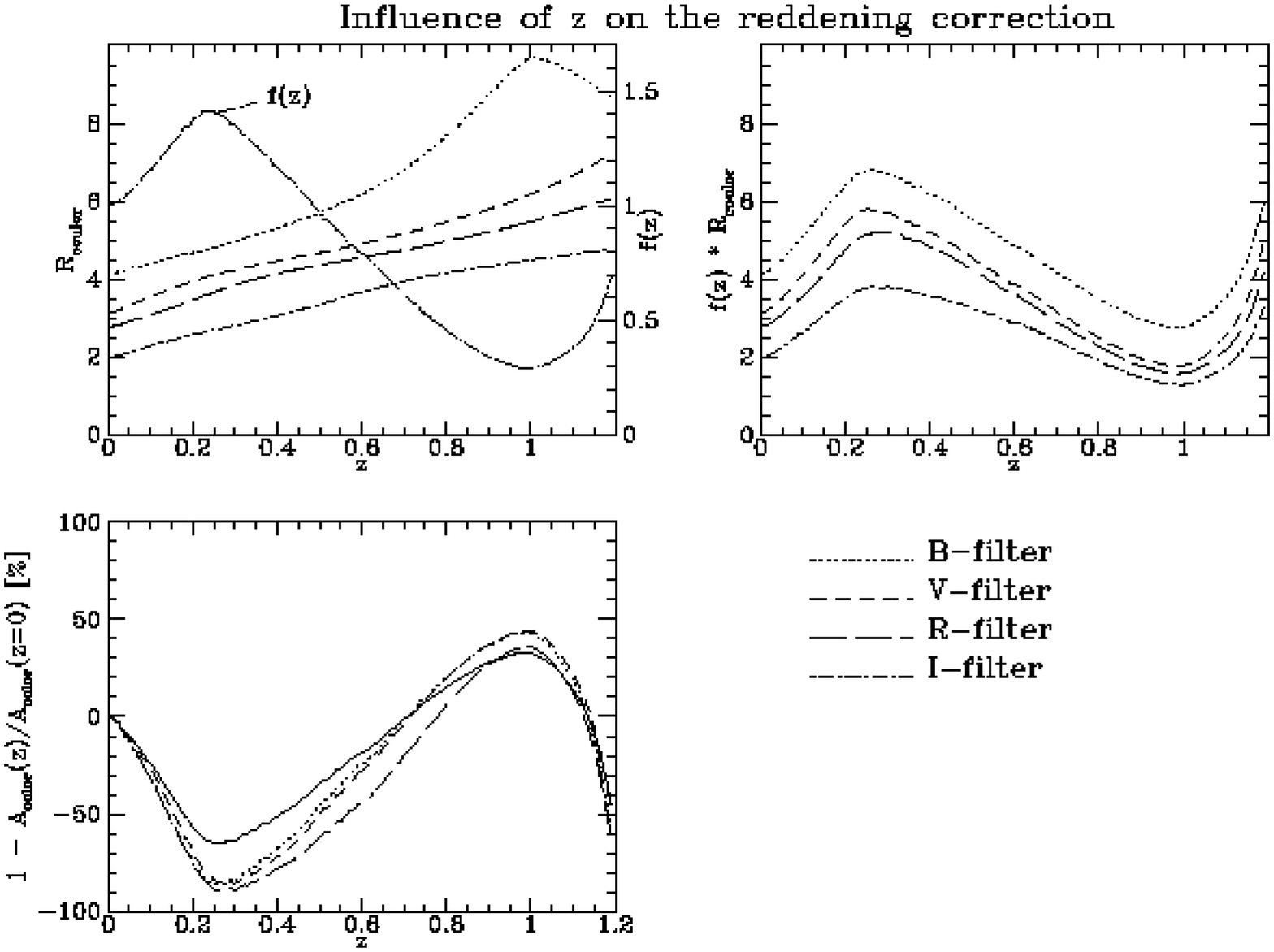,width=12.6cm,rwidth=9.5cm,angle=270}
\figure{14}{Influence of the cosmological red shift z on the value for the  extinction 
coefficient $R_{color}$ (eq. 1) and the correction factor $f(z)$ (upper left) and
the combined effect on the relation between color excess and extinction (upper right).
In addition, the  error is given if the z-correction is neglected (lower graph).}
\endfig
 The cosmological expansion  has two main effects -- the frequency shift and the stretching of the         
spectrum. The former causes 
 a change in the shape of the light curves because the overall energy
distribution of supernovae varies strongly during the first month -- namely,
line blocking at shorter wavelengths  increases with time. Therefore,
 post-maximum declines in B and V become steeper with increasing z (Fig.~11).
 Note that the frequency shift does not imply   that observed magnitudes 
can be translated  directly from one color to another,
because  magnitudes do not provide an absolute
measure for the fluxes, but brightnesses  normalized to an A0V star.
The stretching of the spectra  causes 
an overall dimming of the flux over a given filter range.
 Note that our corrections (Figs. 12 \& 13) are 
consistent with the empirical k-corrections of Hamuy et al (1993a).
 
From the models, 
 the overall flux distribution is given. Therefore,
 k-corrections are 
 consistent with the information from the light curve and spectra.
 In principle, we must not expect uncertainties related to the k-correction
to grow with distance, as it is inherent for all purely statistical 
methods where the k-correction must be estimated from local supernovae.
 This may  be regarded as an          
advantage for using models because the k-corrections show  large individual variations
of the order of 0.3  ... 0.4 mag at about maximum  light for z=0.5  even for `normal' bright
 supernovae (Fig. 13, see also Hamuy et al. 1993b). Note that, before day 5 to 10, the
k-correction     decreases  for all models but the Helium detonations. Except for the latter models,
\ni is concentrated towards the center and the radioactive heating of the photosphere is 
 delayed. In  Helium detonations, \ni in the outer layers keeps the temperature high
from the very beginning.
 
 For distant supernovae, the correction for interstellar reddening 
depends on the spatial distribution of the absorbing interstellar material.
In light of the determination of $q_o$ and the high accuracy needed, 
it may be  useful to look at the possible influence of z.
 The cosmological expansion has two main effects: a) a photon, emitted at a 
redshift z, sees the absorption law at a redshift $z'$ of the absorbing material, and
b) the differential absorption between colors changes. Thus,
 the extinction $A_{color}$ due to the interstellar medium at a redshift $z'$ is given
 by
 $$ A_{color}~ = ~ R_{color,z'} ~f_c(z) ~E_{B-V}  \eqno(1)  $$
\noindent
where $ R_{color,z'=0} $ is the ratio between $E_{B-V}$ and the extinction.
 The factor $R_{color,z'}$ takes into account the cosmological redshift $z'$
of  the absorbing material (Fig. 14) and
 $E_{B-V}$ is the measured color excess.
The function $f_c(z) $ corrects for the differential change with $z'$ between the B and V. 

 In the general case of n  clouds at redshifts $z'_1 ... z'_n$, eq. 1 becomes 
 
 $$ A_{color}~ = \sum_{i=1}^n R_{color,z'_i} ~f_c(z'_i) ~E_{B-V,i} \eqno(2)   $$
 
\noindent
where $E_{B-V,i} $ gives the relative contribution of a particular cloud.
 In principle,     $E_{B-V}$ must be determined independently, e.g. by 
the relation between the line absorption from UV-lines and their redshifts.
 In practice, 
 the  likely cases are that the absorption happens within our
galaxy ($z'=0$) and within the parent galaxy ($z'=z$) (or a linear superposition of both).
The contribution from our own galaxy can be measured independently.
 Obviously (Fig. 14), the influence of z on the reddening correction must be taken into account because  
the resulting, systematic error is comparable to the size of the reddening correction itself. For example,
an $E_{B-V}=0.05^m$, if determined 
for a supernovae at z=0.2 ... 0.4,  would translate 
into an error of $0.15^m$ in $M_V$. Note the possible errors for 
$q_o$.

\section{Comparison of theoretical and observed light curves}
 It is  necessary to confront our parameterized models with 
observations because not all scenarios may be realized in nature.
 The main aim is to get some insight into the physics of the explosion
and progenitor evolution.
 Individual comparisons  help to answer questions about the flame 
propagation and the progenitors of SN~Ia, i.e. whether they have
masses close to the Chandrasekhar limit, originate from low-mass
objects, or may be understood in the framework of the merging scenario.
 It may be interesting to ask for the 
spread among all models that  are found to resemble observations.
 
In the following comparison we have considered the full set of  SNe~Ia
explosion models (see Table~1). The following supernovae are       analyzed:    
  SN~1937C, SN~1970J, SN~1971G, SN~1972E, SN~1972J, SN~1973N,
 SN~1974G, SN~1975N, SN~1981B, SN~1983G, SN~1984A, SN~1986G,
 SN~1988U, SN~1989B, SN~1990N, SN~1990T, SN~1990Y, SN~1990af,
 SN~1991M, SN~1991T, SN~1991bg, SN~1992G, SN~1992K, 
 SN~1992bc, SN~1992bo, and  SN~1994D.
This sample consists of  almost all SNe~Ia for which very good light curves have been
 published   or which are interesting for distance determinations.
  The observational data were mainly selected from the
compilations of Ciatti \& Rosino (1978), Barbon et al. (1989a), and
Cadonau \& Leibundgut (1990), Filippenko et al. (1992ab), and
from the  CTIO supernovae search 
(Phillips et al. 1987, Hamuy et al. 1993a, 1994, Wells et al. 1994). 
 We refer to these compilations for a detailed discussion of
the individual measurements and for references to the individual
observers.  If additional data are used, we will refer
to the original literature.
 Note that this sample includes six SNe~Ia measured at Asiago in the 
1970s (SN~1970J, SN~1971G, SN~1972J, SN~1973N, SN~1974G, SN~1975N). 
 Due to problems  with the absolute calibration of the B color,
the apparent brightness $m_B$ may be underestimated
by up to several tenths of a magnitude (Turratto 1995, 
private communication). In principle, an increase of the brightness in B leads to an overestimate
of distances.   However, at the same time,
the interstellar reddening is systematically underestimated, which has the opposite
effect. Therefore, we assume symmetric, 
large error bars for the SNe~Ia observed at Asiago (M\"uller \&
H\"oflich 1994). 

 The criterion determining the ``best" theoretical model
for a given observed supernova is  the best overall agreement between
the shape of the calculated and the observed monochromatic light curves.
Thus, the ``best" theoretical model is not determined by the
distance-dependent absolute maximum magnitude of the supernova, but by
light curve characteristics, for example, the pre-maximum behavior, the
width of the maximum, the rate of the steep post-maximum decline, and
the rate of the subsequent linear decline. 
 
 Before we quantify these criteria,
we have to address the question of fitting an individual model to a given
set of observations and to determine the following parameters:
a) the time of the explosion $t_o$, b) the distance d, and c) the interstellar
reddening $E_{B-V}$. In general,
 we base our fitting procedure on the B and V colors 
because they are very 
 sensitive to the interstellar reddening,
a consistent set of data is available, and filter functions are probably
closest to the standard system.
 
 A horizontal shift of the theoretical light curves 
provides information on the moment of the explosion.
 A vertical shift of provides information on the monochromatic 
distance module, i.e.
 
 $$ m_{color} -  M_{color} = 5 log {d \over 10pc} -5 + A_{color} \eqno(3) $$
 
\noindent
where d is the distance,  $ A_{color} $ is the interstellar extinction, and
{\sl color} stands for  B, V, R, or I.
The interstellar reddening law ($R_{color} = A_{color} / E_{B-V}$) for our
Galaxy gives $R_B=4.0$ where its frequency dependence can be taken from Savage \& Mathis
(1979).     The general applicability of these values is debated.
From a statistical analysis of SNe~Ia, J\"oeever (1982) derives
$R_B=1.8$.  Capaccioli \etal (1990) derive a similar value but with
a large uncertainty ($R_B = 1.7 \pm 1.4$).            Della Valle \&
Panagia (1992) conclude from a detailed discussion of the previous
studies and from their own study of SNe~Ia that $R_B = 3.35 \pm 0.75$
($2\sigma$), which is consistent with the value derived for our Galaxy.
 However, all supernovae-based estimates rely on the invalid assumption
of SNe~Ia being standard candles. The differences with respect to the Galactic values
are likely produced by this incorrect assumption. Therefore, we use the Galactic value and
extinction laws for all wavelengths from the UV to the IR range.
If our estimate of $E_{B-V}$ is smaller than the uncertainties inherent
in our models (see above), we neglect the intrinsic reddening within 
the host galaxy and  use the  value for the Galactic extinction
 towards the parent galaxy 
$\widetilde E_{B-V}$ given by  Burstein \& Heiles (1984).

 In previous comparisons, we have fitted theoretical and observed
light curves by visual inspection (HKM91, KMH92,
MH94, HMKW1995b) to find the best overall
agreement. This could be justified because 
 the non-homogeneity of the 
light curve data overshadowed the benefits of a sophisticated fitting procedure.
 In this paper, a more quantitative approach is used because several 
well observed LCs became available. In principle, the most straightforward 
way would be the `classical' procedure where each observational
point has the same statistical weight and the standard deviation $\chi $ is minimized, i.e.
 
$$ \tilde \chi =\sqrt{{\sum_{j=B,V}\sum_{i=1}^{n(j)} {\chi_{i,j}^2}\over (n_{B}+n_{V})
 (n_{B}+n_{V}-1)}} \eqno(4)$$
 
with $$\chi_{i,j} =  |M_{model,color}-M_{obs,color}| \eqno(5) $$
 
\noindent
where $M_{model,color}-M_{obs,color} $ is the distance modulus at the time i
for a given distance d and 
interstellar reddening and $n_B$ and $n_V$ are the number
 of observations in V and B, respectively.
 
 The direct approach becomes questionable if the coverage of the LCs is  incomplete,
 if the data-points are unevenly distributed and if errors both in models and 
and measurements are neglected. To overcome the former problems, we minimize the time average of
the time integrals of the standard deviation, i.e.
 
 $$ \tilde \chi \rightarrow  \bar \chi = \sum_{j=B,V}\int_{t=t1_j,t2_j} F(\chi(t)) dt \eqno(6)$$.
 
\noindent
where $t1_j$ and $t2_j$ are the dates of the first and last observation of color j. To reconstruct 
 $F(\chi(t))$, we rectify the irregularly sampled data set $[t_j,\chi_j ]$ by means
 of Wiener filtering based on a fast algorithm (Rybicki and Press  1995). 
 This  allows for a consistent solution within the Newton-Raphson iteration for
 $t_o, ~E_{B-V} $ and $d$. Wiener filtering requires the definition of a correlation length $w_i$.
The $w_i$ of the LC varies strongly with time.
 Consequently, any reconstruction would  be effected
by the assumed correlation length. However, the correlation length
is  expected to vary only slightly if we consider the change of 
the standard deviation.
The correlation length $w_i$ of $\chi_i $ can be assumed to be constant. The remaining
errors are small because 
 the resulting function $F(\chi)$   depends only in second order on $w_i$.
 Our ansatz is equivalent to the assumption that $w_i$, for the reconstruction of the observed  
LCs, is given by the theoretical LCs. Obviously, for successful models, this approach
is  valid. For unsuccessful models,  our procedure may give 
wrong fitting parameters (i.e. $t_o$, $M-m$ and $E_{B-V}$) but those models
are excluded due to their large $\bar \chi $ (see below).

 Following the discussion on the accuracy of the monochromatic colors based on the models,
the error             
 $\bar\varepsilon_{color}$ is assumed to be $0.05^m$ and $0.1^m$ for V and B, respectively. 
After
maximum light, a corresponding linear increase in the error
 is taken to be $0.0025$ and $0.007 mag/day$, respectively.
The error $\varepsilon_{obs,color}$ is estimated from the scatter in the observations and the accuracy given
by the specific observation.
 
 For each model, $t_o$, d and
 $E_{B-V}$ are iteratively determined by  a Newton-Raphson scheme
to minimalize $\bar \chi$. Two more restrictions 
are introduced. In several cases, the supernova was discovered on photographic
plates or in the R color. Then, in the fitting procedure, the time of the explosion is forced to
be prior to the  first detection. If the galactic contribution to the reddening 
were known, this value  provides a lower bound for $E_{B-V}$. The infrared 
light curves are used for consistency checks.
 
 Now, we want to address the question of defining acceptable models or, more appropriately, of the defining
criteria to reject  certain models. As mentioned earlier, the shape of the early LCs
is  very model dependent. We demand the following criteria: a)  post-maximum declines $dM_{15}(V)$
and $dM_{60}(V)$ are consistent (see Table 2); b) If pre-maximum
LCs are observed, the slope up to maximum must be consistent; c) The maximum of 
$\chi_{i,B/V}$ is within  $3 \sigma$ error bars; d) The infrared colors are consistent
with the observations within the error bars, e) $\bar \chi $ is smaller than  $\bar\chi_{obs,B,V} $;
and e) photospheric  expansion velocities, as inferred from spectra, agree with the models.

The dependence of the photospheric
radius and of the photospheric expansion velocity $v_{ph}$ as a
function of time can also be used to test explosion models  and,
in particular, it can be used to separate ``non-standard" from "standard"
models (HMK93, KMH93, H95). 
 In some cases (SN1988U, SN1994D), detailed NLTE-spectra are constructed
and compared with observations. A detailed spectral analysis
of 27 SNe~Ia for different dates is  beyond the scope of this paper.
 Detailed studies of spectral peculiarities
even of elements such as Ti are needed to get good spectral fits. In 
turn, this would  call for  good spectral time coverage 
 (H95) which is rarely available.
To include the information of the photospheric velocity,
we have used a somewhat more simple approach for the majority of supernovae.
The velocity $v_{ph}$ can be measured directly by the Doppler shift of spectral
lines. A comparison based on the velocities thus  determined 
should only be regarded as a guideline, for the following reasons.  Line
photons are emitted at larger radii than continuum photons.
Consequently, different lines form in different layers above the
photosphere.  Secondly, the line forming region is not only determined
by temperature and density, but also by the radial profile of the
abundances.  For example, the change of
$v_{ph}$ implied by the Si~II line is less than that inferred from the
Mg~II line (see, \eg SN~1989B in Fig.~16).  This effect is more
pronounced during later epochs when the ejecta become transparent.
Thus, most weight should be given to the early epochs near maximum light
when the line forming region is compact.  In addition, the expansion
velocities are based on measurements of the Doppler shift corresponding
to the minimum flux of a certain feature but  the result may be 
influenced by NLTE-effects
 (H\"oflich, 1990).  Finally, in our models,
the photospheric velocity is taken at a Rosseland optical depth of 2/3;
however, at some frequencies, the line forming region is located at smaller
radii. From our detailed NLTE-analysis of  SN1994D, the
total error must be expected to be of the order of 10 to 15 \%.
 
Often, more than one theoretical model reproduces the
observations within the uncertainties (see Table~3). Sometimes 
because the overall structures are rather similar (N32/W7) or, sometimes, they are 
based on the same explosion mechanism.  For instance, in case of SN1972E, both delayed detonation 
models M35 and N21 allow for a quantitatively similar fit.
 Following Table 1, this must be interpreted as follows:
this specific supernova can be understood within the delayed detonation scenario
from a C/O white dwarf close to the Chandrasekhar limit and a central density of about
$2.8...3.5~10^9 g~cm^{-3}$, and the transition from the deflagration to the detonation 
happens at the high densities of about $3...5~10^7 g ~cm^{-3}$.

\subsection{Previously analyzed supernovae}
In  previous studies,  the following set of  supernovae
have been analyzed and discussed in detail: SN~1937C, SN~1970J, SN~1971G, SN~1972E,
 SN~1972J, SN~1973N, SN~1974G, SN~1975N, SN~1981B, SN~1983G, SN~1984A, 
SN~1986G,  SN~1989B,  SN~1990N, SN~1991T, SN~1991bg, SN~1992bo, SN~1992bc, and
SN~1994D (H\"oflich et al. 1992, Khokhlov et al. 1992,
MH94, H\"oflich et al.1994, H95).
 The early systematic studies of the first 13 objects  were based on
a subset of 11 models with \ni  masses above 0.52 $ M_\odot $. As for the rest of this sample,
subluminous SNe~Ia have been considered, but the mechanism of Helium detonations 
was not. For  SN~1991bo, SN~1991bc, and SN~1991bg, only pulsating
delayed detonation models were considered in HKW95.         
Therefore, all SN~Ia  are        re-analyzed with our full 
set of models, including Helium detonations using our new
quantitative fitting procedure to overcome one of the main points of criticism
of the earlier works.
 
 In general, our previous results are confirmed - namely, the changes in
 the time of the        explosion, distances,  reddening corrections and models ranges 
 remained well within the former error bars.
 Yet, as a tendency, delayed
detonation models with a somewhat smaller central density (M-series) provide better
fits than the old N-series. For several  bright and slightly subluminous SNe~Ia,
Helium detonations of sub-Chandrasekhar mass white dwarfs between $0.6$ and $0.9 M_\odot $
cannot be ruled out from the light curves 
but they can be excluded 
by early spectra or the minimum velocity of Si (e.g. SN1972E).
 In the corresponding model, the photosphere would be formed in or just below the 
Ni-rich layers. Test calculations based on our NLTE-code confirmed
this result, i.e. the synthetic spectra did not come close to the observations. 
 
 In the following, we want to discuss three supernovae of the old sample 
in more detail because either our new models provided a better understanding of the object 
 or the data sets have been improved.
 
 In our earlier study of SN1991bg, we found agreement between the subluminous
pulsating delayed detonation model PDD5 with $0.12 M_\odot $ of Ni. 
Harkness was able to reproduce the spectra at about maximum light by his  LTE models
(Wheeler et al. 1995). However,
there was some quantitative disagreement in the B light curve, and the IR-emission at maximum
light was systematically too dim  by about $0.15^m$ to $0.2^m$ in I and R, respectively. We find
a somewhat better agreement with the new model PDD1c which produces slightly less Ni
($0.10 M_\odot$).   Still, a possible problem persists. We need  a rather large 
 reddening. Despite the fact that, for low photospheric temperatures, the intrinsic color is very sensitive
 to the temperature, we do not think that a further reduction of the overall energy distribution could
solve the problem. Instead, we favor the following two `physical' solutions.
The extra extinction may be produced by dust-formation 
 in subluminous SNe~Ia. In principle, dust can form at early phases. Whether it can persist
 is under investigation (Dominik et al., 1995). Another possibility
for   very red colors at maximum light is selective line blanketing, mainly caused 
by Ti II lines (Branch,
private communication). To  answer these questions, detailed spectral analyses by 
NLTE models are under way.

 As for the distance moduli, the differences with respect to our previous  works are always 
well within the  error range, i.e. less than $0.1^m$, with the exception of SN1972E.
  For this supernova, the very first observation in the B filter implied a very rapid
rise and a slow decline, ruling out all models but the 
delayed detonation model N21 in which a large amount of Ni is produced which, therefore, is
located close to the surface.
 Recently,  Leibundgut (1993, private communication) re-analyzed 
this observation and found the first measurement 
to be about $0.5^m$ brighter. This, in turn, allows better fits by models with a slower rise. 
Although SN1972E is still at the bright end of SN~Ia, this reduces the distance by about
$10...15 \%$.

 \begfig 0.5cm
\psfig{figure=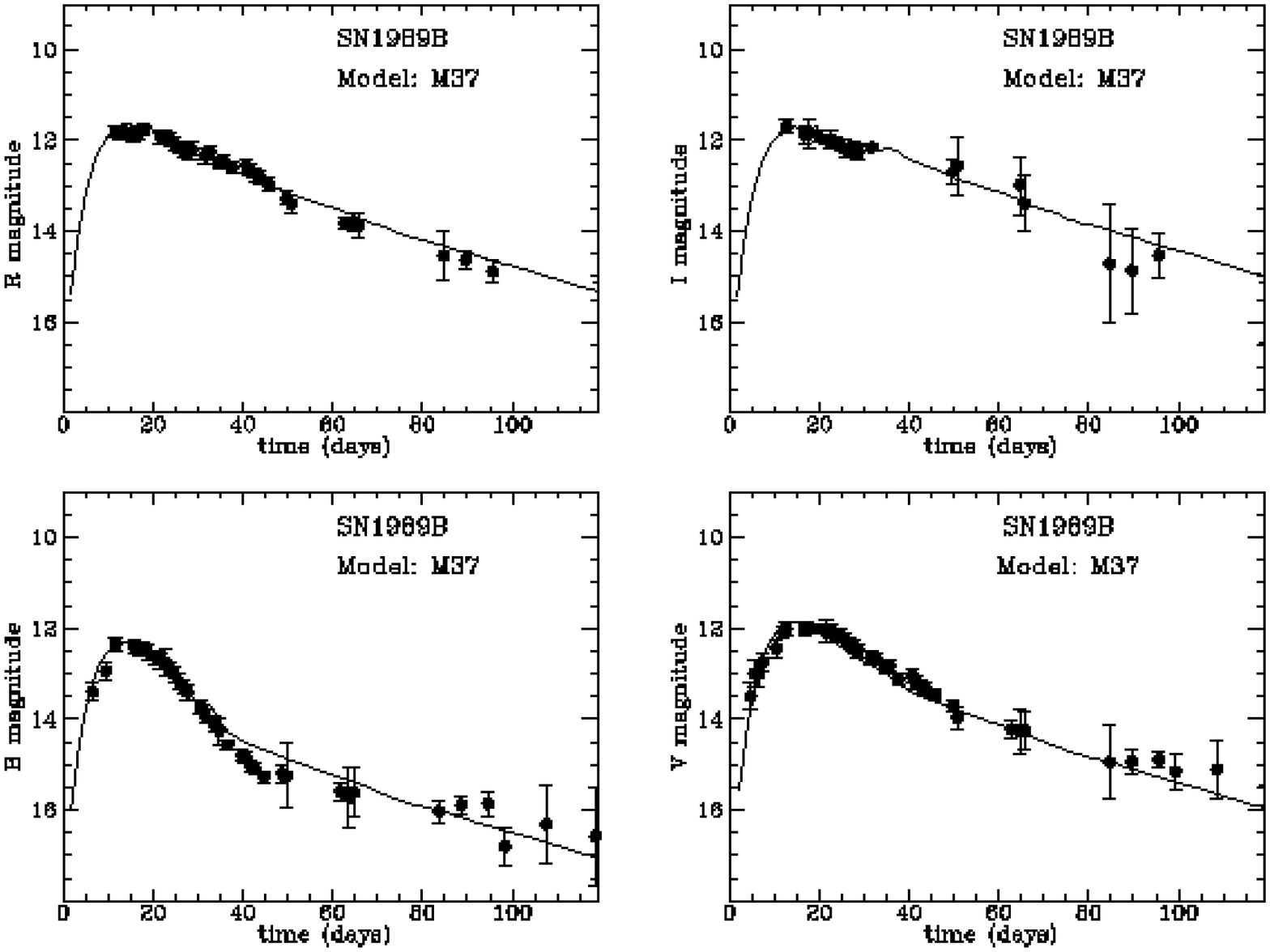,width=12.6cm,rwidth=9.5cm,angle=270}
\figure{15}{
Monochromatic light curves of SN~1989B 
 band compared with the calculated light curve of the
delayed detonation model M37. The 2 sigma-error ranges are
given by Wells et al. (1994)
} 
\endfig
%
%
 \begfig 0.1cm
\psfig{figure=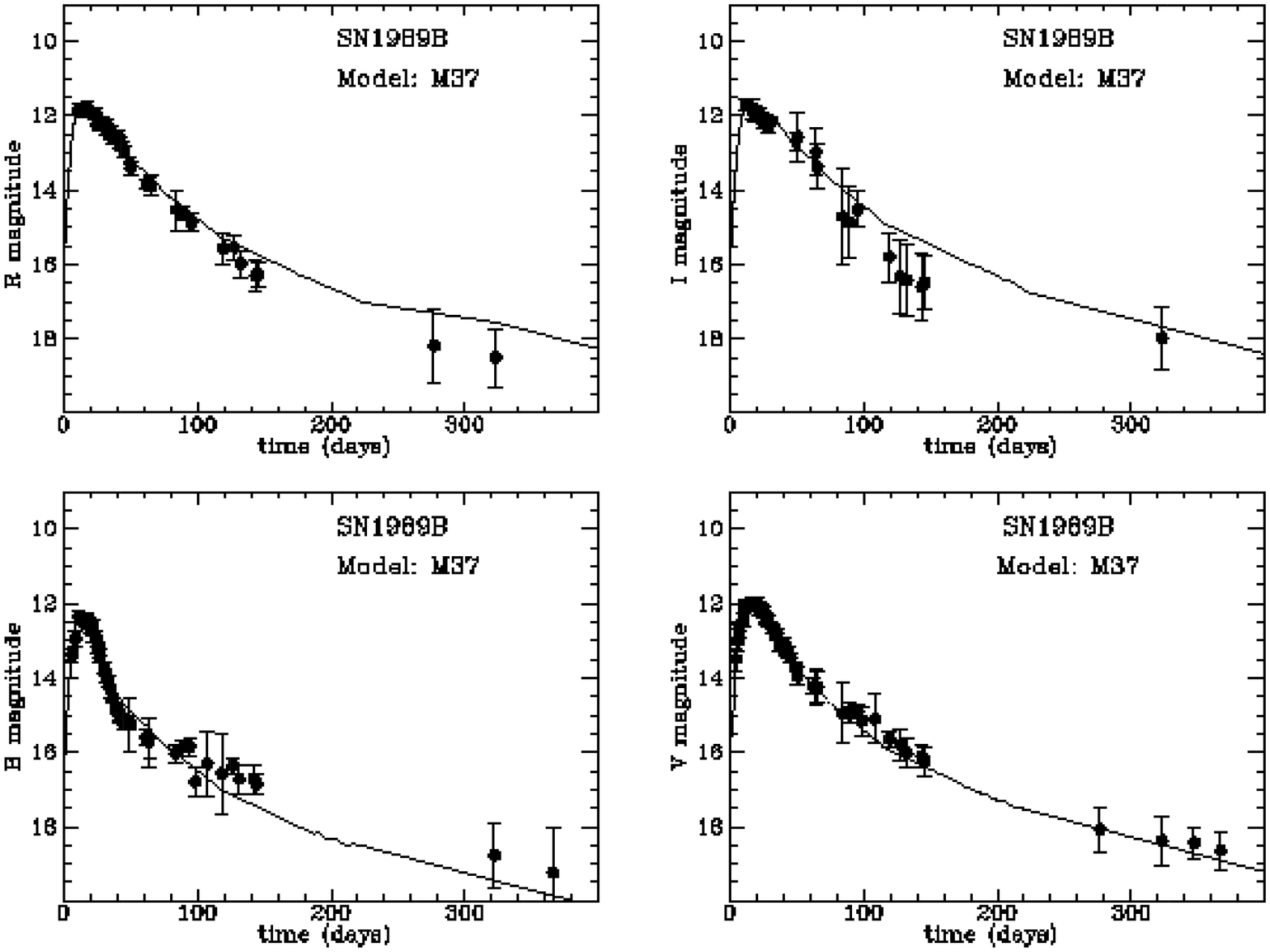,width=12.6cm,rwidth=9.5cm,angle=270}
\figure{16}{Same as 15  but including the late times.}
\endfig
%
%
 \begfig 0.1cm
\psfig{figure=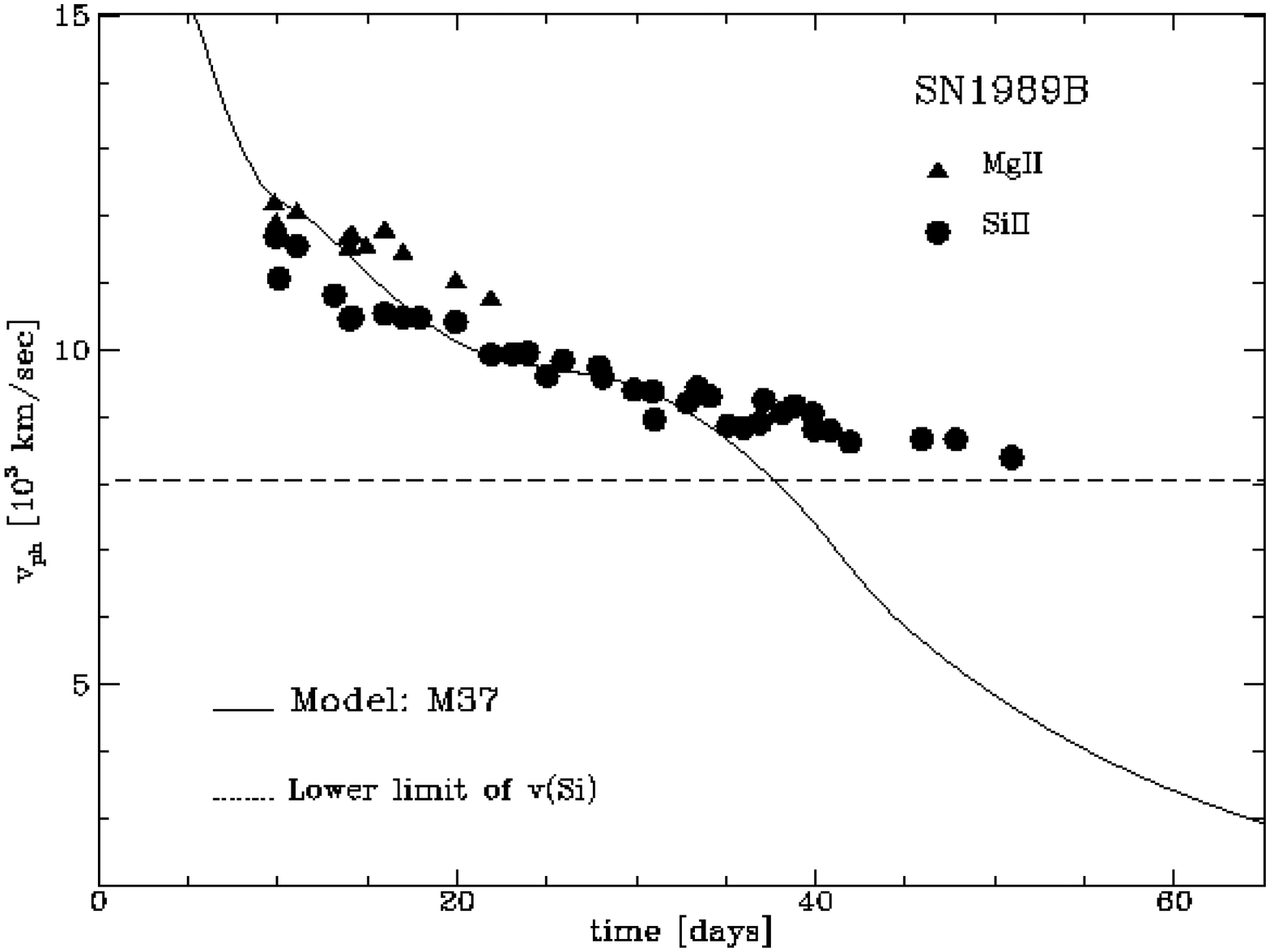,width=12.6cm,rwidth=9.5cm,angle=270}
\figure{17}{Photospheric velocity of the delayed detonation model M37 in 
comparison with the expansion velocities implied by the line shifts of Mg II
and SiII. The dotted line is the velocity at which, in M37, the Si abundance has dropped
to 10 \% of its maximum value.}
\endfig
 One of the earliest detections relatively to the time of maximum light
 of any Type I~a supernova is  SN1989B. 
The first observation ($m_V \approx 17^m$) on SN~1989B in NGC~3627 is
available from a pre-discovery image taken by Marvin \& Perlmutter
(1989) at JD~2447549.5.  This epoch is about 15-16~days before $V$
maximum ($V_{max} = 12.0^m$) implying that the rise time to maximum is
very well known.  Unfortunately, measurements of $V$ obtained by
different authors show discrepancies of $\approx 0.4^m$ (Barbon \etal
1990). Recently, a very  homogeneous data set was published 
that  provides an excellent
coverage both of optical and infrared light curves and spectra (Wells et al. 1994).
The photometry is mainly based on observations at CTIO with its well-defined set of filters.
 The light curves and photospheric expansion velocities can be
 reproduced by the delayed detonation model M37 and, somewhat less successful by
M36 (Figs. 15-17). After about day 120, the decay of the light curves follows
 the radioactive decay of $^{56}Co $. Note that, at late times, the theoretical LCs  
are  uncertain because of our approximate treatment of the 
energy distribution. The observational error bars  increase strongly
for later times. Thus, the good agreement with the late time light curves
 must not be given great weight.
 Early on, the photospheric velocity (Fig. 17) can be measured by the
SiII and MgII lines at $6355 \AA $ and $4481 \AA$, respectively. 
The velocities indicated by the MgII line
and of the SiII line   stop following the receding
photosphere at day 25 and 35, respectively,  because of the decrease of 
the Si and Mg abundances.
The best agreement between calculated and observed monochromatic light
curves can be obtained for SN~1981B if an interstellar reddening of $E_{B-V} = 0.45^m$ is
assumed.  This value is significantly larger than the color excess
resulting from the intergalactic extinction towards the parent galaxy
$\widetilde E_{B-V} = 0.01^m$ (Burstein \& Heiles 1984). It must be attributed to the 
intrinsic extinction of the parent galaxy which is highly inclined.
Our value is comparable to the
estimates of Wells et al. (1994), but it is significantly smaller than the estimates
of  Barbon \etal
(1990; $0.70^m \ldots 0.95^m$ with a mean value of $0.81^m$) 
 because their value is determined using a standard light curve.  The
distance to NGC~3627 resulting from our models is $8.7\pm 3$~Mpc.
\subsection{ 
 SN~1988U 
}
 SN~1988U was discovered on   August 9, 1986 at $V\approx 22.05^m$ in
 a faint member of  the distant cluster AC118 at a redshift of z=0.31
 (Norgaard-Nielsen et al. 1989). Subsequently, this object was monitored
in V and R over a period of about 2 months down to about $24.4^m$.
Based on the spectrum taken about 10 days after the discovery, this
object was clearly identified as a Type Ia event (Norgaard-Nielsen et al. 1989).
 Up to now, it represents the most distant
SN~Ia for which good measurements including spectra
 have been published. Although the database does
not exactly match our criterion of a well-observed SN~Ia, we have included
this object to demonstrate  the advantage of a simultaneous analysis of 
light curves and spectra for the distance determination of faint SNe~Ia.
 \begfig 0.5cm
\psfig{figure=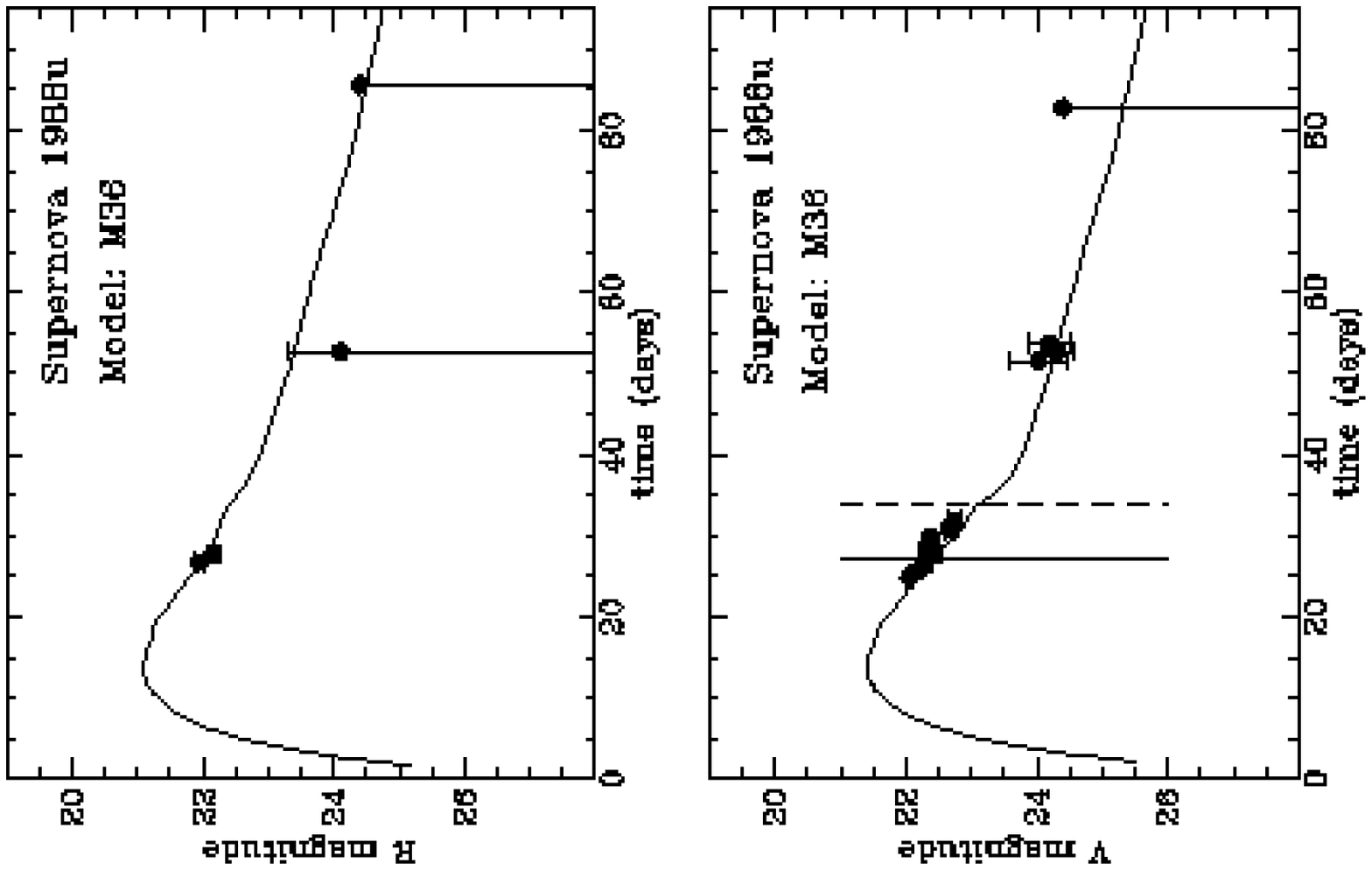,width=12.6cm,rwidth=9.5cm,angle=270}
\vskip -8.1cm
\psfig{figure=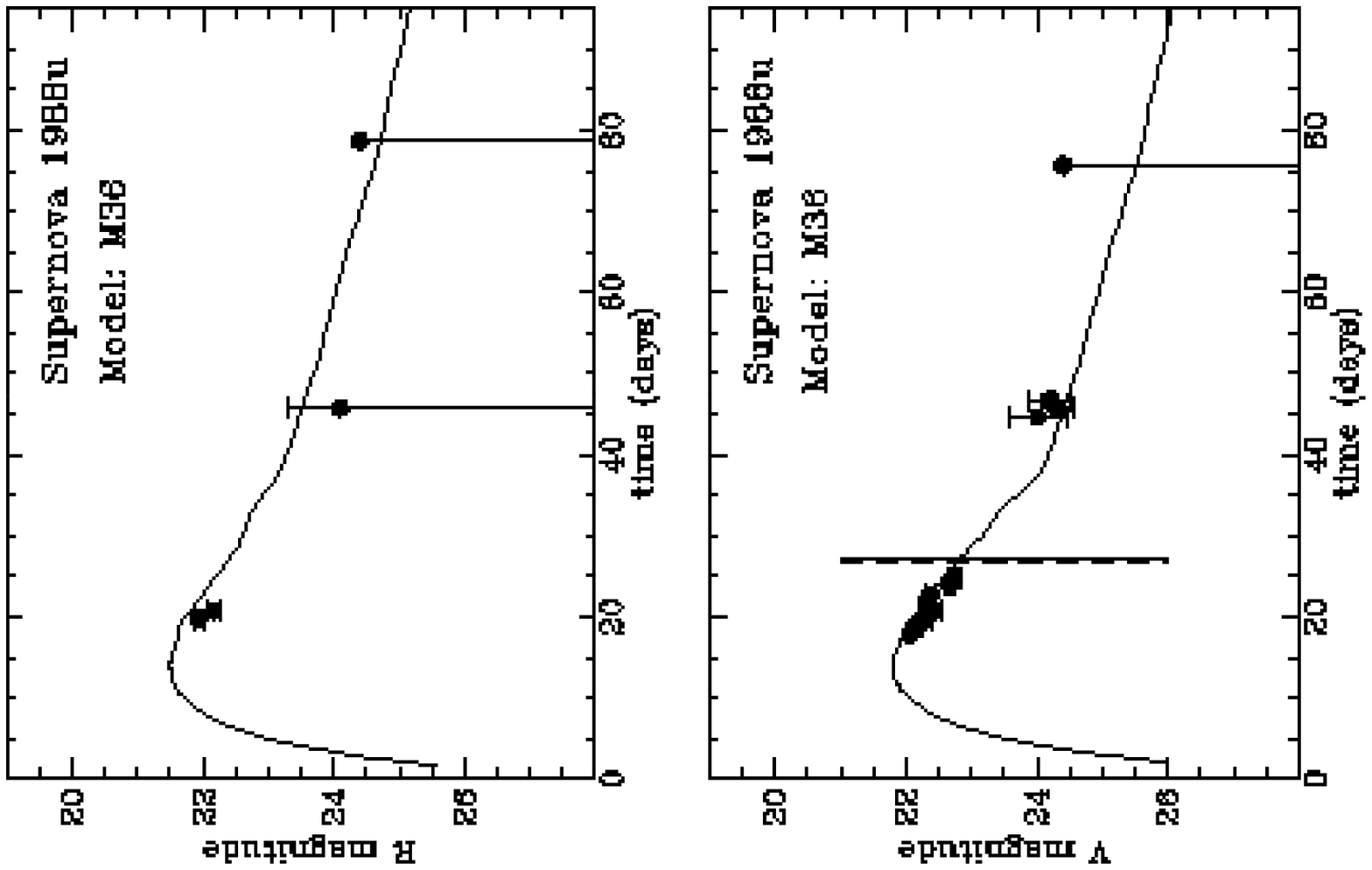,width=6.3cm,rwidth=9.5cm,angle=270}
\figure{18}{
Monochromatic V and R light curves of SN~1988U           
in comparison with the calculated light curve of the
delayed detonation model M36. The dashed line marks the time
when the photospheric velocity of the models corresponds to the observed line shift
(thin line). The right and left plots correspond to different time shifts. Both      
fits of the LC reproduce the observations, but the right fit can be ruled out by      
the spectrum.
}
\endfig
 We assumed a redshift of z=0.31 and corrected our light curves accordingly.
From the post-maximum decline, and the observation at about
day 40 after the explosion, all bright delayed detonation 
models (i.e. N21/M35) and  `normal bright' envelope and
pulsating delayed detonation models can be ruled out. Very slowly
declining models are in conflict with the upper limit for the brightness at late times.
 On the other hand, both subluminous and Helium detonation models decline too fast past 
maximum light to allow for an agreement between the measurements at about day
20 and 40. Additionally,  for subluminous models,
the photosphere has receded too far to be consistent with the observed line shifts
of Si II at 6355 \AA .
 This leaves the classical deflagration model W7 and the delayed 
detonation models N32/M36 as good candidates (Fig. 18). 
 An interstellar extinction $E_{B-V} $ of $0.055^m$ for AC118 (Knude  1977) is consistent     
with our upper limit  of $0.12^m$.
 The  advantage of a simultaneous analysis of light curves and spectra is
evident: Judging from the light curves,
both fits are equally good, but  the time of the explosion is not well
defined. This effect alone
 causes  a $30\% $ uncertainty in the distance. However, the right fit
is inconsistent with  the photospheric velocity. We get a distance of 
$1.44 \pm 0.25 Gpc$.
\subsection{ 
 SN~1990T 
}
SN~1990T was discovered by Antezana in the galaxy PGC 63925 on July 21, 1990
shortly after maximum light. Accurate photometric measurements in B, V, R, and I are
available spanning  a period of about 4 months 
(Hamuy et al. 1995). Two spectra, 5 and 26 days after the discovery, are taken.
At the first date, the line shift of the SiII feature at $6355 \AA$ indicates
a photospheric expansion velocity of $\approx  8700 km/sec$.
%
 \begfig 0.1cm
\psfig{figure=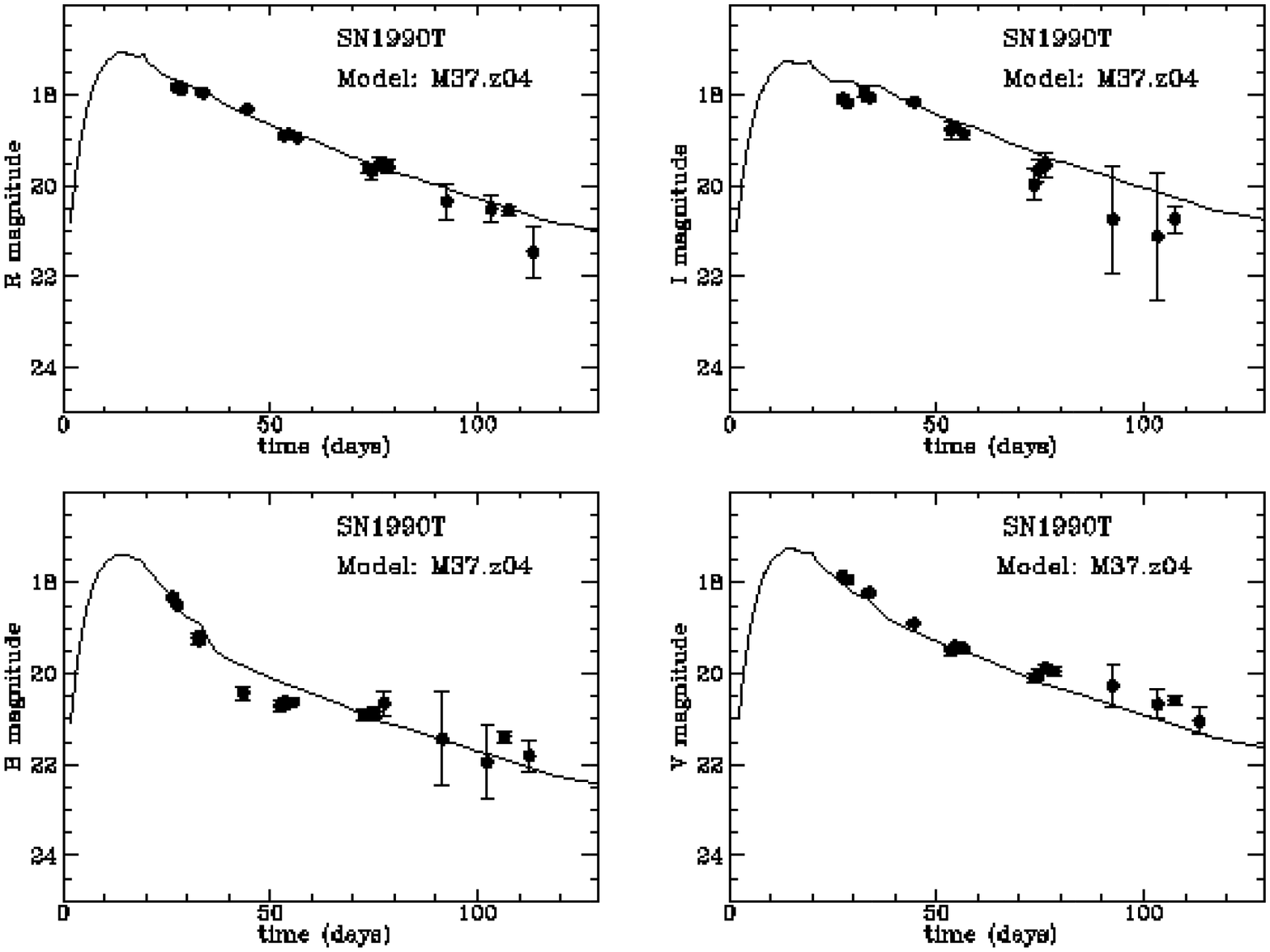,width=12.6cm,rwidth=9.5cm,angle=270}
\figure{19}{Monochromatic light curves of SN~1990T 
 filter band compared with the calculated light curve of the
delayed detonation model M37.}
\endfig

 The optical and infrared light curves can be well reproduced by the delayed
detonation models with a Ni production between 0.43 to 0.60 $M_\odot$ (Fig.~19).
From the fits, the time of the discovery can be determined to be about two weeks after
maximum light. The expansion velocity is consistent with the previous estimate but
hardly provides further restrictions on the models. However, our analysis favors
models with Ni masses  at the lower end for `normal' bright supernovae. 
 Due to the cosmological red shift, the B color of the models
is rather uncertain because, for slightly subluminous models, the corresponding
wavelength range is very sensitive to line blanketing. This explains
some of the problems at about day 50. 
 The V,R and I colors provide best fits for $E_{B-V}=0.0^m$ but B
calls for $E_{B-V}=0.1^m$.   Based on the intrinsic brightness of
models M38 to M36, the distance to PGC~5128 is $180 \pm 30$~Mpc. 
\subsection{ 
 SN~1990Y 
}
SN~1990Y was found by Wischnjewski (Hamuy et al. 1995)
 in an anonymous elliptical galaxy on August 22, 1990 
about two weeks after maximum light. BVRI photometry is available over  the subsequential
period of 3 months (Maza et al. 1994, Hamuy et al. 1995). One week after the
discovery,  spectra were taken (Della Valle et al. 1990). The Doppler shift of the Si II
line at 6355 \AA ~ corresponds to a photospheric expansion velocity of about 9000 km/sec.
%
 \begfig 0.1cm
\psfig{figure=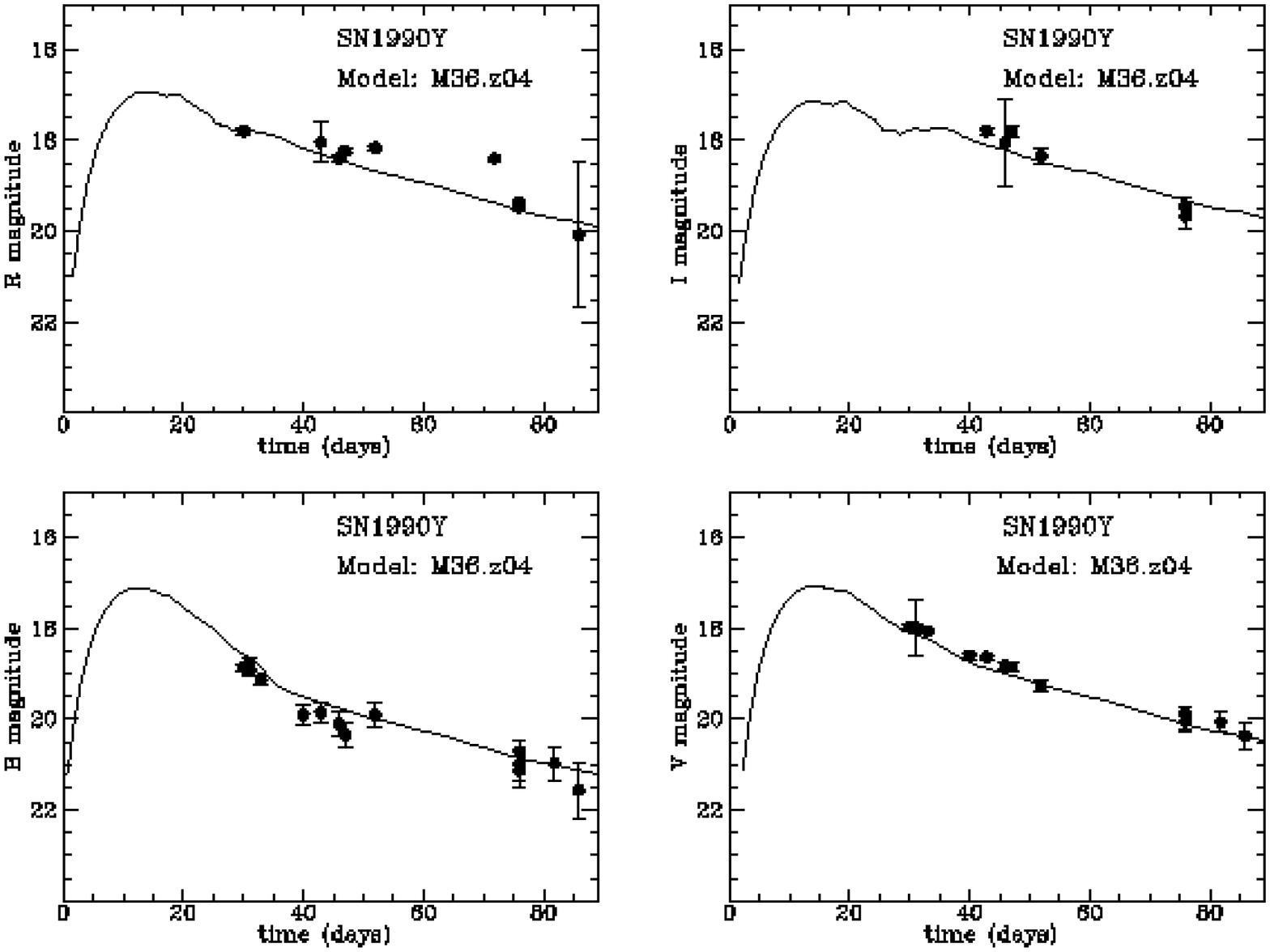,width=12.6cm,rwidth=9.5cm,angle=270}
\figure{20}{Monochromatic light curves of SN~1990Y 
compared with the calculated light curve of the
pulsating delayed detonation model M36 at a redshift of 0.04.}
\endfig

 Due to the lack of measurements, there is hardly any restriction on the type of the
explosion, i.e. pulsating, delayed detonation, and deflagration models can 
reproduce the light curves (Fig.20). However,
    strongly subluminous models can be ruled out based on the
infrared colors and the high expansion velocities of Si II seen two to three weeks
past maximum light. The colors are compatible with little 
interstellar reddening.  Based on the intrinsic brightness of
models, the distance is $195 \pm 45$~Mpc. 

\subsection {SN1990af}
%
 We find that the B and V light curves can be reproduced within the error bars 
by the classical deflagration model W7 and the delayed detonation models N32/M36/M37 
if we correct for the cosmological expansion and an interstellar reddening of
$E_{B-V}$ of $0.05^m$ (Fig.~21).  SN~1990af is well within the range of normal bright
supernovae. Subluminous models can be  ruled out because of their  
narrow maxima.
 \begfig 0.1cm
\psfig{figure=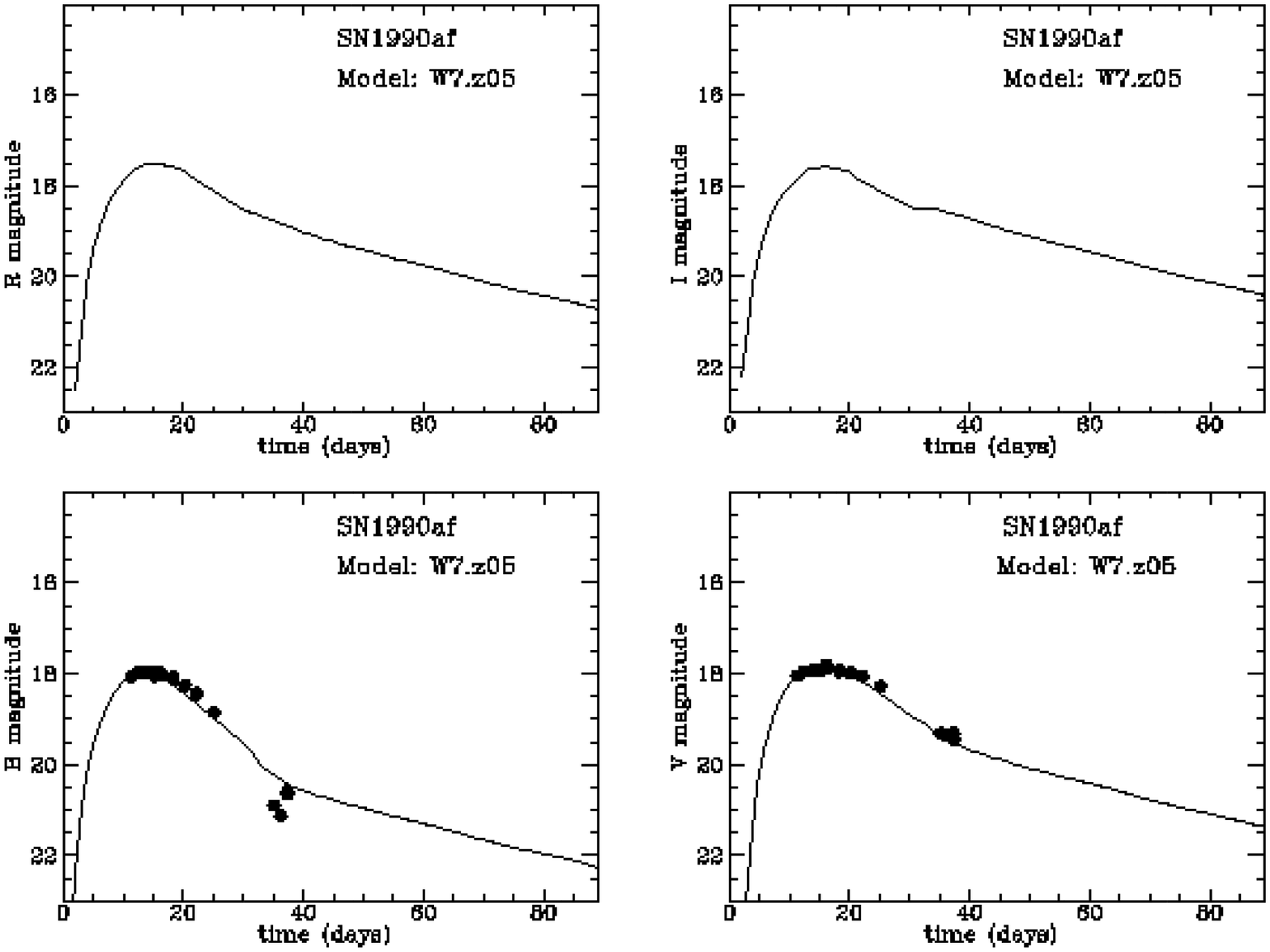,width=12.6cm,rwidth=9.5cm,clip=,angle=270}
\figure{21}{Monochromatic light curves of SN~1990af
compared with the calculated light curve of the
pulsating delayed detonation model W7.}
\endfig
SN~1990af was discovered in an anonymous elliptical  galaxy at z=0.0503 on October 24, 1990 
well before maximum light. 
 BV photometry is available over the following month.
 Spectra taken right after the discovery imply  a 
 photospheric expansion velocity   of  12,040 km/sec (Hamuy et al.  1995).

 In our models,
 z=0.05 causes a sufficiently large change in the shape of the light curve whereas
Hamuy et al. (1993b) found that their k-corrections based on calibrated templates 
are insufficient.
 Our agreement does not
necessarily imply that our k-correction is more reliable 
due to uncertainties in our models for
 wavelengths shortwards the B band.
    The distance is  determined to be 
 $265 \pm 85$~Mpc. The huge error range is attributed            
to the fact 
that only a few of our models can be ruled out and, consequently,
the successful models  cover  the entire  brightness strip for
 'normal' SNe~Ia.
\subsection{ SN~1991M}
SN~1991M  was discovered in the spiral galaxy IC1151 on  March 12, 1991
by the Berkeley Automated Supernovae
Search (Pennypacker et al. 1991). First colors were reported by Hook and
McMahon (1991) about 3 days later. 
V, R, and I photometry were published by
Ford et al.(1993), mainly obtained at the  Leuschner, Whitin, and
VanVleck observatories starting about 9 days after the discovery.
 Note that the very small error bars (Ford et al., 1993)
seem to be somewhat  optimistic. In many instances,
measurements  differ by several sigma even if taken at the same time.
 \begfig 0.1cm
\psfig{figure=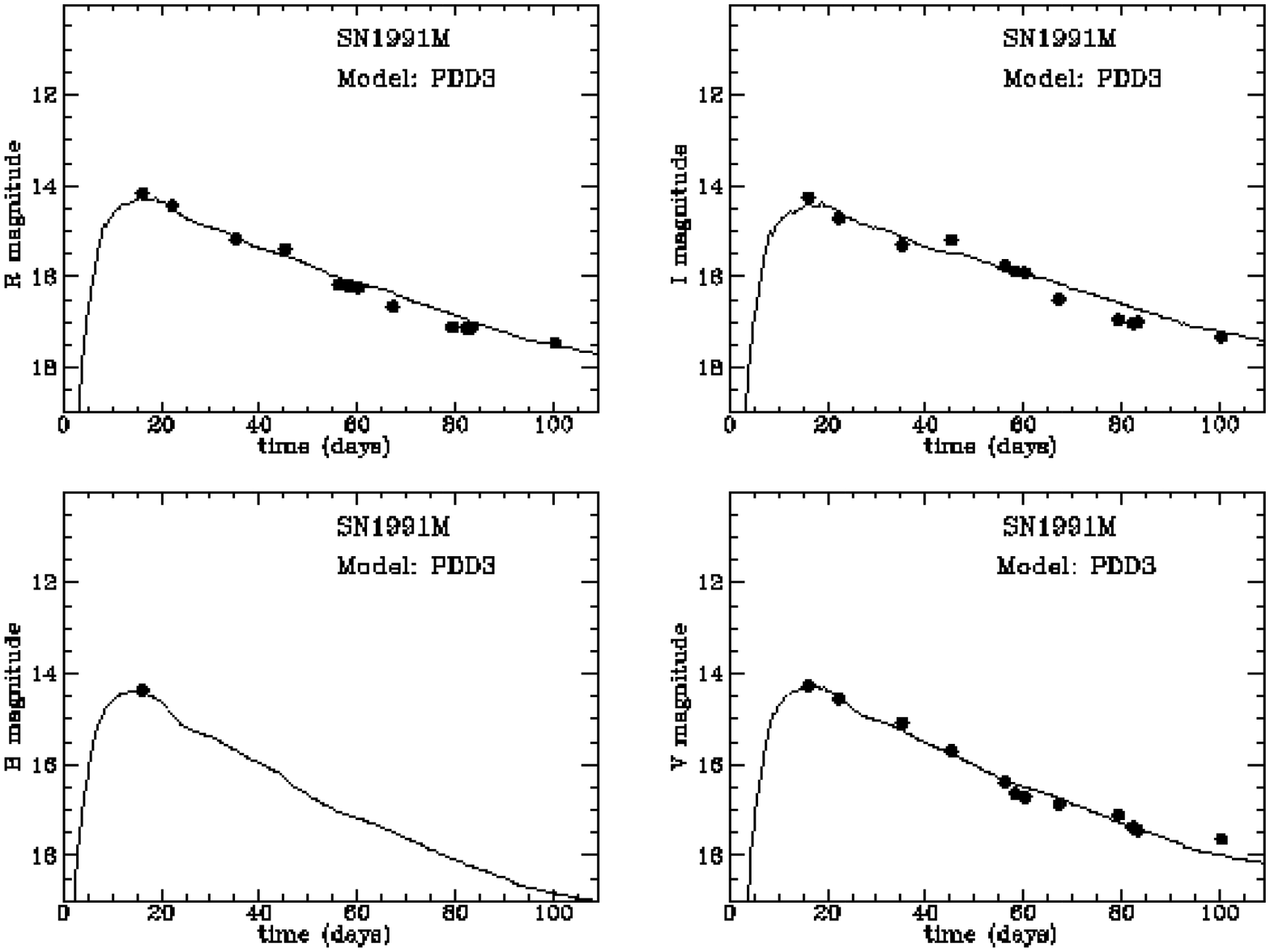,width=12.6cm,rwidth=9.5cm,angle=270}
\figure{22}{Monochromatic light curves of SN~1991M 
compared with the calculated light curve of the
pulsating delayed detonation model PDD3.}
\endfig
 In the infrared, this may be explained by different transmission 
functions.
We assume an observational  error of $0.1^m $ and $0.2^m$ for the V and the
infrared colors, respectively.
 The rather slow post-maximum decline in V    
cannot be reproduced by delayed detonation models in which the transition to the detonation
occurs at low   densities, but requires M35.
 In principle, the bright delayed detonation model N21
reproduces the post-maximum decline, but its IR flux is too small.           
Similar small declines can also be produced  by strongly pulsating delayed detonation models.
The lack of measurements at and before maximum light and of spectra disallow  distinguishing
between the alternative models.
In conclusion, both the optical and IR light curves can be reproduced by
 strongly pulsating model PDD3 (Fig.~22)  or the bright delayed detonation model
M35. The secondary IR maxima in I is too weak in the models. 
A possible explanation may be that the transmission function of I is somewhat broader compared to Bessell's (1990)
calibration, and, consequently,
 the Ca II-IR triplet emission
contributes to the observed flux
in I (HKW95). According to our fits, $E_{B-V} =  
 0.12^m$. This  value should be taken
with caution because it is  based on one B measurement. This uncertainty accounts for a large
error in the distance of $41 \pm 10$~Mpc.

\subsection{ 
 SN~1992G 
}
SN~1992G  was discovered by Shunji Sasaki, Hasaki-machi, and Ibaraki
 on February 9, 1992 in the spiral 
galaxy NGC 3294 at a photographic magnitude of about $14^m$ (Kosai 1992).
 A similar brightness was found on a photographical plate from February 7, 1992 by Sasaki (Kosai
 1992) suggesting that SN1992G was close to maximum light at that time.      
 Our analysis is based on the V, R, and I photometry of
Ford et al. (1993) which provides  excellent coverage of the  
subsequent three months. The B and V colors at about 2 and 4 weeks after the 
discovery have been determined photographically by Tsvetkov (1994).
%
 \begfig 0.1cm
\psfig{figure=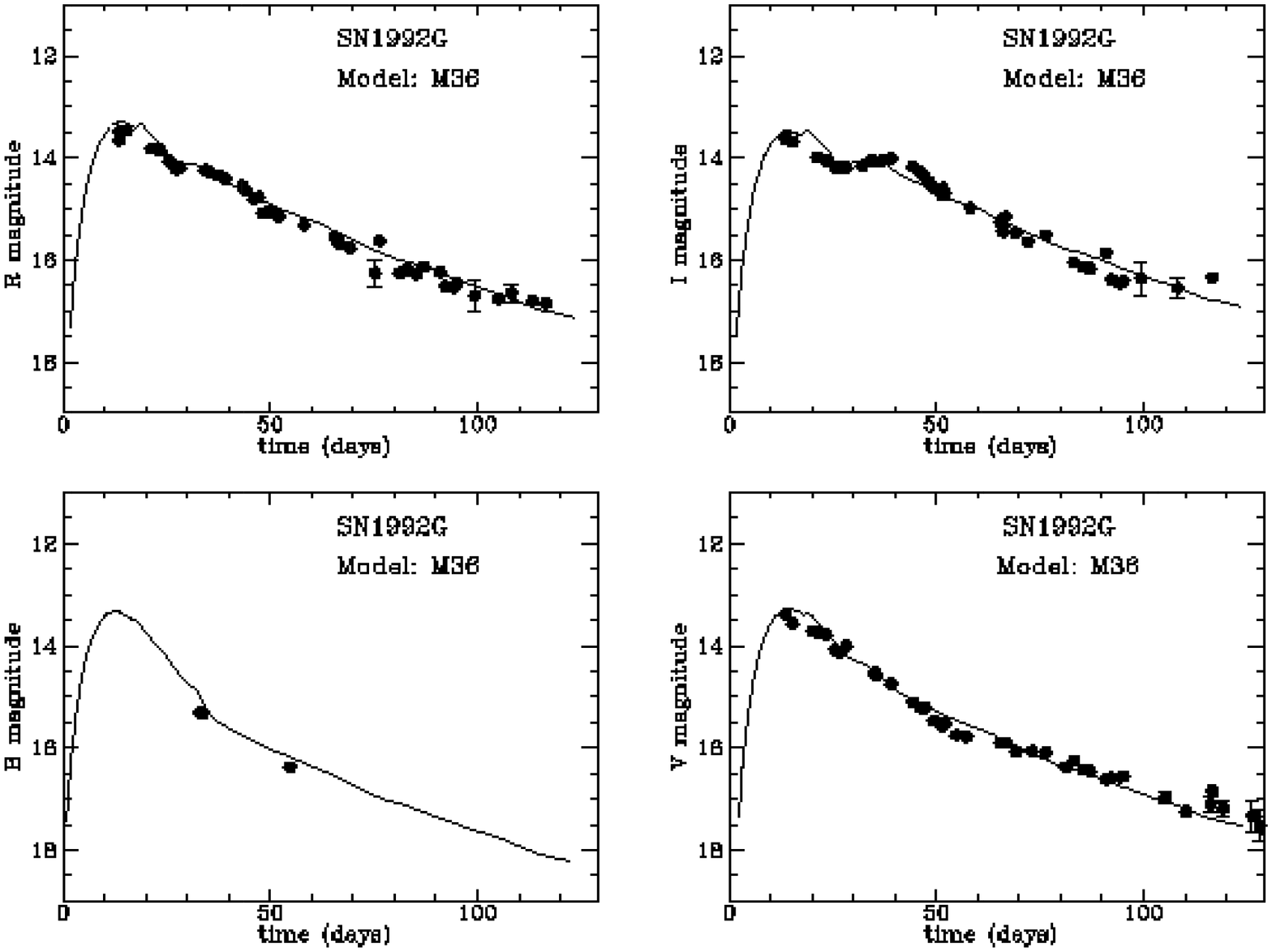,width=12.6cm,rwidth=9.5cm,angle=270}
\figure{23}{Monochromatic light curves of SN~1992G 
compared with the calculated light curve of the
 delayed detonation model M36.}
\endfig
 On February 21, spectra by Iijima, Turatto and Cappellaro  (1992) were taken.
The SiII feature indicated an expansion velocity of about 10000 km/sec.
  The  light curves and  the expansion velocity can be well reproduced by our 
delayed detonation  model M36, although the secondary maximum in I is  too
weak. The discrepancy may be explained by differences in the I filter
(see last section). Besides the strong secondary IR-peak, the observations 
also can be fitted 
by the pulsating delayed detonation model and even  Helium detonation models.
 Clearly, the different models could be distinguished from
the early light curves because the models predict very different slopes
 (see Fig. 5-6).
 Our fits are consistent with a reddening $E_{B-V}=0.05^m$, but this value
is  based on  photographic B colors.
        The distance to  NGC 3294 is $21 \pm 6$~Mpc.
\subsection{ 
 SN~1992K 
}
SN~1992K  was found on March 3, 1992 in  the SBb galaxy ESO- 269-G57 on a photographic
plate  by Antaezana (Hamuy et al. 1994).
 Starting about three days after the discovery, this object was 
continuously monitored during the following 5 months both spectroscopically and in the BVI colors
(Hamuy et al. 1994).
 Spectra taken right after the discovery 
 indicate a photospheric expansion velocity  of 9085 km/sec.
 Both the spectral peculiarities and the red continuum suggest that  SN~1992K is almost
a twin of the strongly subluminous supernova 1991bg (Filippenko et al. 1992ab, Leibgundgut et al. 1993). 
 
 \begfig 0.1cm
\psfig{figure=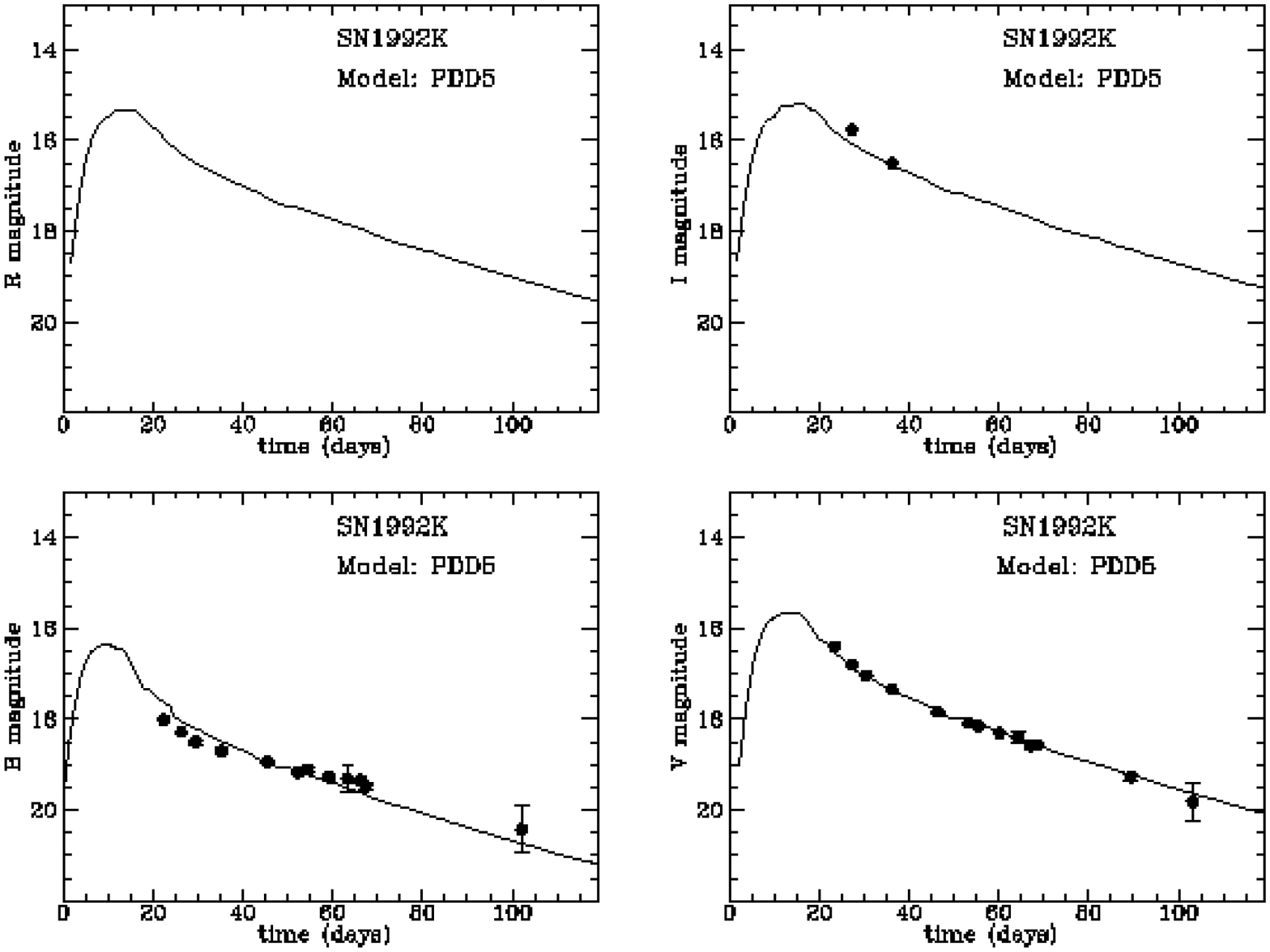,width=12.6cm,rwidth=9.5cm,angle=270}
\figure{24}{Monochromatic light curves of SN~1992K 
compared with the calculated light curve of the
pulsating delayed detonation model PDD5.}
\endfig

 We find that the B and V light curves can be reproduced within the error bars 
by the same strongly subluminous models as SN1991bg. However, PDD5 tends to
 provide a  better fit than  PDD1c.
 The intrinsic  reddening is found to be $0.18^m$, i.e.  larger than  the Galactic
reddening correction in the direction of the host galaxy ($0.04^m$, Burstein \& Heiles 1982). 
    Based on the intrinsic brightness of
models, the distance is $43 ^{+15}_{-8}$~Mpc.  For subluminous SNe~Ia, the brightness
depends sensitively on the actual amount of \ni . This explains the rather large error bars
for this  well-observed object.

\section {Explosion mechanism and type of the galaxy}
It is     interesting to relate the  type of the host galaxy with the explosion
scenario. 
 Here, we  distinguish ellipticals from spirals and irregular
systems. Because, in general, ellipticals consist of stars with main sequence masses
less than $\approx 3 M_\odot$ but, in the latter groups,  massive stars
are also present, this may provide important information on the progenitor 
evolution.
 
 In total, about 18 spirals, 5 ellipticals and 2 irregular galaxies
are in our sample (see Table 3). For the  galaxy in AC 118, no classification is
available. The classical deflagration model W7  and delayed detonation hereafter
 (hereafter group I)  are simultaneously able to reproduce the light curves of
several SNe~Ia. The same is true for pulsating models and envelope
models (group II).  In spirals and irregular galaxies, both
groups are represented, but group I is  more frequent 
than II (12:7). For ellipticals, the relation is reversed (1-2:3). SN~Ia of group II
are highly favored. The possible two candidates (SN~1990Y and SN~1990af) 
of group I in ellipticals are in anonymous galaxies at about 200 to 270 Mpc making
the classification of the type of the galaxy rather uncertain.
 We note that, for bright supernovae,  the mean brightness of  
group II is systematically fainter by about 0.1 to $0.2^m$. 
\begtabsmall       
\table{3}{
 List of observed Type~Ia supernovae used in our sample.
  Columns 2 to 7 give the parent
galaxy, its type,  peculiar velocity $v_z$ according to the MERCG catalogue (Kogoshvili 1986),
the distance modulus of the host galaxy and 
 the color excess $E_{B-V}$ of the SNe~Ia as determined from our models.
In column 8, we give
 names of those theoretical models (see Table~1) which can reproduce
the observed light curves. Models in bracketts do not fullfil our criterion but are
close. These models have been excluded for M-m.}
\baselineskip=12pt
\hline 
\+ Supernovae  &  in galaxy ~~~&Type~~~& $v_z~~~~~$  & M-m~ &D$ [Mpc]~~~~~ $   &  $E_{B-V}$ &  acceptable models                \cr
\hline                     
\+ SN~1937C     & IC 4182  & Sm & 326  &28.3  &   4.5$\pm 1 $   & 0.10 & N32,W7,DET2  \cr 
\+ SN~1970J     & NGC 7619 & dE & 4019  &34.0  &$ 63 \pm 8$     & 0.01 & DET2ENV2/4, (PDD3)  \cr  
\+ SN~1971G     & NGC 4165 & Sb & 1703  &32.8 &$ 36 \pm 9 $    & 0.04  &  M37, M36,  W7, N32 \cr 
\+ SN~1972E     & NGC 5253 & I  & 407  &28.0 &4.0$\pm 0.6 $   & 0.04 & M35,N21,M36 (HeD12 $^2$)   \cr 
\+ SN~1972J     & NGC 7634 & SBO& 3374  &33.6 &$ 52 \pm 8 $    & 0.01 & W7,M36/37,N32,DET2   \cr 
\+ SN~1973N     & NGC 7495 & Sc & 5126  &34.2 &$ 69 \pm 20 $   & 0.08 & N32, M36, W7, (HeD10/11)   \cr
\+ SN~1974G     & NGC 4414 & Sc & 715  &31.2 &$ 17.5 \pm 5 $  & 0.0  & N32,M36,W7,DET2 (HeD10/11)  \cr 
\+ SN~1975N     & NGC 7723 & SBO& 1823  &32.2  & $28 \pm 7 $    & 0.18: &  PDD3/6/9/1a\cr  
\+ SN~1981B     & NGC 4536 & Sb & 1647  &31.4  & 19 $\pm 4 $    & 0.05 &M35, N21  \cr 
\+ SN~1983G     & NGC 4753 & S  & 1126  &31.4 &$ 15 \pm 4 $    & 0.29 & N32, W7 (M36)\cr 
\+ SN~1984A     & NGC 4419 & Ep& 1187  &31.1  &$ 16 \pm 4 $   & 0.14 & DET2ENV2, PDD3/6) \cr 
\+ SN~1986G     & NGC 5128 & I  & 530  &28.1 &  4.2$\pm 1.2 $ & 0.83 & W7, N32, (M37/8)  \cr 
\+ SN~1988U     & AC 118$^1$ &  - & 91480  &40.8 &$1440 \pm 250 $   & 0.05 &     M36, W7, N32  \cr 
\+ SN~1989B     & NGC 3627 & Sb & 726  &29.7 &$ 8.7 \pm 3 $   & 0.45 & M37, M36\cr 
\+ SN~1990N     & NGC 4639 & Sb & 971  &31.5  &$ 20 \pm 5 $    & 0.05 & DET2ENV2/4, PDD3/1a \cr
\+ SN~1990T     & PGC 63925 & Sa & 11980  &36.2   &$ 180 \pm  30 $    & 0.1 & M37, M38  \cr 
\+ SN~1990Y     & anonym.   & E & 11680  &36.4  &$  195\pm  45 $    & 0.05 & W7, N32, M36/37, PDD3/6/1c \cr 
\+ SN~1990af    & anonym.   & E & 14989  &37.1 & $265\pm  85 $    & 0.05& W7, N32, M36 \cr 
\+ SN~1991M     & IC 1151  & Sb & 2188  &33.10 &$ 41 \pm 10 $   &  0.12    & M35,  PDD3\cr
\+ SN~1991T     & NGC 4527 & Sb & 1727   &30.4 &$ 12 \pm 2 $    & 0.10 & PDD3/6/1a. DET2ENV2  \cr 
\+ SN~1991bg    & NGC 4374 &  dE & 954  &31.3  &$ 18  \pm 5 $    &  0.25      & PDD5/1c \cr
\+ SN~1992G     & NGC 3294 &  Sc & 1592  &32.4 &$ 29 \pm  6 $   &  0.05 & M36, M35, PDD3, HeD10 \cr 
\+ SN~1992K     &ESO269-G57 & SBb    & 2908 &33.2 & $43 ^{+15}_{-8} $   &  0.18 & PDD5/1a,(M39, HeD2)\cr 
\+ SN~1992bc    & ESO-G9   &  S  & 5960 &34.6 &$ 83 \pm 10 $   &  0.04 & PDD6/3/1c   \cr 
\+ SN~1992bo    & ESO-G57  &  S  & 5662 &34.5 &$ 79 \pm 10 $   &  0.03 & PDD8  \cr 
\+ SN~1994D     & NGC 4526 &  S0 & 487 &31.1 &$ 16 \pm 2  $   &  0.00 & M36, (W7, N32)   \cr
%
\hline 
\noindent {$^1$ the host galaxy is member of the cluster 118} 
 
\noindent {$^2$ can be ruled because spectra indicate Si at higher        
velocities}
\endtab         
 
 Although our sample is 
still too small to draw final  conclusions, comparisons of this kind may be regarded as
a promising way to get a link to the stellar evolution of the progenitors.

\section {Distance determinations, H$_o$ and  q$_o$}
 The large absolute brightness of SNe~Ia makes them very attractive
candidates to determine extragalactic distances and the cosmological constants.
 The use of SNe~Ia is based on the assumption that their absolute brightness
is known. Unfortunately, SNe~Ia are not standard candles,
but their individual brightness in V may differ by several magnitudes.
 To overcome this problem, several groups have developed methods to allow
for a sorting out of subluminous supernovae on basis of spectral peculiarities
(Fisher et al.  1995) or by using correlations between the shape of the 
blue or visual light curves and the apparent magnitude (Hamuy et al.  1995; 
Ries, Press \& Kirshner 1995). Both approaches allow for the use of SNe~Ia or a subset as 
standard or quasi-standard candles. The first approach has the 
advantage that the measurements can be  corrected for interstellar  reddening because     
a small variation of the intrinsic colors can be assumed for this subgroup
(Miller and Branch, 1994).  Unfortunately, it crucially depends on
the homogeneity hypothesis for 'normal' bright SNe~Ia both in terms of the brightness
and  the intrinsic color. Therefore, the criterion for disregarding observations
must be very strict. In general, the Oklahoma group finds values for $H_o$ of
 about 50 to 60 $km/Mpc~sec$ with some  preference towards the first value
(e.g. Branch \& Miller 1993, Miller \& Branch 1990,1993; but see Fisher et al. 1995).
 Recently, a second approach
has been used by two groups based  on observational data and statistical analyses of 13 
light curves mainly measured  by the CTIO-group (Hamuy \etal  1995). The connections between
the post-maximum decline in B (Phillips \etal  1993) or the shapes of the
visual light curves (Riess \etal  1995)  and the absolute brightness at 
maximum light have been investigated, respectively.                       
  Empirical relations have been found that allow for 
an excellent representation of the observational properties.          
 The relation between the absolute brightness and the
 light curve shapes can be described by a one parameter family within an accuracy
of $0.21^m$ over the entire range of brightnesses observed, even without
any correction for interstellar reddening (Riess et al.  1995).
 This makes SNe~Ia `quasi'-standard candles. 
 Both groups   derive $H_o$ to be 65 and 67  within a one sigma error of
5 and 7, respectively  (Hamuy \etal  1995, Riess \etal  1995).
 
 The biggest 
 advantage  of  purely empirical methods is their straightforward applicability and
their independence from any physical assumption concerning explosion mechanisms.
 One of the disadvantages of empirical methods is their reliance on the accuracy of
secondary distance indicators both for calibration purposes and for testing the relations.
 For a few cases, the period-luminosity relations 
of $\delta $-Cep  stars has been used but, in general, 
the less accurate surface brightness 
fluctuation for galaxies have been used.
 For illustration of possible problems,
  several modern
P(L) relations for $\delta $-Cep  stars are given in Fig. 25. They show a spread
of about $0.5^m $.
Whether it is related to metallicity effects, mixing of stars pulsating in the first overtone and
fundamental mode or based on systematic, observational errors        
calls for further investigations. Another disadvantage is the sensitivity to selection effects.
Based on a larger sample of 35 light curves,  
the uniqueness of the relation
between absolute brightness and shape of light curves has been questioned at the level of
$\approx 0.5^m$ (Hamuy et al. 1995b). Another problem may be that 
 a correction for the interstellar reddening  is not yet taken into account in
the latter methods.  Reddening corrections will 
be included in the near future because, apparent from  a larger set of well-observed  SN~Ia, there
also exists a strong relation between the intrinsic color and the
absolute brightness (Riess \& Kirshner, private communication). From 
the theoretical point of view,
 the intrinsic color and the brightness must be expected to depend on details
such as the  metallicity of the progenitor,
mixing processes etc. because of the strong influence of metallicity on  U and B, but, 
we cannot rule out the existence of a strong relation between intrinsic brightness and color.
 A related problem is the k-correction 
which must be applied for distant SNe~Ia
at $ z \geq 0.025$. Here, the intrinsic blue and UV wavelength ranges  are shifted into the 
V filter range of the  observer. Both from the observations (Hamuy et al.  1993a) and
our models (sect. 3) it is clear that 
 the k-correction in V varies significantly even for `normal' SNe~Ia.

A different, independent way to determine distances is the use of models (MH94) or 
of theoretical relations such as Arnett's law (Arnett et al. 1985)
to obtain the intrinsic brightness. 
 Theoretical approaches do not rely on secondary distance scales although they can be 
tested by those (see next section), and they overcome most of the problems mentioned 
above. However, various uncertainties  should be kept in mind, in particular,
reliance  on the assumption of SN~Ia being standard candles
(see sect. 2 and 3).
 
 Our method is based on fitting observations of individual 
SNe~Ia.  Both the interstellar reddening and k-correction are consistent
 because the overall spectral energy distribution is given by the 
 model.  Doing this, we can determine a parameter range  
 which reproduces the observed monochromatic light curves
(and spectra) or which cannot be excluded from the observations.
  Consequently,
this provides a measure for the allowed ranges of distance for the host galaxy 
and hence values H for the Hubble constant. 
These individual
values are shown in Figure~26.  The error bars for H reflect the
uncertainties in the distance determination arising from model
variations, observational uncertainties, and extinction corrections
(Table 3).
 Our procedure does not rely on the assumption of a particular 
model or even  mechanism because we have shown  (e.g. sect. 3, HKM93, KMH93)
that the intrinsic brightness is related to some other observable quantities 
almost independently from the scenario.

 \begfig 0.5cm
\psfig{figure=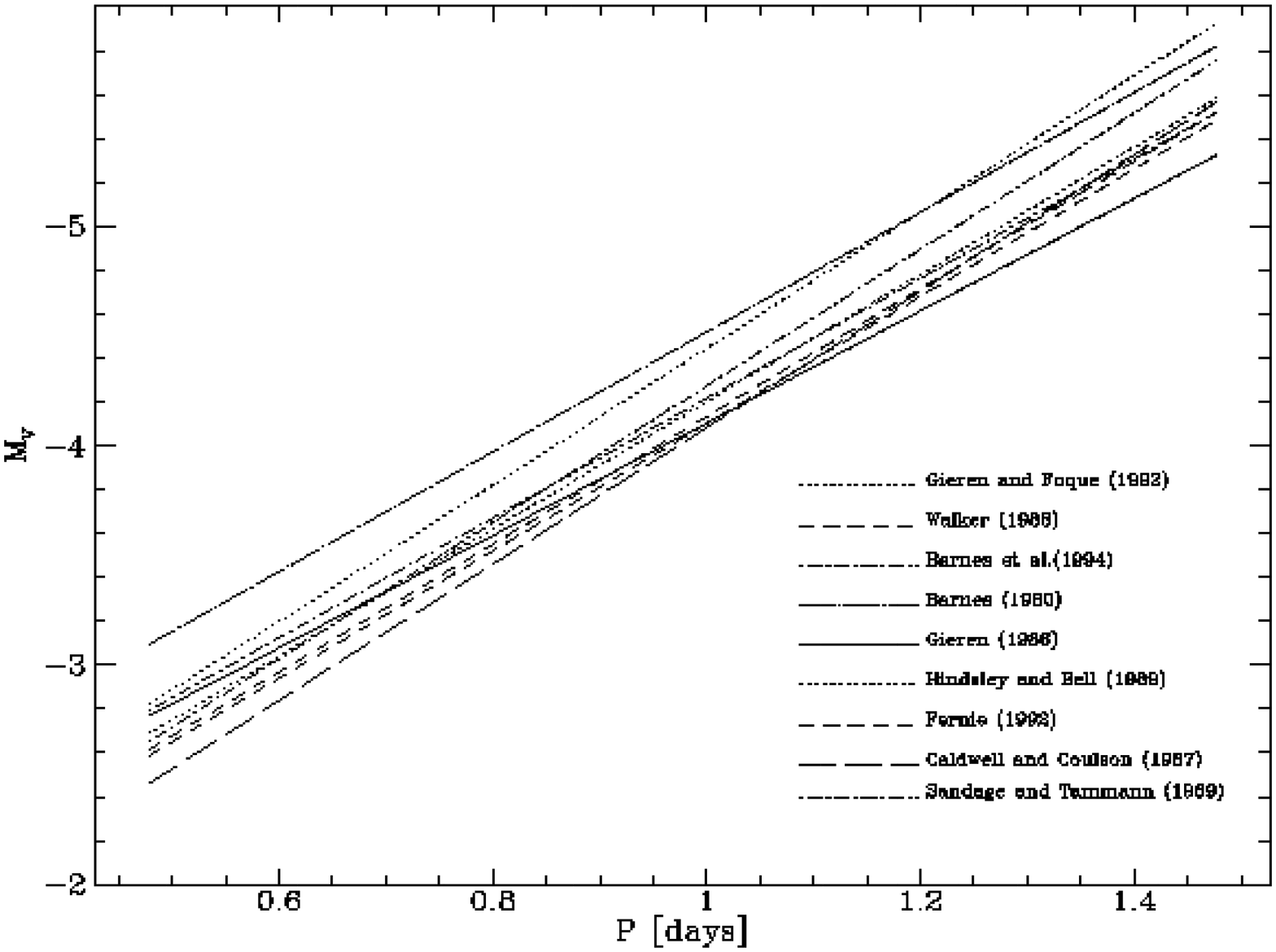,width=12.6cm,rwidth=9.5cm,angle=270}
\figure{25}{Period luminosity relations of $\delta $-Cep stars as given
by various authors. The spread of up to  0.5$^m$ translates into a distance
change of 25 \%.}
\endfig

The individual values H for the Hubble constant $H_o$
 show a wide variation in our sample ranging
from 30 to 127~$km~ s^{-1}~Mpc^{-1}$.  The spread is caused by nearby SNe~Ia with
$v_o \leq 1000$~km/s which, in several cases,
 are inconsistent with the uncertainty
range of mean value \Hbar .
  The scatter of the individual H values for more distant
supernovae ($v_o > 1000$~km/s) is consistent within the statistical error  of 
the mean Hubble constant \Hbar and with  a common value for $H_o$.
 We note that  \Hbar  hardly depends on the exclusion of close supernovae.
Both facts imply that all supernovae with $v_o \leq 1000$~km/s are not
yet in the Hubble flow and that peculiar, random motions  dominate the local velocity  
field.                 
  The remaining systematic deviation is comparable with the dipole-field in the velocity v
found by COBE (Stevens, Scott \& Silk 1993) and the orientation of our residual values agrees
roughly with the COBE results.                
Unfortunately, our sample of high quality data is 
too small ($ \approx 10$) to provide a reliable determination of the size  of the dipole field, but,  
as a tendency, it seems somewhat smaller than implied by COBE.
 We estimate the uncertainty by, in quadrature, the sum over all distant supernovae of  
 the statistical  (table 4) and  systematic errors. Based  on
the comparison between detailed NLTE and LC calculations (see section 2), we find $H_o $ to 
be $67.0 \pm 9$~$km~s^{-1}s~Mpc^{-1}$ within a 2 $\sigma $ error. Alternatively, we can determine the 
error  if we define the 2 $\sigma $ range such that 95 \% of our supernovae are within this
range. If we assume that the individual errors are Gaussian, 
we get the same error range as given above.
It must be remembered that the errors are non-Gaussian, but the consistency between the 
two different error estimates implies that our error range is realistic.
 Our new value of $H_o$ is
 in good agreement with our previous estimate of $66 \pm 10 ~km/s/Mpc]$ that  was  
based on a smaller  group of observations (50 \%) and a subset (30 \% ) of  models
(MH94). Note that our value of $H_o$ is hardly effected by possible problems with 
the six supernovae observed at Asiago because their small statistical weight (due to the 
large error bars) and the fact that the corresponding values  are close to $H_o$.
 Recently, it was suggested that $H_o$ may be as small as $30 km~ sec^{-1} Mpc^{-1}$  on
large scales (Bartlett et al 1995). From our analysis, this suggestion must be excluded, at least
for scales smaller than 200 Mpc. In our sample with $z\leq 0.3$ ,
 $\Delta H_o / \delta v $ is about  $1.~10^{-4} [Mpc^{-1}] $ which is 
consistent with no deviation from a linear Hubble-relation within  
the statistical errors (Table 4). 
\begtab         
\table{4}{Arithmetic mean \Hbar of the individual values H for Hubble constant $H_o$
based on $n$ galaxies (column~8) with expansion velocities larger
than $v_o$ (column~1).  Column~3 gives the standard deviation $2 \sigma_{Gauss}$
of H, if a Gaussian distribution is assumed (see text).
 $\Delta H_o $ is the mean distance between  H and \Hbar and
$\Delta H_1 $ denotes the maximum distance between the error ranges of H and \Hbar. 
  If \Hbar lies within the uncertainty range of H, $\Delta H_1$ is  zero. 
In columns 6 and 7,
$ \delta H_o / \delta v $ and its standard deviation are given for $ z \leq 0.1$. The
values of the Hubble constants and the velocities are given in km/sec~Mpc and km/sec.}
\hline
\+ $v_o$~~~~~~~ & \Hbar~~~~~~~ & $2 \sigma_{Gauss} ~~~~~~$ & $\Delta H_o~~~~~~~~$ & 
 $\Delta H_1$~~~~~~~~~&$\delta H_o/ \delta v $~~~~
& $\sigma (\delta H_o / \delta v)$~&~~~~~ n \cr
\hline 
\+\z5000. &  67.0& 4.   & \z5.0 & \z0. &-6.E-5 & 8.E-5 & \z9 \cr
\+\z1000. &  65.7 & 5.  &  \z7.5 & \z0. &-2.E-5 & 1.1E-4 & \z20 \cr
\+   \z0. &  67.6 & 7. &  \z13. & \z20. &-8.0E-5 & 2.0E-4 &  \z26 \cr
\hline
\endtab

 \begfig 0.5cm
\psfig{figure=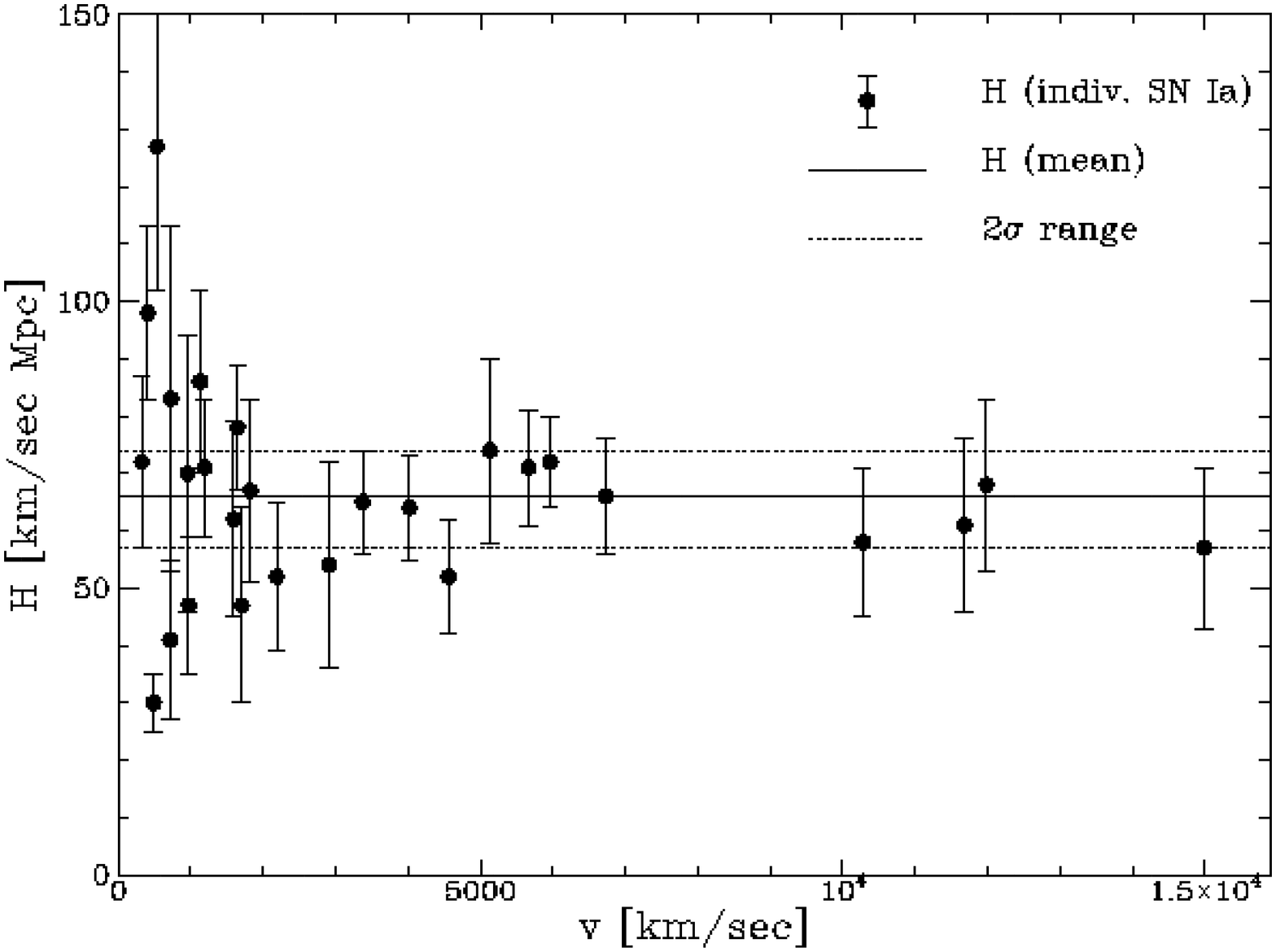,width=12.6cm,rwidth=9.5cm,angle=270}
\figure{26}{Hubble values H are shown based on individual distances
 (see Table 3). SN1988U at v=91500 km/sec gives $H= 64 \pm 10$  [km/sec Mpc].
}
\endfig

Finally, we want to  stress that our value for the Hubble constant agrees well with that
indicated by SNe~II based on the Baade-Wesselink method for SNe~II. Based on sophisticated
atmospheres, which take the dilution of the radiation field by Thomson scattering
 properly into account, $H_o$ was determined to be 
 $65 \pm 12 $ (H\"oflich 1990, 1991) and 
$73 \pm 7 km/sec Mpc$ (Schmidt et al. 1994) with 1 $\sigma $ errors.

 Up to now, Euclidean space was assumed. Undoubtedly, this
 provides a very good description up to cosmological redshifts of $z \leq 0.1 $. 
 In light of the very distant supernova 1988U at $z \approx 0.31$, it 
 is     tempting to constrain the value of the deceleration parameter $q_o$.
 If we assume a Robert-Walker metric and set the cosmological constant to zero,
 the relation between the distance modulus, the Hubble constant, and $q_o$ can
be approximated by 
$$ m-M \approx 25 - 5 log(H_o) + 5 log (cz) + 1.086 (1-q_o) z  \eqno(7)$$ 
for $ z \leq 1.5 $ (e.g. Weinberg 1972).
 Formally, $q_o$ can be determined
independently from the value of $H_o$ because only flux ratios are measured by $q_o$.
 The uncertainties in the measurements enter by the probability functions of 
the individual measurements.
 However, the same errors enter the determination of $q_o$ and $H_o$. To get a 
 conservative estimate, we assume that the errors in  $H_o$ and in the individual
distance of the high red shift supernova are additive.
 Using our value for $H_o$, SN1988U at  z=0.31
(Norgaard-Nielsen et al. 1989) and our distances from Table 3, we get 
 $q_o = 0.7 \pm 0.5 $.

\section{Discussion and Conclusions}
To cover the entire set of scenarios for SNe~Ia, Helium detonations of low-mass 
white dwarfs are  presented, discussed in some detail,
 and included in our study with parameters close to 
those suggested by Woosley and Weaver (1994). Both  bright and subluminous
SNe~Ia can be produced by Helium detonations ranging from $V=$ $-18.5^m$ to $-19.7^m$.
 By fine tuning the mass of the WD and mass of the helium layer atop, Helium detonations
may be able to produce LCs somewhat dimmer
by up to one magnitude if we use our $V(M_{Ni})$
relation (Fig. 8).  The lower limit is given by the critical amount of He
 to be burned in order to trigger a central detonation in carbon.
 The models follow the normal pattern
previously found for other models-- namely,
  the absolute brightness in V at maximum light is rather insensitive to the model
as long as the mass of \ni exceeds $\approx 0.4 M_\odot$.  Helium detonations
 tend to be only
marginally dimmer ($\approx 0.1^m$)  compared to the other scenarios. 
 The differences to WW94 may be 
 a consequence of differences in  the opacities. In our calculations, 
the temperature dependence
of the opacities has been taken into account, whereas WW94
 assumed the opacities to be  independent of chemistry, density, and 
temperature (see sect. 2).
 The distinguishing features of Helium detonations are  blue 
 maxima  both for normal and subluminous
supernovae, a rapid increase of the luminosity
caused by \ni heating, followed by a phase of slow rise to maximum light 
  and a 
fast  post-maximum decline. The fast decline is caused by the rapid increase of the
 escape probability of gamma photons originating in the outer \ni 
and the fast decline rate of \ni .
 
 The different explosion scenarios can generally be distinguished based on differences in
the slopes of the early light curves
 and the photospheric expansion velocities.
 For all models with a \ni production  $\geq 0.4 
M_\odot $, $M_V$ ranges  from -19.1 to $-19.7^m$. 
 The peak brightness strongly declines with \ni for $\leq 0.4 M_\odot $ of \ni.
 As a general tendency, the post-maximum declines
are related to $M_V$,  but there is a significant spread in the decline rate 
 among models with similar
brightess.  For all models but the Helium detonations, the colors become very red for  
small  $M_{Ni}$.
 
 The change of the apparent light curves due to the cosmological expansion has been discussed. Overall, LCs              
 become dimmer with increasing red shift z because the energy is spread over a wider frequency range.
The shapes of the light curves are significantly altered. In particular, the post maximum
decline  in the optical wavelength bands becomes steeper and the ratio between maximum light 
and the radioactive tail increases  with z. 
 
We find that a further z-correction is crucial for the interstellar reddening if  z $\geq 0.1 $.
 If a local extinction law is applied, estimates for the extinction become systematically wrong
by up to a factor of two, depending on the red shift of the absorbing cloud. Even 
reddening  as little as
 $E_{B-V}=0.05^m$ in a host galaxy at z=0.3 translates into an error of about 10\% in  
distance. 
  
 We presented a quantitative method for fitting  data to models based on Wiener filtering 
(Rybicki and Press, 1995). 
The reconstruction technique is applied to the standard deviation from the theoretical LC to overcome
problems with measurements  distributed e unevenly in time.
The direct reconstruction of the light curves
would have to deal with strongly variable correlation lengths. Our ansatz 
is equivalent to the assumption that  the correlation length
of the LC is given by the theoretical model to be fitted. By minimizing the error, 
the time of the explosion, the distance, and the reddening correction can be determined.
 Sadly, we find that the result of all these efforts agree pretty well with our
previous fits based on `eye-balling'.
 
Observed monochromatic light curves of 27 SNe~Ia  are  compared with 
theoretical light curves.
According to our results, normal bright, fast Type~Ia supernovae (\eg SN~1971G,
SN~1994D) with rise times up to 15~days (17~days) for the blue
(visual) light curve can be explained by delayed detonation with
 different densities $\rho_{tr}$ for the transition from a deflagration to
a detonation. For PDDs, the density $\rho_{tr}$ stands for the density
at which the detonation starts after the
first pulsation. Typically, $\rho _{tr} $ is about $2.5~10^7~g/ccm$.
 Central densities of the initial WDs  range from 2.2 to 
   3.5 $10^9$ g/ccm. As a tendency, models at the lower end of this range tend
to give somewhat better fits. 
This may be explained by a high
accretion rate, some variation in the chemical composition or by an additional trigger 
mechanism for the explosion (a little He atop?).
We note that the classical deflagration W7 (Nomoto et al. 1984)
provides similar good fits in several cases because its structure resembles
some of the delayed detonation models.
 
The ``standard" explosion models are unable, however, to reproduce rise
times to blue (visual) maximum longer than 15~days (16~days), provided
the progenitor is a C/O white dwarf of about 1.2 to $1.4~\ms$ (KMH92).     
  In fact, slow rising and declining light curves have been
observed (\eg SN~1990N, SN~1991T) which require models with an envelope of
typically 0.2 to $0.4~\ms$. The envelope can be produced during a strong pulsation
or during the  merging of two white dwarfs.
  The lower
value should not be regarded as a physical limit, because  it is likely
that a continuous transition from models with and without an
envelope exists. Note that a unique feature of this class of ``non-standard" models is
very high photospheric expansion velocities ($v_{ph} \approx
16,000$~km/s) at about maximum light, which drop rapidly to an almost
constant value between 9000 and $12,000$~km/s.  This ``plateau" in
$v_{ph}$ lasts for 1 to 2~weeks depending on the envelope mass 
(KMH93).  In fact, there is some evidence for the plateau in
$v_{ph}$ from the Doppler shift of lines of SNe~Ia which show a slow
pre-maximum rise and post-maximum decline and for which spectra have
been obtained (\eg SN~1984A, SN~1990N) (MH94).
 
Strongly subluminous supernovae (SN1991bg, SN1992K, SN1992bc) can 
be explained within the framework of pulsating
delayed detonation models with a low transition density. 
 In particular, the models become systematically redder and the post-maximum
decline becomes steeper with       decreasing 
brightness (Table 2) in agreement with observations. However, we do not get 
unique relations between these different quantities.
 The evolution
of the photospheric expansion velocity $v_{ph}$ (HKW95) and,
 in particular, its steady decline, is  consistent with observations.
 We must also note that we  need a rather high intrinsic reddening 
 for SN1992K and SN1991bg. Whether this can be explained by 
selective line blanketing, dust formation, or foreground clouds, or a combined effect 
 is under investigation.
 The latter possibility must be regarded as unlikely, because, at later phases, the colors
of these two supernovae are close to those of bright SNe~Ia.
Note that some of the `classical' delayed detonation models (M312, M39) also produce
strongly subluminous LCs, but these do not fit  any of the measurements.     
 
Our Helium detonation models are rather unsuccessful in reproducing observations,
mainly due to the rather steep post-maximum  declines  for normal bright supernovae. 
 For strongly subluminous supernovae,
the blue color at maximum light and the strong IR-maximum are both in contradiction to the
observations. 
Further problems may rise for  late time spectra because,
for subluminous models, a large fraction of the entire \ni expands at rather
high velocities.  
 Quantitatively, multi-dimensional effects may alter the LCs mainly due to a higher       
escape probability  for photons compared to the one-dimensional models.
 To test this, multi-dimensional
radiation transport must be coupled with expansion opacities. Simple remapping of
a 3-D structure would probably miss the main effect.  However, the basic features of the 
LCs and spectra (sect. 2 \& 3) are  not expected to change 
because they are inherent to the outer \ni . 
In fact, the amount of nuclear He-burning  is      expected to be even larger in order to
 trigger the central C/O-detonation because
 the compression wave may be less focused compared to the 1-D case.
 
 Our findings with respect to the explosion scenario can be concluded as follows.
 Models with masses close to the Chandrasekhar limit provide
the best agreement with the observations.
 Delayed-detonation and deflagration models similar
to W7 and pulsation or merging scenarios are required.
 We also have to emphasize that the current models are parameterized. A more
 realistic treatment, e.g. of the burning front,
 may quantitatively change the parameter range of successful models. 
 We do not expect qualitative  changes on the large scale structures of
SNe~Ia because of the time scales.

 We found some evidence for  a relation between the type of the explosion
and of the host galaxy. In elliptical galaxies, SN~Ia  with shell-like structures
are highly favored. These may be understood within the pulsating and merging scenarios.
 Because ellipticals consist on low mass stars only, this may provide a hint to the
progenitor evolution; however, a larger sample of observations is needed to confirm this trend.

Based on our light curves, we have also determined the individual
distances of the parent galaxies of the analyzed SNe~Ia. Our method           
does not rely on secondary distance indicators and allows for a consistent
treatment of interstellar reddening and the interstellar redshift.  The
advantages of a consistent inclusion of information from the spectra has 
been demonstrated for SN1990Y and SN1988U. The advantages are  obvious if the data
sets are rather sparse as  can be expected for distant SNe~Ia.      
 We find $H_o $ to be $67 \pm 9 km/sec Mpc$ within a 95 \% confidence level.
This value agrees well with our previous analysis based on a subset of observations
and models ($66 \pm 10 km~Mpc^{-1} sec^{-1}$, MH94) and is consistent with the result
of Shigeyama et al. (1992) which is based on W7 only.
 A strong variation of the local value can be 
ruled out at least on  scales below 200 Mpc. From SN1988U, the deceleration
parameter $q_o$ is $0.7 \pm 0.5 $. More and 
 more distant supernovae are required to provide better limits. These
measurements are currently under way by the Berkeley collaboration (Perlmutter et al. 1995),
or may be provided as a by-product of deep sky surveys for galaxies if 
designed accordingly. 
  
 We want to mention some other  determinations of $H_o$ which are based on independent,
purely statistical methods and primary distance indicators. The agreement with the value given above 
 may indicate that, at least for SN~Ia, the values obtained by methods which do not
treat SNe~Ia  as  standard candles are converging 
to similar values of $H_o$. Recently, Hamuy et al. (1995) found
$65 \pm 5$,  Riess et al. (1995) give $67 \pm 5$,  and 
Fisher et al. (1995) get a value of $60 \pm 10 km/Mpc sec$.
From our models, both the empirical relations between
the $M_V/dM(15)$ like-relations and the ansatz to deselect subluminous supernovae seems 
to be justified, but we expect a  dispersion  of $\approx 20 \%$ for individual supernovae.

Note that, at maximum light in V, our models predict bolometric corrections 
$BC= M_{bol} -M_V$ between $-0.05^m$ to $0.4^m$
 \footnote{$^1$}{ e.g. $BC (PDD5)=0.4^m$,
 $BC (W7, N32)=0.2^m$, $BC (DET2ENV2) = 0.1^m$, $BC (M35)=-0.03^m$, 
$BC (M36)=-0.05^m$}. 
Much larger  values previously reported in literature  
 ($BC(SN~1981B)=0.57^m$ and ~$BC(SN~1992A)=0.74^m$ at and five days after maximum light,
 respectively,  Nugent et al. 1995a) have been corrected recently by $\approx -0.5^m$ (Baron 1995, 
private communication). Since BC is a critical issue for the determination of $H_o$ from models, 
 we include a more detailed discussion in the   appendix.
 
 A comparison of our estimates for $m_V-M_V$  with those determined
from $\delta $-Cep  stars of the host galaxies provides an additional test for the
reliance of our models and, in particular, the 
monochromatic fluxes. From HST observations of 
$\delta $-Cep stars in IC4182, NGC5253, and NGC4536,
 distance moduli have been found  to be  $28.47 \pm 0.08$, $28.10 \pm 0.07$, and 
$31.17 \pm 0.20$ mag, respectively (Friedman et al. 1994;
Sandage et al. 1994; Tammann 1995) which  
compare well with our estimates of $28.3 \pm 0.25$, $28.0 \pm 0.15$, and 
$31.4 \pm 0.21$ mag, respectively (see Table 3; H\"oflich et al. 1993b, M\"uller \& H\"oflich  1994).

 Finally, we have to bring up several open questions 
which are currently under investigation.
 
 Some well observed SNe~Ia show spectral pecularities.
 Firstly, strong iron lines are observed in premaximum spectra    
of SN1991T (Filippenko  al., 1992a) which may favor helium detonations rather than delayed detonations.
However, both the very long rise
time and the slow decline are in contradiction with helium detonation models and
 a significant amount of radioactive \co and \ni              
in  the outer layers can be excluded from late time spectra because the 
Fe and Co lines indicate velocities $\leq 10,000 km/sec$ (Spyromillo et al., 
1992) without any evidence for \co moving at high velocities ($\approx 14,000 ... 17,000 km/sec$).
 Note that, at late times,
only the local energy deposition due to the $\beta ^+$ decay contributes to the light curve and,
then, both the inner and the outer \co should contribute to the luminosity according
 to their mass ratio.
 A detailed study based on NLTE-model atmospheres is needed, to test
 whether the iron lines can be understood as a consequence of ionization 
equilibria or by an enhanced metallicity in the progenitor (H\"oflich et al. 1995c). 
Secondly, there is some evidence for 
Helium  in SN1994D (Cumming, 1994ab) but, evidently, no enrichment of iron group
elements as in SN1991T (H95).
 The He in SN1994D may be accumulated from the surrounding accretion
disk.  Alternatively, both helium and very little \ni in SN1994D and some  \ni in SN1991T
  may be explained
as following within the scenarios for white dwarfs close to the Chandrasekhar limit:
If the accretion rate drops below the critical rate needed for steady burning,
a little Helium can be accumulated atop the outer layers (e.g. 0.01 $M_\odot$).
 Close to the Chandrasekhar limit, the central density is very sensitive to
the mass. Even this small  increase in mass will cause a contraction and may
trigger a deflagration wave starting in the center,                              
 or  the Helium may detonate and this
may trigger a deflagration rather a detonation in the center.           
 We note, that neither of these scenarios would affect the determination of 
$H_o$ because a little additional energy hardly increases the brightness and even
classical helium detonations are comparably bright.
 
 A  second complex of open questions is related to the very red color observed
in SN1991bg and SN1992K. Whether this can be understood by selective line  
blocking or dust formation is currently under investigation.
 
\bigskip
\noindent
\noindent
 {\sl Acknowledgement:}{PAH would like to thank R.P. Kirshner and his
 group, and R. Hix  for many useful discussions 
and S. Woosley for valuable comments
during the Texas meeting in Munich.  We also thank K.~Nomoto for supplying his
 explosion model W7 in machine readable form. Special thanks go to 
J.C Wheeler for carefully reading the manuscript and valuable  
comments and to
 F.K. Thielemann for providing his marvelous network and many discussions and,
last but not least, Al Cameron for his support including the computer
equipment.                           
 We also want to thank the referee D. Branch, E.M\"uller and B. Press  for useful 
discussions.  We also want to thank G. Avrett, H. Uitenbroek and B. Kurucz
for helpful discussions on the bolometric correction and the solar spectrum.
 This work has been supported by 
grant Ho1177/2-1 from the Deutsche Forschungsgemeinschaft.
}

\vfill
\eject
\noindent
 {\bf Appendix: Remark on  Bolometric Corrections}
 
\noindent
  Following the usual nomenclature, we  define the 
 bolometric correction BC as 
 
 $$BC= M_{bol} -M_V.\eqno(A1)$$
 
 We note that the quantity BC does not depend on the wavelength, but the
use of the V  band  allows for transformation of the bolometric luminosity into the
absolute brightness by  relations given in    literature (e.g. Lang 1980).
 
 In light of recent discussions in literature on the correct size of BC 
at maximum light in V (Nugent et al. 1995b, Branch, private communication),
it may be useful to 
investigate possible systematic uncertainties of our models which  typically  give 
values between 0. and $0.2^m$ for normal bright supernovae.
 
 \begfig 0.5cm
\psfig{figure=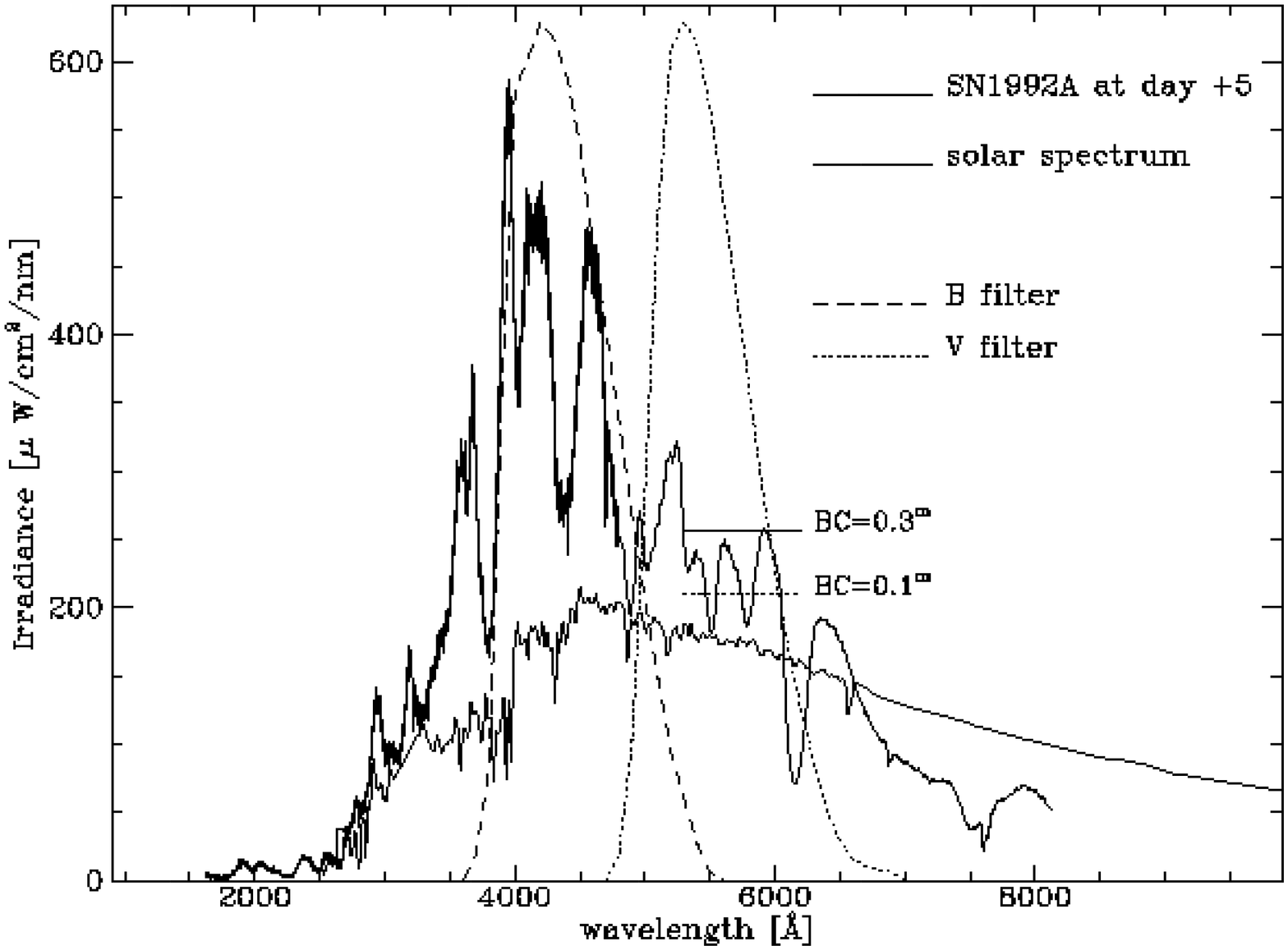,width=12.6cm,rwidth=9.5cm,angle=270}
\figure{A1}{Comparison of the solar flux distribution (thin line, Kohl, Parkinson \& Kurucz 1992, Neckel \& Labs 1984,
Labs \& Neckel 1970, White 1977, Avrett 1992)
with the observations of SN1992A at about 5 days past maximum light (thick line, Kirshner et al. 1993).
 Both spectra are normalized to the bolometric irradiance of the sun measured in $\mu W /cm^2/nm$.
 In addition, the B and V filter functions are given  according to Bessell (1990).
The horizontal lines at about 5500 \AA ~give the mean flux level in V which would be required for
 BC of $0.1^m$ (dots) and $0.30^m$ (dashed dotted, new value of Nugent 1995 (private communication) based
on spectral fits (Nugent et al. 1995b)).
}
\endfig

 Based on our NLTE-analysis  of the blue SN1994D (H95) with $BC \approx 0.^m$, 
Branch (private communication) concluded that we may 'waste' UV flux and systematically
underestimate BC.  Admittedly, our calculated UV flux 
is rather uncertain due to the approximate treatment of line blanketing and scattering,
 etc., although we include several  effects which, generally, are omitted 
 (Kirshner et al. 1993, Harkness et al. 1995, Baron et al. 1995) -- namely the fact that 
diffusion time scales at the photosphere are comparable to the expansion time scales.
However, despite these uncertainties, strong limits on the systematic errors can 
be derived from the energy conservation.
  At maximum light, the UV flux is smaller than $\approx 20 \%$ of the total flux 
in all models. This implies a strict upper limit of $0.20^m$ for the error in BC,
if we assume that  the entire UV flux is converted into optical photons. 
To provide a larger BC, flux from the U and
 B bands  has to be converted. This happens soon after maximum light 
because  line blanketing is rapidly increasing in U and B.
 Consistent with the observations of SNe~Ia,
the color indices become redder with time and our predicted BC  increases    
 to $0.1$ to $0.5^m$ on time scales of one to two  weeks depending on the model.
 Note that, from the observations, a strong change of BC must be expected because the   
bolometric, B and V light curves do not peak at the same time. Consequently, the use
of BC is very error-prone when it comes to the determination of absolute fluxes at 
maximum light. In contrast, using $M_V$ and fitting the entire LC provides a robust procedure 
to determine absolute fluxes (e.g. in V)  because it relies on the known total
energy release by radioactive decay and the well measurable shape of the LC.
  
 Another estimate of the correct value of BC can be obtained directly from observations.
 Some pre-maximum IUE observations are available for SN1990N 
and SN1981B, and very accurate HST measurements are available for SN1992A at 
about 5 days after the B maximum  (Kirshner et al. 1993, Nugent et al. 1995a). The observed spectrum 
of SN1992A is given in comparison to the observed solar spectrum in Fig. A1.
 Both spectra are normalized to the same bolometric luminosity. Therefore, a comparison
of the flux in the V band provides BC. In other wavelength bands (e.g. B),
the difference between the fluxes is given by the sum of BC and the intrinsic color indices (e.g. B-V) compared   
to the solar values.
 For the normalization of the spectrum of SN1992A, we assume that the IR-flux of SN1992A becomes small
above 1 $\mu m$. The corresponding uncertainty  in BC is $\approx 0.1^m$. We note that,
 historically, the bolometric corrections are  normalized
to a G2V star but, nowadays, $BC(Sun)= -0.07^m$ of the sun
is used as reference point (Lang 1980, Avrett 1992).
 Only 8 \% of the solar radiation 
is emitted in the UV and the solar flux peaks in the optical wavelength band (Fig. A1).
Therefore, BC is negative for all main sequence stars. 
  The UV flux of SN1992A is comparable to
the sun, making it impossible to produce a large BC correction by redistributing UV photons.
Instead, the flux in B is increased at the expense
of the IR-flux producing the blue color index in SN~Ia ($B-V (SN1992A)= 0.1^m$ vs. $B-V (Sun)=0.67^m$).
  Physically, the reduced IR-flux in supernovae is caused
by  the smaller free-free emission as a consequence of the lower densities at the photosphere.
 Apparently, the gain in V is very moderate. Most of the flux difference in the B band can be attributed to the
differences in the intrinsic colors. Both
for the B and V band, BC is between $0.$ and $0.2^m$ which is consistent
with our models for 'normal' bright supernovae.
 
 Note that the new value of BC $0.30^m$ for SN1992A 
(Nugent 1995, private communication) seems to be slightly too high compared 
to the observations (Fig. A1). However, this discrepancy should not be overvalued because, for 
atmospheres, small inconsistencies in the physical and numerical approximations
will cause errors in the radial distribution of the density, abundances, and temperature,
which enter quadratically the absolute flux (H95).

\vfill
\eject
\vfill
\eject
\heading{References}

\journal {Arnett  W. D.}{1969}{Ap. Space Sci.}{5}{280}

\inbook  {Arnett W.D.}{1982}{Supernovae}{}{Dordrecht}{D.Reidel}{221}

\journal {Arnett W.D., Branch D., Wheeler  J.C.}{1985}{Nature}{314}{337}
 
\inbook {Avrett G.}{1992}{Proceedings of the Workshop on the Solar Electromagnetic
Radiation Study for  Solar Cycle 22}{R.F. Donnelly}{ NOAA ERL}{Springfield}{20}
 
\journal {Barbon  R., Ciatti  F., Rosino  L.}{1973}{A\&A}{29}{57}
 
\journal {Barbon  R., Benetti  S., Cappellaro  E., Rosino  L., Turatto  M.}
         {1990}{A\&A}{237}{79}
 
\journal {Barbon  R., Cappellaro  E., Turatto  M.}{1989a}{A\&AS}{81}{421}
 
\journal{ Barbon  R., Iijima  T., Rosino  L.}{1989b}{A\&A}{220}{ 83}
 
\journal {Barnes  T. G. I., Moffett T.J., Gieren W.P.}{1994}{ApJ}{405L}{51}
 
\journal {Barnes  T. G. I. }{1980}{JAtSc}{37}{2002}
 
\journal {Baron E., Hausschild P.H., Branch D., Austin S., Garnavich P., Ann H.B.,
Wagner R.M., Filippenko A.V., Matheson T., Liebert J. }{1995} {ApJ}{441}{170}
 
\infuture {Bartlett J.G., Blanchard A., Silk  J., Turner M.S.}{1995}{Nature}{submitted and
astro-ph/9407061}
\journal {Benz W., Thielemann F, Hills J.G. }{1989}{ApJ}{342}{986}
 
\journal {Bessell M.} {1990,} {PASP} {102} {1181}
 
\journal {Branch  D.}{1981}{ ApJ }{248}{1076}
 
\inbook  {Branch D., Tammann G.A}{1992} {Type Ia Supernovae as Standard
 Candles} {S. van den Bergh}{Berlin/Heidelberg/New York }{Springer Press, Ann.Reviews}{267}
 
\journal {Branch D.}{1993}{ApJ}{405}{L5}

\journal{ Burstein  D., Heiles  C.}{1984}{ApJS}{54}{33}

\journal{ Cadonau  R., Leibundgut  B.}{1990}{A\&AS}{82}{145}
 
\journal {Caldwell  J. A. R., Coulson  I. M. }{1987}{AJ}{93}{1090}                                    

\infutbook{ Canal  R.}{1995}{ Proc. of Les Houches Session 
LIX}{ J. Audouze, S. Bludman, R. Mochovitch, J. Zinn-Jutin}{Paris}{ Elsevier 
Science Publishers}{in press}
 
\journal { Capaccioli  M., Cappellaro  E., Della Valle  M., D'Onofrio  M.,
Rosino  L., Turatto  M.}{1990}{ApJ}{350}{110}

\privcom {Challis P.}{1994}{private communication}
 
\journal{ Ciatti  F., Rosino  L.}{1978}{A\&AS}{34}{387}
 
\journal{Collela  P., Woodward  P.R.}{1984}{J.~Comp.~Phys.}{54}{174}
 
\journal {Cowan J.J., Thielemann F.-K., Truran J. W.}{ 1991}{Phys. Rep.}{208}{267}

\journal {Cristiani  S., Cappellaro, E., Turatto  M., Bergeron  J., Bues  I.,
Buson  L., Danziger  J., Di Serego-Alighieri  S., Duerbeck  H.W.,
Heydari-Malayeri  M., Krautter  J., Schmutz  W., Schulte-Ladbeck 
R.E.}{1992}{A\&A}{259}{63}
 
\circular {Cumming R.J.} {1994a} {IAU-circular} { 5953}
 
\circular {Cumming R.J.}{1994b}{IAU-circular} {5951}
 
\journal{Della Valle  M., Panagia  N.}{1992}{AJ}{104}{696}
 
\infuture { Dominik  C., H\"oflich P., Khokhlov A.}{1995} {ApJ} { in preparation}
 
\journal {Fernie  J. D }{1992}{AJ}{103}{1647}                                    

\journal {Filippenko  A.V., Richmond  M.W., Matheson  T., Shields  J.C.,
Burbidge  E.M., Cohen  R.D., Dickinson  M., Malkan  M.A., Nelson  B.,
Pietz  J., Schlegel  D., Schmeer  P., Spinrad  H., Steidel  C.C.,
Tran  H.D., Wren  W.} {1992a} {ApJ} {384} {L15}

\journal{Filippenko  A.V., Richmond  M.W., Branch  D., Gaskell  C.M.,
         Herbst  W., Ford  C.H., Treffers  R.R., Matheson  T., Ho  L.C.,
          Dey  A., Sargent  W.L., Small  T.A., van Breugel  W.J.M.}
         {1992b} {AJ}{104}{1543}

\infuture {Fisher  A., Branch  D., H\"oflich P., Khokhlov A., Wheeler J.C}{1995}{ApJ Let} {in press }
 
\journal{Ford  C.H., Baker  M.L., Filippenko  A.V., Treffers  R.R., Paik  Y., Benson  P.J.}
{1993}{AJ}{106}{1101} 
 
\journal{Freedman, W. L.;Madore, B. F.; Mould, J. R.;Hill, R.; 
Ferrarese, L.;Kennicutt, R. C.; Saha, A.; Stetson, P. B.; Graham, J. A.;
Ford, H.; Hoessel, J. G.; Huchra, J.; Hughes, S. M.;
Illingworth, G. D.}{1994}{Nature}{371}{75}
 
\journal 
{Frogel J.A., Gregory B., Kawara K., Laney D., Phillips M.M.,
Terndrup D., Vrba F., Whitford, A.E.}{ 1987}{ ApJ}{ 315}{ L129}

\journal {Gieren W.P., Barnes  T. G. I., Moffett  T.J.}{1993}{ApJ}{418}{135}
 
\journal {Gieren W.P., Foque P.} {1992} {AJ} {106} {734}
 
\journal {Gieren W.P.} {1986} {ApJ} {306} {25}
 
\journal {Hachisu I., Eriguchi Y., Nomoto K.}{1986a}{ApJ}{308}{161}

\journal {Hachisu I., Eriguchi Y., Nomoto K.}{1986b}{A\&A}{168}{130}

\journal {Hamuy M. \etal} {1993a} {AJ} {106} {2392}

\journal {Hamuy M. \etal} {1993b} {PASP} {105} {787}

\journal {Hamuy M. \etal} {1994} {AJ} {108} {2226}
 
\infuture {Hamuy M., Phillips M.M, Maza J., Suntzeff N.B., Schommer R.A., Aviles A.} {1995} {AJ} {in press}

\infuture {Hamuy M., H\"oflich P., Khokhlov A., Wheeler J.C.} {1995b} {ApJ Let} {in preparation}

\journal {Hansen  C. J., Wheeler J. C.}{1969}{Ap. Space Sci.}{3}{464}

\journal {Hindsley  R.R., Bell R.A.} {1989} {ApJ} {341} {1004}
 
\book{H\"oflich, P.}{1990}{Habilitation Thesis, Ludwig Maximilians Univ.}
{M\"unchen}{published as MPA-90 563}
 
\journal {H\"oflich  P., Khokhlov, A., M\"uller, E.} {1992} {A\&A} {259} {549}
 
\inbook  {H\"oflich P.}{1991}{Supernovae}{S. Woosley}{New York}{Springer-Press}{415}

\journal {H\"oflich  P., Khokhlov  A., M\"uller  E.} {1991a} {A\&A} {259} {243}

\journal {H\"oflich  P., Khokhlov  A., M\"uller  E.} {1991b} {A\&A} {248} {L7}
 
\journal {H\"oflich  P., M\"uller  E. \&  Khokhlov  A.} {1993} {A\&A} {268} {570}
 
\infutbook {H\"oflich P., M\"uller, E., Khokhlov A.,}{1993b}{Supernovae and Supernovae Remnants}
{D. McCray, Z. Wang \& Z. Li}{Xiang}{Cambridge University Press}{in press}

\journal {H\"oflich  P., Khokhlov  A., Wheeler  J.C.} {1995} {ApJ} {in}  {press}
 
\journal {H\"oflich P.}{1995}{ApJ}{443}{533}
 
\inbook {H\"oflich P., M\"uller E., Khokhlov A., Wheeler C.J.}{1995b}{$17^{th}$ Texas Symposium on
Relativistic Astrophysics}{Tr\"umper}{Munich}{Springer}{in press}

\infutbook{Ho\"flich P., Khokhlov A., Nomoto K., Thielmann F.K., Wheeler C.J.}{1995c}
{Type Ia Supernovae}{Barcellona}{R. Canal, J. Isern, P. Ruiz-Lapuente}{Kluver}{Amsterdam}{in preparation}

\circular{Hook  I., McMahon  R.} {1991} {IAU-Circular} {5218}
 
\journal {Hoyle  F.,   Fowler  W. A.}{1960}{Ap. J.}{132}{565}

\journal{ Iben  I.Jr., Tutukov  A.V.}{1984}{ApJS}{54}{335}
 
\circular {Iijima T., Turatto M., Cappellaro E.}{1992}{IAU-Circular}{5456 \& 5458}

\journal {Ivanova  I. N., Imshennik  V. S.,     Chechetkin  V. M.}{1974}
          {ApSS}{31}{497}
 
\journal {J\"oeever  M.}{1982}{Afz}{18}{574}
 
\journal {Karp  A.H., Lasher  G., Chan  K.L., Salpeter E.E.} {1977}{ApJ}{214}{161}

\inbook {Kogoshvili  K.} {1986/1993}{Selected Astronomical Catalogs CD Vol. I}{L.E. Brotzman \&
S.E. Gessner}{Greenbelt}{National Space Science Data Center}{}

\journal {Khokhlov  A.}{1991a}{A\&A} {245} {114}

\journal {Khokhlov  A.}{1991b}{A\&A} {245} {L25}

\journal { Khokhlov  A., M\"uller  E., H\"oflich  P.}{1992}{A\&A}{253}{ L9}

\journal {Khokhlov  A., M\"uller  E.,   H\"oflich, P.} {1993} {A\&A} {270} {223}
 
\journal {Kirshner, R. P. et al.}{1993}{ApJ}{415}{589}

\journal {Knude  J.K.} {1977}{ApJ.Let.}{18}{115}
 
\book {Kohl J.L., Parkinson W.H., Kurucz B.L.}{1992}{Center and Limb Solar Spectrum in 
High Spectral Resolution from 225.2 nm to 319.6 nm}{National Bureau of Standards}{Boulder}

\journal {Kosai, J.} {1992} {IAU-circular}{5452}{}
 
\journal {Labs D., Neckel H.}{1970}{Sol. Phys.}{15}{79}

\inbook {Lamla  L.}{1982} {Landold B\"ornstein, New Series B 2a}{}{}{} {54}
 
\book {Lang C.}{1980}{Astrophysical formulae} {Springer Press}{Berlin/Heidelberg/New York}
 
\journal {Leibundgut  B., Kirshner  R.P., Filippenko  A.V.,
          Shields  J.C., Foltz  C.B., Phillips  M.M., Sonneborn  G.}
          {1991}{ApJ}{371}{L23}

\journal {Leibundgut  B. \etal}{1993}{AJ}{105}{301}

\journal {Livne E.}{1990}{ApJ}{354}{53}

\journal {Livne E., Glasner A.S.}{1990}{ApJ}{361}{L244}
 
\circular { Marvin  H., Perlmutter  S.}{1989}{IAU-Circular}{4727}

\journal {Maza  J., Hamuy  M., Phillips  M.M., Suntzeff  N.B., Ayiles  R.}
         {1994}{ApJ}{424} {L107} 
 
\journal {Miller  D.L., Branch  D.} {1990} {AJ}{103} {379}
 
\book {Mihalas D.}{1978}{Stellar atmospheres}{Freeman}{San Francisco}
 
\journal {Miller  D.L., Branch  D.} {1990} {AJ}{100} {530}
 
\inbook {M\"uller  E., H\"oflich  P.}{1991}{SN1987A and Other
Supernovae}{I.J. Danziger and K.~Kj\"ar}{Garching}{ESO}{379}

\journal {M\"uller  E.,   H\"oflich  P.}{1994} {A\&A} {281} {51} 
 
\journal {Neckel H., Labs D.}{1984}{Solar Phys.}{90}{205}

\journal {Nomoto  K., Sugimoto  S., \& Neo  S.} {1976}{ApSS}{39}{L37}
 
\journal {Nomoto  K., Sugimoto  D.}{1977}{PASJ}{29}{ 765}
 
\inbook {Nomoto K.} {1980} {IAU-Sym. 93}{D. Sugimoto, D.Q. Lamb \& D. Schramm}{Dordrecht}{Reidel}{295}

\journal { Nomoto  K.}{1982}{ApJ}{253}{ 798}
 
\journal{Nomoto  K., Thielemann  F.-K., Yokoi  K.}{1984}{ApJ}{286}{ 644}
       
\infutbook { Nomoto  K., Yamaoka  H., Shigeyama  T., Iwamoto  K.}{1995}{Supernovae 
and Supernova Remnants}{R. A. McCray, Z. Wang \& Z. Li}{Cambridge}{Cambridge University
Press}{ in press}
 
\journal {Norgaard-Nielsen  H.U., Hansen L., Henning E.J., Salamanca A.A.,
 Ellis R.S., Warrick, J.C.}{1989}{Nature}{339}{523}

\journal {Nugent P., Baron E.,Hauschild P., Branch D.}{1995a}
{ApJ Let.}{ 441}{L33}
 
\infuture {Nugent P., Branch D., Baron E., Fisher A., Vanghan T.}{1995b}
{Phys. Rev. Lett.}{ submitted}

\inbook {
Paczy\'nski  B.}{1985}{in: Cataclysmic Variables and Low-Mass X-Ray
Binaries}{D.Q. Lamb, J. Patterson}{ Reidel}{Dordrecht}{ 1}
 
\circular {Pennypacker  C. et al.}{1991}{IAUC}{5207}

 \journal  {Perlmutter C. et al.}{1995}{ApJ Let} {440}{95}

\journal {Phillips  M. M.}{1993}{ApJ}{413}{L108}

\journal{Phillips  M.M., Phillips  A.C., Heathcote  S.R., Blanco  V.M.,
         Geisler  D., Hamilton  D., Suntzeff  N.B., Jablonski  F.J., Steiner 
         J.E., Cowley  A.P., Schmidtke  P., Wyckopf  S., Hutchings  J.B.,
         Tonry  J., Strauss  M.A., Thorstensen  J.R., Honey  W., Maza  J.,
         Ruiz  M.T., Landolt  A.U., Uomoto  A., Rich  R.M., Grindlay  J.E.,
         Cohn  H., Smith  H.A., Lutz  J.H., Lavery  R.J., Saha  A.}{1987}
        { PASP}{90}{592}

\journal {Pskovskii  Yu.P.}{1970}{Astron. Zh.}{47}{994}

\journal {Pskovskii  Yu.P.} {1977} {Sov. Astr.} {21} {675}
 
\journal {Riess A.G., Press W.H., Kirshner R.P.}{1995}{ApJ}{438}{L17}
 
\journal {Ruiz-Lapuente  P., Jeffery D., Challis P.M., Filippenko  A.V.,
Kirshner R.P., Ho L.H., Schmidt B.P., Sanchez F., Canal R. }{1993}{Nature}{365}{728}

\infuture {Rybicki G.B., Press W.H.} {1995} {Phys.Rev.Let.} {in press}
 
\journal {Sandage  A., Tammann  G.A.}{1969} {ApJ} {157}{683}
 
\journal {Sandage  A., Tammann  G.A.}{1993} {ApJ} {415}{1}
 
\journal {Sandage A, Saha A., Panagia N., Panagia N., Macchetto F.D.}
{1992}{ApJ}{401} {L7}
 
\journal {Sandage A, Saha A., Tammann G.A., Labhardt L., Schwengler H., Panagia N., Macchetto F.D.}
{1994}{ApJ}{423} {L13}
 
\journal {Schaller G., Schaerer D., Meynet G., Maeder A.}{1992}{A\&A Suppl}{96}{269}
 
\journal {Schmidt  B.P. et al.}{1994} {ApJ} {432} {42}
 
\journal {Shigeyama T., Nomoto K., Yamaoka H., Thielemann F.K. }
 {1992} {ApJ}{386}{L13}
 
\journal {Spyromilio J., Meikle W.P.S.,
Allen D.A., Graham J.R.}
{1992} {MNRAS} {258} {53p}

\journal {Stevens  D., Scott  D., Silk  J.} { 1993} {PhRvL} {71} {20S}

\inbook{Thielemann  F.-K., Nomoto  K., Hashimoto  M.}{1994}
{Supernovae}{ Les Houches}{S. Bludman  R., Mochkovitch  J., Zinn-Justin}{Elsevier}{Amsterdam}{629}
 
\inbook{ Thielemann  F.-K., Arnould  M., Truran  J.W.}{1987}
{ Advances in Nuclear Astrophysics} {E. Vangioni-Flam}{Editions fronti\`eres}{Gif sur Yvette}{525}

\journal {Tsvetkov, D.Yu} {1994}{Astron.L.}{20}{374}

\journal {Van den Bergh  S., Pazder J.}{1992}{ApJ}{390}{34}
 
\journal {Walker, A.R.} {1988} {PASP} {100} {949}

\journal{Webbink  R.F.}{1984}{ApJ}{277}{355}
 
 \book {Weinberg, S.} {1972}{Gravitation and Cosmology: Principles and Applications of the 
General Theory of Relativity}{New York}{John Wiley \& Sons}

\journal {Wells  L.A. + 45 coauthors} {1994} {AJ} {108}{2235}  

\journal {Wheeler  J. C., Harkness  R .P.}{1990}{Rep. Prog. Phys.}{53}{1467}

\infuture {Wheeler J. C., Harkness R .P., Khokhlov A., H\"oflich P.}
{1995}{Phys. Rep.}{in press}
 
\book {White O.R.}{1977}{The SSolar Output and its Variation}{Colorado Associated Univ. Press}{Boulder}

\journal {Woosley  S.E., Weaver  T.A.}{1986}{ARAA}{24}{}

\infutbook{ Woosley  S. E. \& Weaver , T. A.}{1995}{ Proc. of Les Houches Session 
LIX}{ J. Audouze, S. Bludman, R. Mochovitch, J. Zinn-Jutin}{Paris}{ Elsevier 
Science Publishers}{in press}

\journal {Woosley  S. E. \& Weaver, T. A.} {1994}{Ap. J.} {423}{371}

\inbook {Woosley S. E., Weaver T.A., Taam R.E. } {1980} {in: Type I Supernovae}{C.Wheeler}{Austin}{U.Texas}{96}

\journal {Yamaoka H., Nomoto K., Shigeyama T., Thielemann F.}{1992}{ApJ}{393}{55}

\bigskip

\end